\documentclass[a4paper,titlepage,openright]{report}

\usepackage[latin1]{inputenc}

\usepackage{german}
\usepackage{a4wide}
\usepackage[amssymb]{SIunits}
\usepackage{amssymb}

\usepackage{graphicx}
\usepackage{lscape}
\usepackage{enumerate}

\usepackage{url}

\usepackage[dvips]{hyperref}
\usepackage{breakurl}

\begin{document}

\begin{titlepage}

\thispagestyle{empty}

\begin{center} \bf  \large

Trajectories of Rubber Balloons used in Balloon Releases:  \\
Theory and Application
\\*[1cm]
Trajektorien von Gummiballons in Ballonwettbewerben: \\
Theorie und Anwendung
\\*[1cm]
\large
Patrick Glaschke
\\*[1cm]
\end{center}

\begin{center} \bf  \large
Abstract/Zusammenfassung
\end{center}

\vspace{2cm}

\noindent
Balloon releases are one of the main attractions of many fairs. Helium filled rubber balloons
are released to carry postcards over preferably long distances. Although such balloons have
been considered in atmospheric sciences and air safety analysis, there is only scarce literature
available on the subject. This work intends to close this gap by providing a comprehensive
theoretical overview and a thorough analysis of real-life data. All relevant physical properties
of a rubber balloon are carefully modelled and supplemented by weather observations to form a
self-contained trajectory simulation tool. The analysis of diverse balloon releases
provided detailed insight into the flight dynamics and potential optimisations. Helium balloons
are found to reach routinely altitudes above 10~km. Under optimal conditions, they could stay
more than 24 hours airborne while reaching flight distances close to 3000~km. However, external
weather effects reduce the typical lifetime to 2\,--\,5 hours.

\vspace{2cm}

\noindent
Ballonwettbewerbe sind eine der Hauptattraktionen von vielen Volksfesten. Heliumgefüllte Gummiballons
werden mit einer Postkarte gestartet, um eine möglichst große Strecke zurückzulegen. Obwohl solche
Ballons in der Atmosphärenforschung und der Luftfahrtsicherheit eine gewisse Rolle gespielt haben,
gibt es nur spärlich Literatur zu diesem Themenkreis. Diese Arbeit versucht diese Lücke mit einem
umfassenden theoretischen Überblick und einer sorgfältigen Analyse von realen Meßdaten zu schließen.
Alle relevanten physikalischen Eigenschaften eines Gummiballons werden sorgfältig modelliert und
mit Wetterbeobachtungen ergänzt, um ein abgeschlossenes Programm zur Trajektoriensimulation zu
erstellen. Die Analyse verschiedener Ballonwettbewerbe eröffnete detaillierte Einsichten in die
Flugdynamik und mögliche Optimierungen. Es zeigt sich, daß Heliumballons regelmäßig Höhen um
10~km erreichen. Unter optimalen Umständen bleiben sie mehr als 24 Stunden in der Luft und können
dabei fast 3000 km weit fliegen. Jedoch beschränken externe Wettereinflüsse die typische Lebenszeit
auf 2\,--\,5 Stunden.

\end{titlepage}

\newpage
\thispagestyle{empty}

\tableofcontents
\listoffigures

\chapter{Einleitung}

Ballonwettbewerbe sind ein beliebter Programmpunkt auf vielen Feiern und Volksfesten.
Ein Heliumballon wird mit einer Karte und einer Rücksendeadresse gestartet,
um eine möglichst große Entfernung zurückzulegen. Der Erfolg ist dabei nicht gewiß --
zu unberechenbar erscheint das Wettergeschehen, und zu guter Letzt muß die Karte auch
noch gefunden und zurückgeschickt werden! Was aber passiert mit dem Ballon nach dem
Start in der Luft? Welchen Einflüssen ist er unterworfen, und läßt sich eine gute Platzierung
mit einer großen Flugweite vielleicht doch gezielt beeinflussen?

Es lohnt sich, mit einer weiter gefassten Perspektive zu beginnen.
Ballons haben schon früh eine entscheidende Rolle bei der Erforschung der
Atmosphäre gespielt \cite{Broeckelmann1909}. Zuerst wurden Heißluft-- und Gasballons eingesetzt,
um Forschern die direkte Untersuchung der Atmosphäre zu ermöglichen. Später rückten
vermehrt unbemannte Ballons in den Vordergrund, die eine Vielzahl automatischer
Messungen bis hin zu astronomischen Beobachtungen durchführen konnten.
Mit automatischen Meßballons (sogenannten Radiosonden) gewonnene Wetterdaten bilden
noch heute das Rückgrat der täglichen Wettervorhersage \cite{Bruening2002}. Im
Zuge der Entwicklung immer ausgefeilterer
numerischer Wettermodelle und Simulationen zur Verbreitung von Luftschadstoffen
werden Ballons auch zur "`Markierung"' von Luftmassen verwendet, um die
Richtigkeit von Strömungsmodellen \cite{Potempski2006,BallTrack2006} zu überprüfen.
Die Umsetzung kann auf mehrere
Weisen erfolgen: Durch Ballons, die automatisch einem konstanten Druckniveau folgen
\cite{johnson2000lagrangian,lacorata2004evidence,moore1953data,riddle2006trajectory},
durch die Auswertung von regelrechten Ballonwettfahrten \cite{baumann1997validation} bis hin zur Einbindung
von Ballonwettbewerben in die Forschung \cite{sakagami12diffusion,stocker1990characteristics}.
Eine ausführliche Übersicht über den Stand der Trajektorienberechnung ist in \cite{stohl1998computation}
enthalten.

Die Untersuchung der Dynamik von Kartenballons\footnote{Im Folgenden werden die bei Ballonwettbewerben
verwendeten Heliumballons kurz als "`Kartenballon"' bezeichnet.} spielt meist nur eine
untergeordnete Rolle. Spezielle Arbeiten wurden zur Umweltverträglichkeit
von Ballons \cite{Burchette1989} und der maximal erreichbaren Höhe durchgeführt
\cite{Kofoed1992}, wobei aber nicht nach einer umfassende Beschreibung eines
frei fliegenden Ballons gesucht wurde. Die Eigenschaften des Ballongummis sowie der absolut
geringe Auftrieb eines Kartenballons erschweren die einfache Übertragung der charakteristischen
Eigenschaften größerer Ballons.
Eine Ausnahme bildet die Arbeit von Roberts \cite{roberts1995dynamics}, die ein einfaches
numerisches Modell zur Flugweitenabschätzung vorstellt. Ein Vergleich mit realen Kartenballonflügen
oder die Berechnung von Ballontrajektorien erfolgte in dieser Arbeit aber nicht.

Die offenen Fragen sind also noch zahlreich:
Wie weit kann ein Kartenballon fliegen? Wie lange bleibt er in der Luft?
Wie hoch fliegt er? Kann die Flugweite optimiert werden, und wenn ja, wie?
Diesen Fragen soll in der folgenden Arbeit systematisch nachgegangen werden,
um ein umfassendes Bild von dem Verhalten von Kartenballons in der Atmosphäre
zu zeichnen.
Zuerst erfolgt eine detaillierte Untersuchung aller relevanten physikalischen
Eigenschaften eines Kartenballons in Kapitel \ref{ChapBallonP}, gefolgt von einer
Zusammenfassung der wichtigsten Umwelteinflüssen in den Kapiteln \ref{ChapWetter} und
\ref{ChapBoden}. Diese Bausteine werden anschließend in Kapitel \ref{ChapSim} zu
einem Computermodell zusammengefügt, das dann mit Hilfe von Ballonwettbewerben und
Experimenten in Kapitel \ref{BalFlight} geprüft wird.

\chapter{Ballonphysik \label{ChapBallonP}}

\newcommand{\melloc}{melloc\copyright}
\newcommand{\rhoAir} {\rho_{\mathrm{L}}}
\newcommand{\rhoAirZ}{\rho_{\mathrm{L},0}}
\newcommand{\rhoAirRef}{\rho_{\mathrm{L},\mathrm{Ref}}}
\newcommand{\opt}{\mathrm{opt}}

\section{Einleitung}

Die theoretische Beschreibung eines Kartenballons umfaßt alle für die Trajektorienberechnung
relevanten Eigenschaften: Angefangen von der Geometrie des Ballons, dem Materialverhalten
des Ballongummis, über Auftrieb und Aerodynamik bis hin zu dem scheinbar unmerklichen Verlust
des Füllgases während des Fluges. Für die meisten dieser Fragestellungen ist es ausreichend,
alleine den Gummiballon zu betrachten. Die angebundene Karte verändert neben dem Gesamtgewicht
nur die Aerodynamik des Kartenballons. Erst nach dem möglichen Platzen des Ballons wird das
Geschehen ausschließlich durch die frei fallende Karte bestimmt.

Die folgenden Abschnitte gehen detailliert auf die Eigenschaften eines Gummiballon ein,
wobei die Aerodynamik unter Berücksichtigung der angehängten Karte in Abschnitt
\ref{SecAeroDyn} diskutiert wird.

\section{Geometrie}

\begin{figure}
\begin{center}
\includegraphics[scale=0.7,angle=0]{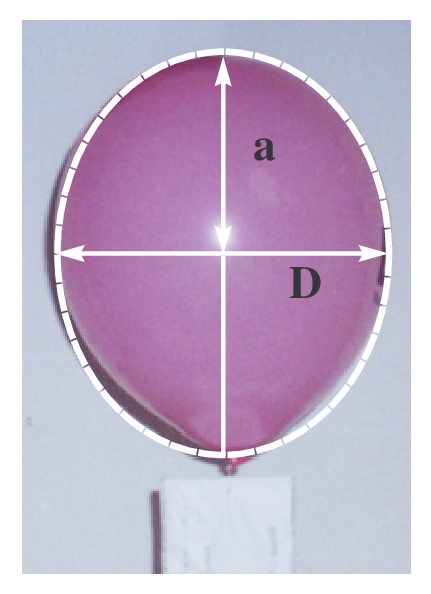}
\end{center}
\caption[Ballongeometrie]{Ballon mit dem Äquivalenzellipsoid mit Durchmesser $D$ und langer Halbachse $a$.\label{BallonG} }
\end{figure}

Die Grundlage für die Beschreibung der Ballongeometrie ist die Kenntnis
des Volumens $V$ und der Oberfläche $A$. Ein Ballon kann grob durch einen prolaten Ellipsoid
mit langer Halbachse $a$ und Durchmesser $D$ beschrieben werden (siehe Abb. \ref{BallonG}).
Wird die Abweichung von diesem Ellipsoid in einem Parameter $\varepsilon$ zusammengefaßt,
ergibt sich das Volumen zu:
\begin{eqnarray}
V &=& \frac{4\pi}{3} \varepsilon R^2a, \qquad   R = D/2 \label{VEllipsoid}
\end{eqnarray}
Bei gleich gefertigten Ballons ist das Verhältnis $a/D$ in etwa konstant und die
obige Formel kann vereinfacht werden zu:
\begin{eqnarray}
V &=& f_V D^3 \label{VDiameter} \\
A &=& f_A D^2
\end{eqnarray}
Mit Hilfe von Ballonphotos wurden die freien Parameter für zwei unterschiedliche
Fabrikate bestimmt:
\begin{center}
\begin{tabular}{|l|llll|} \hline
Mittelung über           &  $\varepsilon$    & $f_V$             & $f_A$             & $1-R/a$ \\ \hline
48 \melloc~32~cm~Ballons &  $ 0,96 \pm 0,02$ & $ 0,63 \pm 0,02 $ & $ 3,71 \pm 0,12 $ & $ 0,20 \pm 0,03 $ \\
10 Volksbank Werbeballons&  $ 0,94 \pm 0,01$ & $ 0,58 \pm 0,01 $ & $ 3,48 \pm 0,05 $ & $ 0,15 \pm 0,02 $ \\ \hline
\end{tabular}
\end{center}
Die beiden Fabrikate unterscheiden sich hauptsächlich in der Abplattung $1-R/a$, die in der letzten
Spalte mit angegeben ist. Die Fehler geben jeweils die Streuung innerhalb der Ballonstichproben an.
Für manche Anwendungen ist es nützlicher, das Verhältnis von Oberfläche zu Volumen anzugeben.
Von allen Körpern hat die Kugel das kleinste Verhältnis:
\begin{eqnarray}
\begin{array}{lcll}
A/V^{2/3} &=& 4,84            & \mbox{Kugel} \\
A/V^{2/3} &=& 5,05 \pm 0,20   & \mbox{\melloc~32~cm~Ballon } \\
A/V^{2/3} &=& 5,00 \pm 0,09   & \mbox{Volksbank Ballon}
\end{array}
\end{eqnarray}
Einen ersten Hinweis auf die elastischen Eigenschaften eines Ballons gibt das Füllvolumen,
das zum Platzen der Gummihülle führt. Um eine erste Übersicht zu erhalten, wurde eine
kleine Stichprobe ausgewählt und bis zum Platzen gefüllt. Während des Füllvorgangs wurden
die geometrischen Abmessungen regelmäßig erfaßt. Die daraus abgeleiteten Platzvolumina sind:
\begin{center}
\begin{tabular}{|llll|} \hline
$ D_{\max}$[cm] & $V_{\max}$[l] & $V_0$[l] & Bemerkung \\ \hline
$34 \pm 1$ & $23 \pm 2$ & 0,1 & Volksbankballon bei 20 $^{\circ}$C \\ \hline
$37 \pm 1$ & $32 \pm 3$ & 0,15& \melloc~32~cm~Ballon bei 20 $^{\circ}$C \\ \hline
$34 \pm 1$ & $25 \pm 2$ & 0,15& \melloc~32~cm~Ballon bei $-10^{\circ}$C \\ \hline
\end{tabular}
\end{center}
Alle Ballons haben sich kurz vor dem Platzen teilweise erheblich deformiert, so
daß die einfache Anwendung der Gleichungen \ref{VEllipsoid} und \ref{VDiameter} nicht
zuverlässig ist. Hinzu kommt, daß die Versuchsbedingungen eine nur ungenaue
Annäherung an die langsame Ausdehnung eines steigenden Ballons sind. Hystereseeffekte
in der Gummihülle und insbesondere die Wechselwirkung mit den fallenden Temperaturen
beim Aufstieg wurden nicht erfaßt.
Aus dem Volumen $V_0$ des ungedehnten (d.\,h. druckfreien) Ballons kann die maximalen
Dehnung des Ballongummis ermittelt werden. Sie liegt im Bereich von 500 \% bis 600 \%
(bezogen auf den ungedehnten Ballon) und ist mit den Angaben in \cite{Burchette1989} vergleichbar.

\section{Innendruck}

\begin{figure}
\begin{center}
\includegraphics[scale=0.4,angle=270]{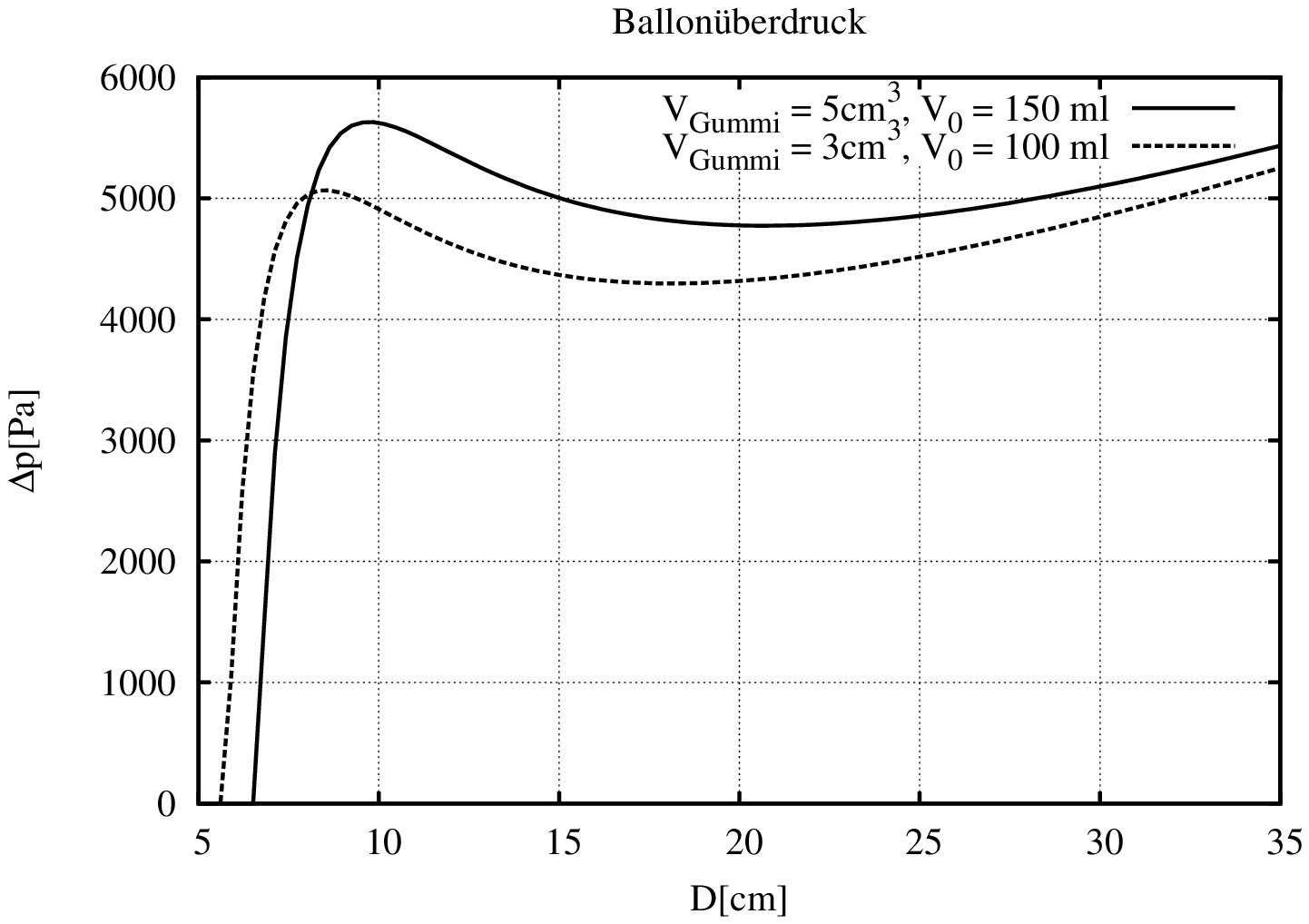}
\end{center}
\caption[Innendruck]{Balloninnendruck in Abhängigkeit des Ballondurchmessers.\label{BallonD} }
\end{figure}

Die Gummihülle eines Ballons übt einen zusätzlichen Druck auf das Füllgas aus. Die Beschreibung
der nichtlinearen elastischen Eigenschaften von Gummi (siehe z.\,B. \cite{mooney1940theory})
erlaubt die Berechnung des Drucksprungs $\Delta p$ als Funktion des Ballonradius $R$. Das Ergebnis
der von \cite{mueller1999physik} durchgeführten Rechnung ist
\begin{eqnarray}
\Delta p(R) &=& \frac{4}{3} \frac{V_{\mathrm{Gummi}}}{V_0} C_1 \left(\frac{R_0}{R}-\frac{R_0^7}{R^7} \right)
\left(1+\frac{C_{-1}}{C_1} \frac{R^2}{R_0^2} \right) \label{GlfDruck} \\
V_0 &=& \frac{4\pi}{3} R_0^3
\end{eqnarray}
mit dem Schubmodul $C_1 \approx$ 0,17 MPa und dem nichtlinearen Parameter $C_{-1}/C_1 \approx 0,1$.
$R_0$ ist der Radius des entspannten (d.\,h. druckfreien) Ballons.
Das Volumen $V_{\mathrm{Gummi}}$ der Gummihülle kann über die Dichte des verwendeten Gummis
aus dem Ballongewicht abgeleitet werden:
\begin{eqnarray}
M_{\mathrm{Gummi}} &=& \rho_{\mathrm{Gummi}} V_{\mathrm{Gummi}}
\end{eqnarray}
Typische Gewichte für die bereits genannten Fabrikate sind:
\begin{center}
\begin{tabular}{|lcll|} \hline
& &  $ M_{\mathrm{Gummi}} $ & \\ \hline
\melloc~32~cm~Ballon   &:& $5,1 \pm 0,2$ & g \\
\melloc~13~cm~Ballon   &:& $0,86$        & g \\
Volksbank Werbeballon  &:& $3,2 \pm 0,3$ & g \\ \hline
\end{tabular}
\end{center}
Die Stichprobe der 13~cm Ballons war zu klein, um eine repräsentative Streuung zu ermitteln.
Das Minimum und Maximum der Druckkurve (siehe Abb. \ref{BallonD}) ist für $C_{-1}/C_1 = 0,1$:
\begin{eqnarray}
r_{\max} &=& 1,476 \times R_0 \qquad  \Delta p_{\max} = 0,745
\times  \frac{4}{3} \frac{V_{\mathrm{Gummi}}}{V_0}  C_1 \\
r_{\min} &=& 3,143 \times R_0 \qquad  \Delta p_{\min} = 0,632
\times  \frac{4}{3} \frac{V_{\mathrm{Gummi}}}{V_0}  C_1
\end{eqnarray}
Mit der bekannten Abhängigkeit des Innendrucks von der Ballongröße kann nun das Volumen eines mit der
Gasmenge $N$ gefüllten Ballons bei einem vorgegebenen Druck $p_0$ und Temperatur $T_0$ berechnet werden.
Unter der Verwendung der Zustandsgleichung für ideale Gase erhält man für die Bestimmung des Ballonvolumens $V$
die Gleichung:
\begin{eqnarray}
(p_0+\Delta p(V) )V = N R T_0
\end{eqnarray}
$R$ ist hier die universelle Gaskonstante. Obwohl der Drucksprung (Gl. \ref{GlfDruck}) selbst keine
monoton steigende Funktion des Ballonvolumens ist, trifft dies auf die gesamte linke Seite zu:
\begin{eqnarray}
\frac{d}{dV} \big( (p_0+\Delta p(V) )V \big) > 0 \quad \forall \quad V>0
\end{eqnarray}
Dies garantiert für alle realisierbaren Kombinationen der freien Parameter jeweils eine eindeutige Lösung
für das Ballonvolumen. Da der Druck über einen weiten Bereich nur wenig variiert, wird er in dieser Arbeit
durch einen konstanten Wert angenähert:
\begin{eqnarray}
\Delta p &\approx& 0,7 \times  \frac{4}{3} \frac{V_{\mathrm{Gummi}}}{V_0}  C_1 \\
& = & 4760 \mbox{ Pa}\times \frac{ V_{\mathrm{Gummi}}/3 \mbox{ cm}^3}{V_0/100 \mbox{ cm}^3}
\end{eqnarray}

\section{Thermodynamik \label{ChapTerm}}

\begin{figure}
\begin{center}
\includegraphics[scale=0.7,angle=0]{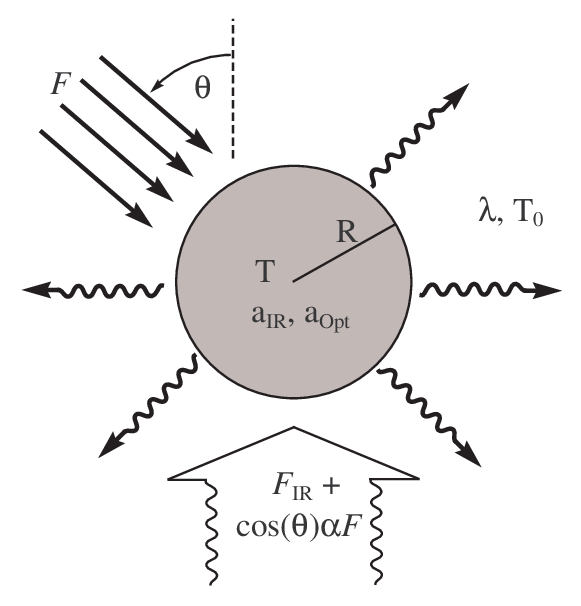}
\end{center}
\caption[Strahlungsgleichgewicht]{Vereinfachtes Strahlungsgleichgewicht, bestimmt durch die
Sonneneinstrahlung $\mathcal{F}$, die diffuse Strahlung
$\mathcal{F}_{\mathrm{IR}}+\cos(\theta)\alpha \mathcal{F}$ von
der Erdoberfläche und die Wärmestrahlung des Ballons.\label{RadiativeEq}}
\end{figure}

Die Temperatur eines Ballons bestimmt das Volumen über die Zustandsgleichung des
Füllgases und die elastischen Eigenschaften der Gummihülle. Während die Temperatur
eines Ballons im Schatten in guter Näherung durch die Lufttemperatur gegeben ist,
muß in der freien Atmosphäre und insbesondere in großen Höhen das Strahlungsgleichgewicht
mit der Umgebung berücksichtigt werden.

Der Arbeit von Nelson \cite{Nelson1962} folgend wird eine einfache Zweibandbeschreibung
(Abb.~\ref{RadiativeEq}) gewählt. Im Bereich optischer Wellenlängen erhält der Ballon
Energie durch die direkte Strahlung der Sonne $\mathcal{F}$ und den durch die Albedo $\alpha$
der Erdoberfläche reflektierten Anteil. Im Bereich infraroter Wellenlängen strahlt der
Ballon selbst Wärmeenergie ab und empfängt Wärmestrahlung $\mathcal{F}_{\mathrm{IR}}$ von der
Erdoberfläche.
Die umgebende Luft der Temperatur $T_0$ trägt über direkte Wärmeleitung bei. In der folgenden
Bilanz wird die adiabatische Abkühlung bzw. Erwärmung des Ballons beim Aufsteigen und Sinken
vernachlässigt:
\begin{eqnarray}
a_{\mathrm{Opt}} \pi R^2  ( 1 +   2 \cos(\theta)\alpha) \mathcal{F}  +
a_{\mathrm{IR}}  2 \pi R^2 \mathcal{F}_{\mathrm{IR}}  &=&
2\pi\lambda_{\mathrm{Luft}} \mbox{Nu} R (T-T_0) +4 \pi a_{\mathrm{IR}} R^2\sigma T^4 \label{EqFluxEq}
\end{eqnarray}
Die Sonneneinstrahlung\footnote{Die Solarkonstante ist 1367 W/m$^2$, Weltorganisation für Meteorologie (1982)}
erfolgt unter dem Zenitwinkel $\theta$, $T$ ist die zu bestimmende Temperatur des Ballons
und $R$ sein Radius. $\sigma$ ist die Stefan-Boltzmann-Konstante.
Im Folgenden wird ein graues Ballonmodell mit Absorptionskoeffizienten $a_\mathrm{IR} = 0,5$
und  $a_\mathrm{Opt} = 0,25$ als plausible Werte angenommen.
Realistische Werte der Albedo $\alpha$ liegen im Bereich $0,2 \dots 0,5$.
$\lambda_{\mathrm{Luft}}$ ist die molekulare Wärmeleitfähigkeit der Luft. Die wirkliche Wärmeleitung ist
um den Faktor\footnote{Die Nusselt-Zahl Nu ist proportional zum Verhältnis der Wärmeleitung einschließlich
Konvektion und Turbulenz zur molekularen Wärmeleitung. Es gilt Nu $>1$.} Nu größer. Eine empirische Formel für die
Nusselt-Zahl einer Kugel ist nach \cite{Kneer2007}:
\begin{eqnarray}
\mathrm{Nu} &=& 2 + (0,4\, \mathrm{Re_{tot}}^{1/2} + 0,06\, \mathrm{Re_{tot}}^{2/3} ) \mathrm{Pr}^{0,4}
\left( \frac{\eta_{\infty} }{\eta_K} \right)^{1/4}
\end{eqnarray}
Der Unterschied der dynamischen Viskosität der Luft im freien Raum $\eta_{\infty}$ zu dem Wert auf der
Kugeloberfläche $\eta_K$ ist für die hier betrachtete Anwendung so gering, daß dieser Korrekturfaktor
vernachläßigt werden kann.
Die totale Reynolds-Zahl $\mathrm{Re_{tot}}$ setzt sich aus der Umströmung der Kugel und dem konvektiven
Anteil, beschrieben durch die Grashof-Zahl Gr, zusammen:
\begin{eqnarray}
\mathrm{Re_{tot}}^2 &=& \mathrm{Re}^2 + 0,4 \mathrm{Gr}  \\
\mathrm{Gr}   &=& \frac{|T-T_0|}{T_0} \frac{g D^3}{\nu^2}
\end{eqnarray}
In der letzten Gleichung ist $g$ die Erdbeschleunigung, $\nu$ die kinematische Viskosität der umgebenden
Luft und $D$ der Durchmesser der Kugel. Bei einer Anströmung mit 1 m/s erhält man für einen Ballon etwa
Nu $\approx$ 100. Für eine einfachere Behandlung wird der Strahlungsfluß aus der Umgebung in Gl. \ref{EqFluxEq}
durch eine effektive Infrarot-Temperatur $T_{\mathrm{eff}}$ ausgedrückt:
\begin{eqnarray}
4  a_{\mathrm{IR}} \sigma  T_{\mathrm{eff}}^4  & = &
2 \lambda_{\mathrm{Luft}} \mbox{Nu} / R (T-T_0) +4  a_{\mathrm{IR}} \sigma T^4 \\
& \approx    & 2 \lambda_{\mathrm{Luft}} \mbox{Nu} / R (T-T_0) +4  a_{\mathrm{IR}} \sigma T_0^4
+ 16  a_{\mathrm{IR}} \sigma T_0^3(T-T_0)
\end{eqnarray}
Umstellen der Wärmebilanz ergibt:
\begin{eqnarray}
T-T_0 & \approx & \frac{T_{\mathrm{eff}}^4 - T_0^4}{\mbox{Nu} \lambda_{\mathrm{Luft}}/(2\sigma a_{\mathrm{IR}} R) + 4 T_0^3}
\end{eqnarray}
Für die betrachteten Temperaturdifferenzen $T-T_0$ können die Strahlungsterme um $T_0$ entwickelt werden:
\begin{eqnarray}
T-T_0 & \approx & \frac{T_{\mathrm{eff}}  -  T_0  }{\mbox{Nu} \lambda_{\mathrm{Luft}}/(8 T_0^3\sigma a_{\mathrm{IR}} R) + 1} \label{SimpledT}
\end{eqnarray}
Die Temperaturdifferenz $T_{\mathrm{eff}} - T_0$ reicht von Werten um $-40$\celsius{} (Nachts, große Höhen)
bis über $+40$\celsius{} (Bodennähe, Sonnenschein). Um den Einfluß auf die Ballontemperatur
abzuschätzen, wird der Faktor in Gl. \ref{SimpledT} ausgewertet.
Die Wärmeleitfähigkeit $\lambda_{\mathrm{Luft}}$ ist unabhängig vom Druck und nur proportional zur
Wurzel der Temperatur, so daß die Wärmeleitfähigkeit am Boden auch für größere Höhen ein guter
Richtwert ist. Mit $R$ = 15 cm ergibt sich die Abschätzung:
\begin{eqnarray}
\frac{\mbox{Nu} \lambda_{\mathrm{Luft}} }{8 T_0^3\sigma a_{\mathrm{IR}} R} &=&
2,8 \times \frac{\mbox{Nu}}{100} \times \left( \frac{293 \mbox{ K}}{T_0} \right)^3
\end{eqnarray}
Der durch die Ballongröße bedingte starke Einfluß der direkten Wärmeleitung reduziert
den Temperaturunterschied zur umgebenden Luft auf Werte um 10 K. Für die
Berechnung des Ballonvolumens und des Auftriebs ergeben sich nur Korrekturen im Prozentbereich,
so daß die Ballontemperatur $T$ gleich der Temperatur der umgebenden Luft $T_0$
gesetzt werden kann. Diese Vereinfachung erspart auch die komplexe Modellierung des Strahlungsflußes
in der Atmosphäre, die bereits mit einfachen Annahmen sehr umfangreich ist \cite{ramanathan1976radiative}.
Bei Tagesflügen fehlen ohnehin genaue Wetterdaten, um den Einfluß der Wolkenbedeckung auf die
Sonneneinstrahlung zu rekonstruieren.

\section{Materialverhalten}

Bei tiefen Temperaturen verliert Gummi seine Flexibilität und erstarrt in einem
Glaszustand. Für einen aufsteigenden Ballon bedeutet dieser Phasenübergang die
Zerstörung der Hülle durch das sich weiterhin ausdehnende Füllgas. Der
Glasübergang erstreckt sich über einen Temperaturbereich und läßt sich
nur näherungsweise durch eine feste Temperatur beschreiben. Besonders
für stark gedehnte Ballons muß ein allmählicher Übergang vom Platzen
durch Überdehnung zum Platzen durch Verglasung angenommen werden \cite{paterson1964effect}.

Zur Modellierung dieses Effekts wird ein einfaches Modell verwendet:
Unterschreitet die Ballontemperatur eine Grenze $T_{\mathrm{Burst}}$, wird
der Ballon als geplatzt betrachtet.
Experimente von Kofoed-Hansen~\cite{Kofoed1992} zeigen deutlich, daß bis $-20$\celsius{}
die elastischen Eigenschaften nicht beeinträchtigt werden.
Burchette~\cite{Burchette1989} gibt die Glasübergangstemperatur mit $-40$\celsius{} an,
während Roberts~\cite{roberts1995dynamics} einen Wert um $-60$\celsius{} annimmt.

Aus den in Kapitel \ref{BalFlight} durchgeführten Simulationen,
auf die an dieser Stelle vorgegriffen wird, ergibt sich die Notwendigkeit
des Glasübergangs bei einer Lufttemperatur von etwa $-55$\celsius{}.
Zusammen mit den in Kapitel \ref{ChapTerm} ausgeführten Überlegungen ist nicht
ausgeschlossen, daß der Ballongummi eine andere Temperatur hat. In diesem
Sinne ist die aus den Simulationen abgeleitete Temperatur hauptsächlich eine
Information über die maximale Höhe, die ein Ballon unter optimalen Umständen
erreichen kann.

\section{Aerodynamik \label{SecAeroDyn}}

Auf einen in $z$-Richtung aufsteigenden Gasballon der Masse $M$ wirkt der Nettoauftrieb
$F_a$ und die Reibungskraft $F_R$
\begin{eqnarray}
M\ddot{z} &=&  F_a + F_R \label{EqSteig1}
\end{eqnarray}
wobei die Reibungskraft durch den Ansatz nach Newton beschrieben wird:
\begin{eqnarray}
F_R &=& -\frac{1}{2}c_w A_q \rhoAir v_z |v_z| \label{Ffric}
\end{eqnarray}
$A_q$ ist die Querschnittsfläche in Bewegungsrichtung, $c_w$ der Widerstandsbeiwert
und $v_z$ die vertikale Geschwindigkeit des Ballons.
Die Lösung der Differentialgleichung \ref{EqSteig1} ist:
\begin{eqnarray}
z(t)   &=& l \ln\left[\cosh\left(t v_{\infty} / l \right)\right] \\
v_z(t) &=& v_{\infty} \tanh\left(t v_{\infty} / l \right)  \label{vRiseEq}
\end{eqnarray}
Die Endgeschwindigkeit $v_{\infty}$ und die Längenskala $l$ berechnen sich aus den Parametern des Ballons zu:
\begin{eqnarray}
v_{\infty} &=& \sqrt{\frac{l F_a}{M}}, \qquad l = \frac{2M}{ c_w A_q \rhoAir } \label{v0RiseEq}
\end{eqnarray}
Mit einer Referenzluftdichte\footnote{$\rhoAirRef = 1,204$ \kilogram \per \cubic\metre, siehe Anhang \ref{UsefulConst}.}
$\rhoAirRef$ kann diese Gleichung weiter vereinfacht werden:
\begin{eqnarray}
v_{\infty} &=& v_{\infty,0} \sqrt{\frac{\rhoAirRef}{\rhoAir}}, \qquad v_{\infty,0} := \sqrt{ \frac{2 F_a}{ c_w A_q \rhoAirRef }}
\end{eqnarray}
Ein Kartenballon erreicht innerhalb von Bruchteilen einer Sekunde die Endgeschwindigkeit $v_{\infty}$.
Das Gesamtgewicht $M$ eines Ballons hängt von den verwendeten Karten\footnote{Ballonpostkarten haben das
übliche Format $105 \times 148$ mm, sind aber mit 80 g/m$^2$ deutlich leichter als Standardpostkarten.}
und Materialien ab.

\clearpage
\noindent
Eine kleine Auswahl bietet die folgende Tabelle:
\begin{center}
\begin{tabular}{|lcll|} \hline
Postkarte                      &:& 3,8 & g \\
Ballonkarte                    &:& 1,6 & g \\
Geschenkband mit Plastikclip   &:& 0,9 & g \\
Papierkarte in PE-Hülle        &:& 2,0 & g \\ \hline
\end{tabular}
\end{center}
Für den Widerstandsbeiwert $c_w$ findet man:
\begin{center}
\begin{tabular}{|ll|} \hline
$c_w$ & Bemerkung \\ \hline
0,44              & Literaturwert Kugel \cite{Schmidt1919} \\ \hline
$ 0,81 \pm 0,02 $ & Ballon alleine beim Aufsteigen \\ \hline
$ 0,93 \pm 0,09 $ & Ballon mit Karte \\ \hline
$0,96$            & Pilotballon nach \cite{Lowell1998}  \\ \hline
\end{tabular}
\end{center}

\noindent
Wenn man die geringe Abweichung eines Ballons von der Kugelform bedenkt, erscheint die deutliche
Abweichung des klassischen Literaturwerts für eine Kugel zunächst erstaunlich. Der Unterschied ist
aber nur zu einem geringen Teil auf die veränderte Geometrie zurückzuführen. Messungen des
Strömungswiderstandes gehen von einem fixierten Prüfkörper aus, was einem im Vergleich zum
umgebenden Medium sehr dichten beweglichen Körper entspricht. Ein steigender Ballon hat aber
offensichtlich eine geringere mittlere Dichte als Luft, so daß die intensive Wechselwirkung
mit der Wirbelschleppe den Luftwiderstand erhöht \cite{Lowell1998,Preukschat1962}.

Um diesen Befund unter realistischen Bedingungen weiter abzusichern, wurde der gefundene \mbox{$c_w$-Wert}
mit den Aufstiegsgeschwindigkeiten von angepeilten Ballons in Bodennähe in Tabelle \ref{SteigData} verglichen.
Zu beachten ist, daß die angegebenen Steiggeschwindigkeiten stark durch die Variabilität des Windes in
Bodennähe schwanken (vgl. Abb. \ref{WindProfil}).
\begin{table}
\begin{center}
\begin{tabular}{|l|c|l|c|c|c|l|} \hline
Datum  & Durchmesser & Höhe & Zeit nach Start & $v_z$ & $v_{z,\mathrm{theo}}$ & Quelle  \\ \hline
30.08.1998 &  6,6\arcminute & 24\degree        & 45 s & 1,3 m/s & 1,32 m/s & Photo  \\ \hline
20.09.1998 & 21,0\arcminute & 31\degree        & 15 s & 1,5 m/s & 1,34 m/s & Photo  \\ \hline
21.07.1999 & 15,6\arcminute & 24\degree        & 25 s & 1,0 m/s & 1,51 m/s & Photo  \\ \hline
05.08.1999 &  9,0\arcminute & 38\degree        & 40 s & 1,7 m/s & 1,45 m/s & Photo  \\ \hline
14.01.2001 & $\approx$ 6\arcsecond & 14\degree & 20 min & 1,8 m/s & 1,14 m/s & Teleskop \\ \hline
\end{tabular}
\end{center}
\caption{Auswertung der Steiggeschwindigkeit $v_z$ aus Ballonstarts. Die vorletzte Spalte enthält
theoretische Werte nach Gl. \ref{v0RiseEq}. Die Ballons sind Teil der in Kapitel \ref{SectStartDyn}
vorgestellten Versuchsreihe\label{SteigData}.}
\end{table}

Platzt der Ballon durch äußere Einflüsse oder ein zu großes Volumen, fällt die
Karte zu Boden. Für eine einzelne Karte
in einer PE-Hülle findet man $v_{\infty,0}=1,0 \pm 0,12 $ m/s.
Die wirksame Fläche $A_{w} = c_w A_q$ beträgt in diesem Fall 326
cm$^2$.

\section{Permeabilität}

Die Gummihülle eines Ballons schließt die Gasfüllung nicht hermetisch ein, sondern
erlaubt einen geringen Gasaustausch, der im Laufe der Zeit zum Schrumpfen des Ballons
führt. Der Gastransport durch die Gummihülle kann in zwei Prozesse zerlegt werden:
Das Gas löst sich zunächst in der Membran (beschrieben durch das Henry-Gesetz) und
diffundiert  anschließend durch die Membran. Die Kombination beider Prozesse liefert
die Stoffmenge pro Zeit $\dot N$, mit der das Gas eine Membran der Fläche $A$ und
der Dicke $d$ durchdringt:
\begin{eqnarray}
\dot N &=& Q \frac{A}{d}\Delta p \label{GlFlux1}
\end{eqnarray}
Getrieben wird der Fluß durch eine Partialdruckdifferenz $\Delta p$. Aus diesem Grund
ist der Gasverlust eines Luftballons deutlich kleiner als der eines Heliumballons, obwohl
die Permeabilitäten selbst für beide Gase vergleichbar sind.
Die Permeabilität $Q$ läßt sich aus der Diffusivität $D$ und der Henry-Konstante $k_H$
berechnen:
\begin{eqnarray}
Q &=& D k_H
\end{eqnarray}
$Q$ ist temperaturabhängig. In Analogie zu Arrhenius-Gleichung gilt die Näherung
\begin{eqnarray}
Q(T) &=& Q_0\exp\big(-T_Q(1/T-1/T_0)\big)
\end{eqnarray}
wobei $T_Q$ die in eine Temperatur umgerechnete Aktivierungsenergie der beteiligten
Prozesse ist. $Q_0$ ist der bei der Temperatur $T_0$ bestimmte Referenzwert. Für die
weiteren Betrachtungen wird $T_Q \approx 1700$ K  (geschätzt nach  \cite{van1946permeability})
verwendet.
Gleichung \ref{GlFlux1} läßt sich für ein ideales Gas leicht mit Hilfe der Konzentration
$n$ schreiben:
\begin{eqnarray}
\dot N &=& \delta \frac{A}{d} \Delta n \label{PermEqu}
\end{eqnarray}
Einsetzen der idealen Gasgleichung liefert die auf die Konzentrationsdifferenz $\Delta n$ bezogene Größe
\begin{eqnarray}
\delta &:=& Q RT
\end{eqnarray}
die weiterhin ebenfalls kurz als Permeabilität\footnote{Um Mehrdeutigkeiten auszuschließen wird in dieser Arbeit
ausschließlich $\delta$ verwendet.} bezeichnet wird. Die Temperaturabhängigkeit ist:
\begin{eqnarray}
\delta (T)   & =& \delta_0 \frac{T}{T_0}\exp\big(-T_Q(1/T-1/T_0)\big) \\
\delta_0 &:=& Q_0 RT_0
\end{eqnarray}
Die Modellierung der Temperaturabhängigkeit der Permeabilität hat einen entscheidenden
Einfluß auf den Gasverlust in großer Höhe und damit die Flugweite. Tabelle \ref{PermTab}
gibt einen Überblick über die Permeabilität von Ballongummi für verschiedene Gase.
Die letzte Spalte enthält Werte für vulkanisierten Naturgummi,
die nur bedingt auf Ballongummi übertragen werden können.

\begin{table}
\begin{center}
\begin{tabular}{|l|l|l|l|l|} \hline
$\delta[10^{-6}$ dm$^2$/h$]$ & \melloc~32~cm & \melloc~13~cm & Daten von \cite{Sarazin2004} & Daten von \cite{van1946permeability}  \\ \hline
N$_2$  & $1,2 \pm 0,2$ & $1,2 \pm 0,7$ & \quad --  & 1,67  \\
O$_2$  & $2,2 \pm 0,9$ & $3,3 \pm 1,7$ & \quad --  & 4,78  \\
Luft   & $1,3 \pm 0,3$ & $1,6 \pm 0,7$ & $0,7$     & \quad  --    \\
H$_2$  & $6,2 \pm 0,5$ & $7,4 \pm 1,0$ & $7,2$     & 10,71 \\
He     &  \quad -- &\quad --           & $3,8$     & 6,31  \\
CO$_2$ &  \quad -- & \, $22 \pm 3$     & \quad --  & 27,53 \\  \hline
\end{tabular}
\end{center}
\caption[]{Permeabilitäten $\delta$ des Ballongummis bei Zimmertemperatur und Normdruck.
Die Volksbankballons sind mit den Daten der \melloc~32~cm~Ballons konsistent. \label{PermTab} }
\end{table}

\section{Diffusion \label{SecDiff}}

\newcommand{\Vt}{\tilde V}
\newcommand{\vt}{\tilde v}

Mit den Permeabilitäten der beteiligten Gase kann der Volumenverlust eines Ballons
als Funktion der Zeit bestimmt werden. In der Praxis sind zwei Gase -- Luft und die Ballonfüllung --
oft ausreichend für eine genaue Beschreibung. Da Luft im wesentlichen ein Gemisch aus Stickstoff und
Sauerstoff ist, bietet es sich aber an, noch ein weiteres Gas hinzuzufügen und einen Ballon mit
drei verschiedenen Füllgasen als Modell zu verwenden. Die Konzentration der drei Gase im
Außenraum sei unabhängig von der Zeit $n_1, n_2$ und $n_3$. Die Stoffmengen der Gase
im Ballon werden durch $N_1, N_2$ und $N_3$ gegeben. Jedes Gas hat eine individuelle
Permeabilität $\delta_i$. Die Zeitentwicklung der Stoffmengen nach Gl. \ref{PermEqu} ist dann:
\begin{eqnarray}
\frac{d}{dt}\left( \begin{array}{l} N_1 \\ N_2 \\ N_3 \end{array} \right) &=&
\frac{A}{d}
\left( \begin{array}{l}
\delta_1( n_1 -  N_1/V) \\
\delta_2( n_2 -  N_2/V) \\
\delta_3( n_3 -  N_3/V) \\
\end{array} \right)
\end{eqnarray}
Die Dicke $d$ der Gummihülle ergibt sich in sehr guter Näherung aus dem Gummivolumen der Hülle
zu $d = V_{\mathrm{Gummi}}/A$. Alle weiteren Parameter wie der Innendruck $p_{\mathrm{innen}}$, der
Außendruck $p_{\mathrm{au"sen}}$ und die Temperatur $T$ werden als konstant angenommen. Zur
Vereinfachung der Gleichung werden neue Variablen $V_i = N_i RT /p_{\mathrm{innen}}$ und
$r_i = n_i RT/p_{\mathrm{au"sen}}$ eingeführt:
\begin{eqnarray}
\frac{d}{dt}\left( \begin{array}{l} V_1 \\ V_2 \\ V_3 \end{array} \right) &=&
-\frac{A^2}{VV_{\mathrm{Gummi}} } \frac{1}{\tilde p}
\left( \begin{array}{ccc}
\delta_1(\tilde p-r_1) & -\delta_1  r_1          & -\delta_1 r_1\\
-\delta_2 r_2          & \delta_2 (\tilde p-r_2) & -\delta_2 r_2\\
-\delta_3 r_3          & -\delta_3  r_3          & \delta_3(\tilde p-r_3)\\
\end{array} \right)
\left( \begin{array}{c} V_1\\V_2\\V_3 \end{array} \right) \label{M} \\
V&=& \sum_i V_i \qquad 1 = \sum_i r_i \\
\tilde p&:=&\frac{p_{\mathrm{innen}}}{p_{\mathrm{au"sen}}} \geq 1
\end{eqnarray}
Zur weiteren Untersuchung wird die Matrix in Gl. \ref{M} diagonalisiert.
Die Eigenwerte $\gamma_i$ sind Lösungen der Gleichung:
\begin{equation}
\sum_{j=1}^3 \frac{\delta_j r_j}{\delta_j-\gamma_i} = \tilde p \label{EigenV}
\end{equation}
Hieraus läßt sich direkt eine Ungleichung für die Nullstellen ablesen:
\begin{equation}
\delta_3 \geq \gamma_3 > \delta_2 \geq \gamma_2 > \delta_1  \geq  \gamma_1 \geq 0
\end{equation}
Die den Eigenwerten $\gamma_i$ entsprechenden Eigenvektoren sind
\begin{eqnarray}
\vt_i &=& \frac{1}{\tilde p}
\left( \begin{array}{l} \frac{\delta_1 r_1}{\delta_1-\gamma_i} \\
\frac{\delta_2 r_2}{\delta_2-\gamma_i} \\
\frac{\delta_3 r_3}{\delta_3-\gamma_i} \end{array} \right)\label{EW1} \label{norm}
\end{eqnarray}
wobei jeder Eigenvektor ein Gasgemisch mit einem Einheitsvolumen beschreibt:
\begin{eqnarray}
1 &=& \sum_{j=1}^3 (\vt_i)_j
\end{eqnarray}
Der Index $j$ bezieht sich auf die einzelnen Komponenten des Vektors $\vt_i$.
Da der Innendruck nur wenig größer ist als der Außendruck, kann der
kleinste Eigenwert $\gamma_1$ über eine Entwicklung nach $\tilde p-1$ bestimmt werden:
\begin{eqnarray}
\frac{1}{\bar \delta} &:=& \sum_{i=1}^3 \frac{r_i}{\delta_i} \\
\gamma_1 &=& \bar \delta  \, (\tilde p-1) + \mathcal{O}\left((\tilde p-1)^2\right)
\end{eqnarray}
$\bar \delta $ kann als die mittlere Permeabilität eines Gasgemisches angesehen werden.
Für den kleinsten Eigenwert $\gamma_1$ hat die durch $\vt_1$ beschriebene Gasmischung praktisch
die Gaszusammensetzung des Außenraums. Dies entspricht einem luftgefüllten Ballon, dessen Gasverlust
ausschließlich durch den Drucksprung der Ballonhülle getrieben wird.
Die Differentialgleichungen für die Eigenfunktionen $\Vt_i$ sind nun:
\begin{eqnarray}
V&=& \Vt_1+\Vt_2+\Vt_3 \\
\frac{d}{dt}\left( \begin{array}{l}\Vt_1 \\ \Vt_2 \\ \Vt_3 \end{array} \right) &=&
- \frac{A^2}{ VV_{\mathrm{Gummi}}}
\left( \begin{array}{l}\gamma_1\Vt_1 \\ \gamma_2\Vt_2 \\ \gamma_3\Vt_3 \end{array} \right)
\end{eqnarray}

\newpage
\noindent
Einsetzen des Ansatzes
\begin{eqnarray}
\Vt_i(t)&=&\Vt_{0,i} \exp(-u(t)\gamma_i)  \label{EFloes}
\end{eqnarray}
reduziert das Differentialgleichungssystem auf die Bestimmung der Hilfsfunktion $u(t)$:
\begin{eqnarray}
\dot u &=& \frac{A(u)^2}{V(u)V_{\mathrm{Gummi}}} \\
t &=& \int_0^{u(t)} V_{\mathrm{Gummi}} \frac{V(u)}{A(u)^2} du \label{LoesU}
\end{eqnarray}
Gleichung \ref{LoesU} stellt $u(t)$ implizit dar. Aus den Eigenfunktionen
in Gl. \ref{EFloes} können dann die Funktionen $V_i(t)$ mit Hilfe der Eigenvektoren in
Gl. \ref{EW1} linear kombiniert werden. Ist nur ein $\Vt_{0,i} \not= 0$, läßt sich
die Lösung explizit angeben:
\begin{eqnarray}
\tau_i &:=& \frac{V_{0,i} V_{\mathrm{Gummi}} }{\gamma_i A_{0,i}^2} \\
\tilde  V_i(t)  &=& \frac{V_{0,i}}{\left(1+t/(3\tau_i)\right)^3}
\end{eqnarray}
Zur Zeit $t=\tau_i$ ist das Volumen etwa auf die Hälfte abgefallen.
Bei einem vernachlässigbaren Drucksprung ($\tilde p=1$) hat das Diffusionsproblem eine Erhaltungsgröße.
Über die Eigenvektoren $w_i$ der transponierten Matrix
\begin{eqnarray}
w_i &=&\left( \begin{array}{l} \frac{1}{\delta_1 - \gamma_i} \\
\frac{1}{\delta_2 - \gamma_i} \\
\frac{1}{\delta_3 - \gamma_i} \end{array} \right)\label{EW2}
\end{eqnarray}
gewinnt man mit $\gamma_1=0$ die explizite Darstellung:
\begin{eqnarray}
\sum_{i=1}^3 \frac{V_i(t)}{\delta_i} & = & \mbox{konst.}
\end{eqnarray}
Die obenstehenden Gleichungen können sofort auf höherdimensionale Matrizen desselben Typs
verallgemeinert werden.

\chapter{Wetter \label{ChapWetter}}

\section{UV-Strahlung und Ozon\label{SectUV}}

Die Sonneneinstrahlung in Verbindung mit Oxidantien in der Luft hat einen
starken Einfluß auf die Haltbarkeit eines Gummiballons. Die UV-Strahlung
zerstört dabei den Gummi nicht nur direkt durch das Aufbrechen der Polymerketten,
sondern vor allem durch die dadurch ermöglichten radikalischen Reaktionen
mit Ozon und Stickstoffoxiden. Einmal gestartet, laufen diese Reaktionen auch
im Dunkeln weiter und führen schließlich zum Zerfall des Gummis bzw. Platzen
des Ballons \cite{Huntink2003}.

Experimente von Burchette~\cite{Burchette1989} bestätigen dieses Szenario, in dem durch
Witterungseinflüsse und Sonnenlicht bereits im Laufe eines Tages der Großteil
der Testballons zerstört wird. Die Zerfallsrate nimmt mit der Spannung des Gummis zu,
so daß besonders stark aufgeblasene Ballons zuerst platzen \cite{Nelson1962}.

In milder Form macht sich die Oxidation des Gummis bereits nach einigen Stunden
durch eine Mattierung der Ballonoberfläche bemerkbar, die von einem
erhöhten Gasverlust begleitet wird \cite{Haberlandt}.

Während die Wichtigkeit der UV-Strahlung außer Frage steht, ist die realistische
Modellierung weitaus schwieriger. Ballonspannung, die Hysterese der Ballonhülle und
die Volumenänderung beim Aufstieg wechselwirken mit der Oxidation des Gummis und den
wechselnden Wettereinflüssen. Ohne kontrolliert durchgeführte Experimente ist es
kaum möglich, eine theoretische Beschreibung zu gewinnen.
Aus diesem Grund wird auf die direkte Modellierung dieses Prozesses zugunsten
einer empirischen Untersuchung der Ballonflugdaten verzichtet.

\section{Regen und Luftfeuchtigkeit\label{SectRegen}}

Regen ist allgegenwärtig. Rund 50 \% aller Tage in Deutschland gelten als Regentage,
haben also einen Gesamtniederschlag von mehr als 0,1 mm in 24 Stunden \cite{klein35others}.
Die Wichtigkeit auch geringer Niederschläge für die Simulation eines Ballons läßt
sich verdeutlichen, indem man die Wirkung eines Regens mit Niederschlagsrate $w$ auf den
Auftrieb mit dem Verlust des Füllgases vergleicht.
Die jeweiligen Auftriebsverluste $\dot F_a$ durch diese beiden Prozesse sind:
\begin{eqnarray}
\dot F_{a,\mathrm{Regen}}&\approx& - g w A_{\mathrm{eff}} \rho_{W} \\
\dot F_{a,\mathrm{Diff}} &\approx&  - g  \rhoAir \delta \frac{A^2}{V_{\mathrm{Gummi}}}
\end{eqnarray}
$A_{\mathrm{eff}}$ repräsentiert die effektive Sammelfläche des Ballons für Regen und
$\rho_{W}$ ist die Dichte von flüssigem Wasser. Die weiteren Variablen entsprechen
der Nomenklatur aus Kapitel \ref{SecDiff}.

\clearpage

\noindent
Auflösen von $ \dot F_{a,\mathrm{Regen}} <  \dot F_{a,\mathrm{Diff}}$ nach $w$ ergibt die
Bedingung für die Dominanz des Regens beim Auftriebsverlust:
\begin{eqnarray}
w & \geq & \frac{\rhoAir}{ \rho_{W} } \frac{A^2 \delta}{A_{\mathrm{eff}} V_{\mathrm{Gummi}}}
\end{eqnarray}
Einsetzen plausibler Werte ergibt eine Größenordnung von 0,14 mm/Tag, also ein sehr leichter
Nieselregen. Eine solche Wasserübernahme könnte bereits bei einem Flug durch Wolken erreicht werden.
Obwohl Ballongummi einen hydrophoben Charakter hat, kann sich genug Wasser in Form großer Tropfen
auf der Ballonhülle sammeln, um den Auftrieb eines Kartenballons vollständig zu neutralisieren und
ihn zu Boden drücken. Eine über den Ballonquerschnitt $A_q$ integrierte Niederschlagsmenge
von 0,1 mm ist dafür bereits ausreichend. Zufälligerweise entspricht
diese Niederschlagsmenge dem Mindestniederschlag von 0,1 mm, um einen Tag als Regentag zu
bezeichnen. Zusätzlich zu der Belastung des Ballons durch anhaftendes Wasser kann auf der
Hülle gefrierender Regen zu einem verfrühten Platzen der Hülle führen und so den Ballonflug
direkt beenden\footnote{Dieses Phänomen ist bei Radiosondenballons bekannt \cite{Weber2010}.}.
Folglich bringt jeder ausgeprägte Niederschlag einen Ballon mit großer Wahrscheinlichkeit
sofort zu Boden.

Leider sind frei verfügbare Daten zu grobmaschig, um
die lokal sehr stark schwankende Niederschlagsverteilung ausreichend aufzulösen
\cite{ahrens2006distance,pathirana2002multifractal}.
Die größte für die Forschung freie Wetterdatensammlung in Europa ist das
{\em European Climate Assessment \& Dataset} \cite{klein35others}.
Auch hier schwankt die Dichte der verfügbaren Stationen sehr stark und ist in vielen europäischen
Ländern zu gering, zudem sind die Niederschlagsdaten nur integriert pro Tag verfügbar.
Versuchsweise wurden diese Wetterdaten über Thiessen-Polygone interpoliert, da die geringe
Stationendichte die Verwendung eines einfachen und robusten Verfahrens erfordert, und mit den
Trajektorienrechnungen (siehe Kap. \ref{ChapSim}) kombiniert. Eine aussagekräftige Auswertung
scheiterte aber an der geringen Auflösung der Daten, so daß dieser Ansatz nicht weiter verfolgt wurde.

\section{Atmosphäre \label{ChapAtm}}

Die Atmosphäre gliedert sich in mehrere verschiedene dynamische Schichten.
Direkt an den Boden grenzt die laminare Unterschicht mit einer Stärke von wenigen
Millimeter, die durch die Viskosität der Luft dominiert wird. Daran schließt
sich die Prandtl-Schicht an, ein turbulenter Luftstrom der in den unteren
Bereichen von der Oberflächenstruktur bestimmt wird. Sie ist 20 bis 60 Meter
dick. Die Eckman-Schicht bildet schließlich den Übergang zu dem durch Druckgradienten
und Corioliskraft geprägten geostrophen Wind.

Das bodennahe Windprofil ist für den Start eines Kartenballons von besonderem
Interesse. Eine oft verwendete Beschreibung ist die Monin-Obukhov-Ähnlichkeitstheorie.
Sie basiert auf einer dimensionslosen Formulierung der relevanten hydrodynamischen
Gleichungen und einer geschickt gewählten Schließungsbedingung. Dieser Ansatz
reduziert die Differentialgleichung für das höhenabhängige Geschwindigkeitsprofil
$v(z)$ auf
\begin{eqnarray}
\frac{K z}{v_*} \frac{\partial v}{\partial z } &=& \Phi_M(z/L) \label{MoninOEq}
\end{eqnarray}
mit der von-K\'arm\'an-Konstante $K$, der Reibungsgeschwindigkeit $v_*$ und der
Obukhov-Länge $L$. Eine kompakte Darstellung mit weiterführender Literatur ist
in \cite{foken2006} enthalten. Die verbleibende universelle Skalenfunktion $\Phi_M$
muß experimentell bestimmt werden \cite{businger1971flux}. Für neutrale Schichtungen
ist $L\rightarrow \infty$ und die Skalenfunktion reduziert sich auf die
Konstante $\Phi_M=1$. In diesem Fall läßt sich Gl. \ref{MoninOEq} einfach integrieren
und man erhält das klassische logarithmische Windprofil:
\begin{eqnarray}
v(z) &=&  \frac{v_*}{K} \ln(z/z_0) \label{vLogProfile}
\end{eqnarray}
Die Integrationskonstante $z_0$ ist die sogenannte Rauhigkeitslänge, die von der
Geländestruktur abhängt und ebenfalls experimentell ermittelt werden muß.

Ein erster Schritt, um die Atmosphäre in großen Höhen zu beschreiben,
sind gemittelte Werte von Wind, Druck und Temperatur als
Funktion der Höhe. Als Grundlage werden die Daten aus Tabelle \ref{AtmData}
für die mittlere Troposphäre über Deutschland verwendet, um ein einfaches und
realistisches Modell zur Hand zu haben.
Diese Daten lassen sich gut durch folgende Funktionen in Abhängigkeit der Höhe $z$ darstellen:
\begin{eqnarray}
p[\mbox{Pa}] &=& 101\,325 \exp\left(-\frac{z}{7,5 \mbox{ km}}\right) \label{pAtmEq} \\
v[\mbox{m/s}] &=& 2,58\frac{z}{1\mbox { km}} + 5,13 \label{VAtmEq} \\
T[^{\circ}\mbox{C}] &=& T_0 -4,223\frac{z}{1\mbox{ km}} -0,187  \left(\frac{z}{1\mbox{ km}} \right)^2 \label{TAtmEq} \\
\rho[\mbox{kg}/\mbox{m}^3] & \approx & 1,26 \exp \left(-\frac{z}{9,2 \mbox{ km}}\right)
\end{eqnarray}
Die Temperatur $T_0$ am Boden ist im Mittel 9\celsius{}, kann aber verändert werden um verschiedene Jahreszeiten
zu repräsentieren. Für einfache Studien des Flugverhaltens sind diese Gleichungen
ausreichend und werden im weiteren als Referenz verwendet. Die Höhenabhängigkeit der Dichte
kann aus dem Druck-- und Temperaturverlauf berechnet werden und ist hier nur für
analytische Abschätzungen mit angegeben.

\begin{figure}
\begin{center}
\includegraphics[scale=0.5,angle=270]{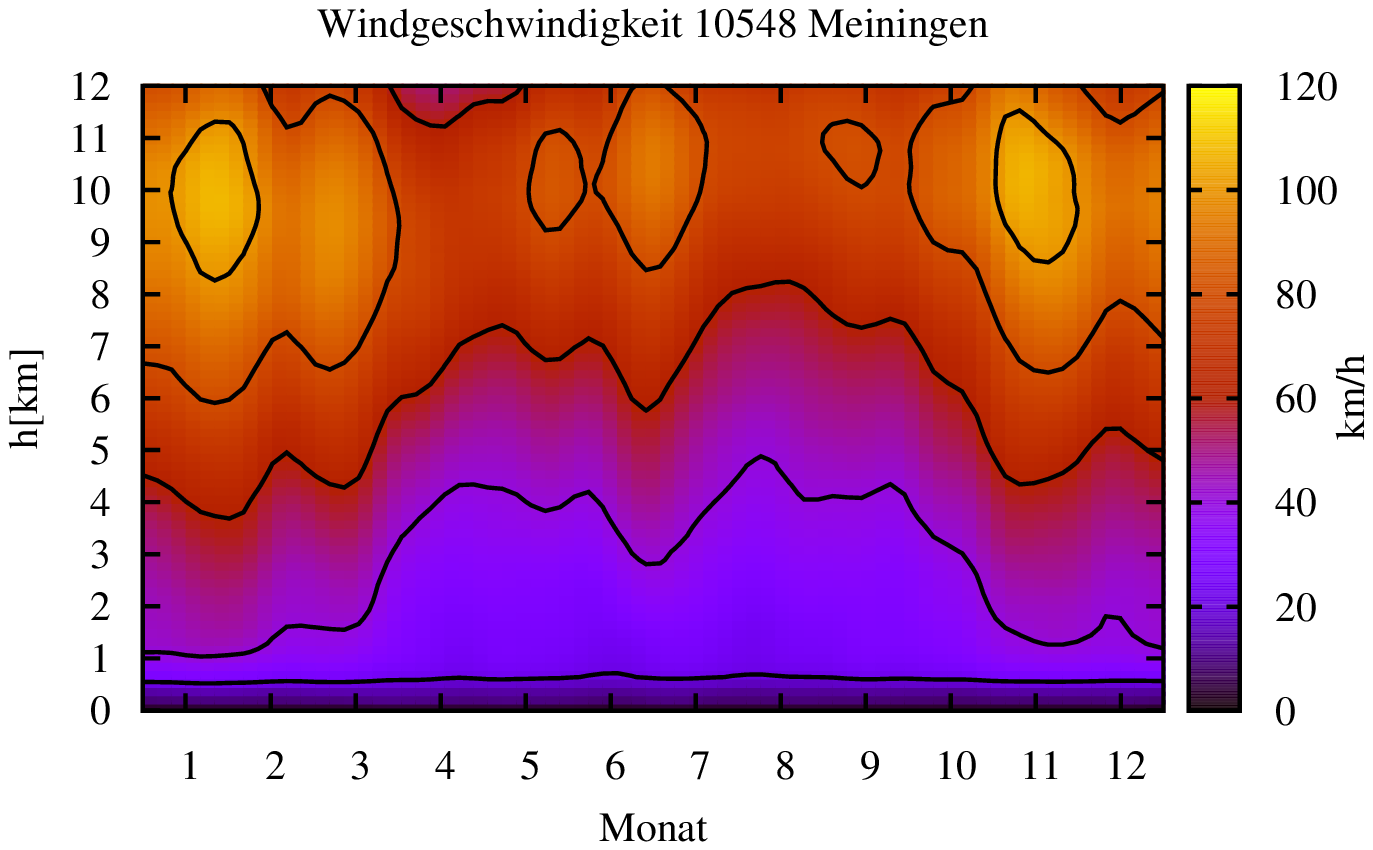}

\includegraphics[scale=0.5,angle=270]{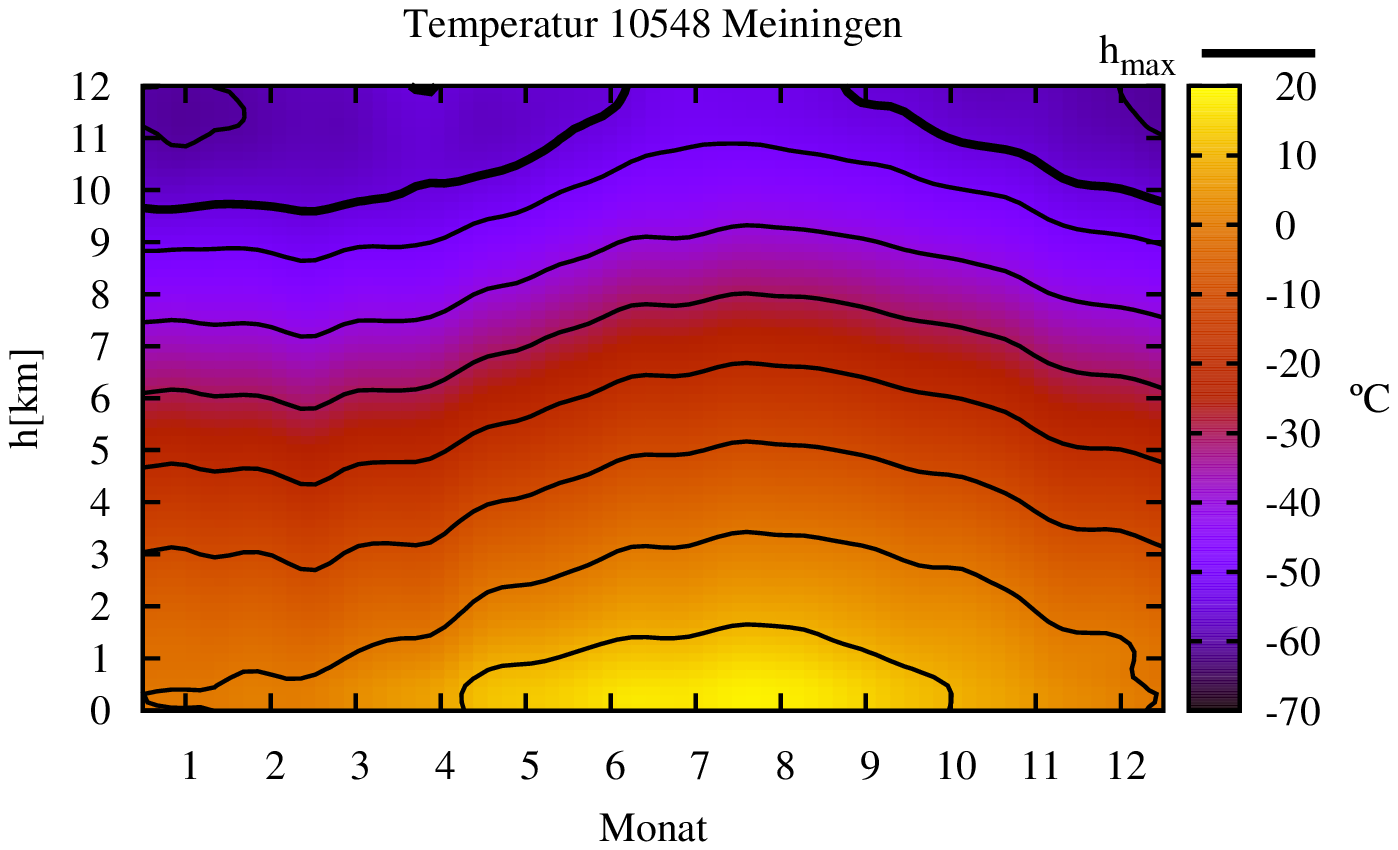}
\end{center}
\caption[Mittlerer Wind und Temperatur Meiningen]{Über die Jahre 2000--2008 gemittelter Wind-
und Temperaturverlauf der Radiosondenstation "`10548 Meiningen"', geglättet mit einem Gauß-Kern
der Breite 7 Tage. $h_{\max}$ markiert die $-55$\celsius-Isotherme. \label{MeiningenD} }
\end{figure}

Die genaue Berechnung von Trajektorien erfordert zwangsläufig bessere zeit-- und
ortsaufgelöste Daten für den betrachteten Zeitraum und Startorte. Als Datenquelle
kommen praktisch nur Radiosondenaufstiege in Frage, die die gewünschten Werte punktuell
als Funktion der Höhe liefern \cite{FederalMetHand}. In dieser Arbeit werden Daten von
\cite{Rawin2010} verwendet, die kostenfrei für Forschung und Lehre zur Verfügung stehen.
Die im HTML-Format
vorliegenden Daten wurden konvertiert und einem einfachen Plausibilitätstest unterworfen,
um Datenfehler zu erkennen. Die Interpolation in der Höhe erfolgt mit einer
Spline-Interpolation dritter Ordnung. Datensätze mit zu wenigen Punkten wurden verworfen,
um Interpolationsprobleme zu vermeiden. Die räumliche Interpolation der Daten $X_i$
an den Stützstellen $\vec x_i$ erfolgt mit einer einfachen inversen
Distanzgewichtung \cite{kahl1986uncertainty}
\begin{eqnarray}
X(\vec x) &=& \frac{ \sum_i X_i w_i }{ \sum_i w_i } \label{IDWEq} \\
w_i &=& \frac{1}{|\vec x_i - \vec x|^4}
\end{eqnarray}
wobei $X$ ein Platzhalter für die zu interpolierende Größe am Ort $\vec x$ ist.
Abweichend von dem oft verwendeten Exponenten 2 in der Gewichtungsfunktion wurde der
Exponent 4 gewählt, der ein glatteres Geschwindigkeitsfeld
liefert und in Vergleichen mit den Ballonflugdaten
geringere Fehler aufzeigt.  Besonders künstliche Maxima der Windgeschwindigkeit an den Stützpunkten
("`Bullseye-Effekt"') werden reduziert. Der Ansatz \ref{IDWEq} gewährleistet, daß die interpolierte
Größe nicht das Maximum der diskreten Werte übersteigen kann. Dies gibt dem
Verfahren eine große Robustheit, bedeutet aber auch eine systematische Unterschätzung
von z.\,B. interpolierten Windgeschwindigkeiten.

Neben der inversen Distanzgewichtung stehen eine Vielzahl von Interpolationsalgorithmen
zur Verfügung (siehe die Übersicht in \cite{li2008review}), die zum Beispiel statistische
Korrelationen oder ausgefeiltere Basisfunktionen verwenden \cite{hannachi2007empirical,Wirth2008}
und dadurch eine höhere Genauigkeit erreichen können.
Da die Interpolationsqualität von Radiosondendaten bereits durch die Dichte der
Messpunkte limitiert ist \cite{stohl1997diagnostic} und komplexe Modelle durch die
fehlende Verfügbarkeit großer Rechenleistung von vornherein ausgeschlossen waren,
wurde die inverse Distanzgewichtung als bester Kompromiss gewählt. Die zeitliche Interpolation
der Daten erfolgt mit einer einfachen linearen Interpolation, um störende numerische Artefakte bei
dynamischen Wettersituationen zu vermeiden.

Eine anschauliche Darstellung der jahreszeitlichen Effekte gibt die Datenstatistik der
Station Meiningen in Abbildung \ref{MeiningenD}. Das Temperaturprofil folgt der Temperatur
am Boden, variiert aber sonst nur wenig. Der Gang der
Windgeschwindigkeit mit der Jahreszeit zeigt eine halbjährliche Periode mit einem
starken Maximum im Winter und einem weiteren Maximum im Sommer, die auf Strahlströme
in rund 10 km Höhe zurückzuführen sind. Die noch deutlich erkennbaren Schwankungen
des gemittelten Windes deuten auf einen zu kurzen Mittelungszeitraum hin.

Die Fluktuation der Windgeschwindigkeit um diesen Mittelwert
wird gut durch eine zweiparametrige Weibull-Dichtefunktion wiedergegeben \cite{pavia1986weibull}:
\begin{eqnarray}
f(v) &=&  \frac{k}{v_0} (v/v_0)^{k-1}\exp\left(- \left(\frac{v}{v_0}\right)^k\right)
\end{eqnarray}
Der Formparameter $k$ ist vom Ort, der Höhe und der Jahreszeit abhängig.
Der jahresgemittelte Wert liegt im Bereich $k=1,8\dots 2,6$, so daß $k=2$ eine realistische
Wahl ist. Für $k=2$ entspricht die Weibull-Verteilung der Rayleigh-Verteilung. Mittelwert E und
Varianz Var sind:
\begin{eqnarray}
\mathrm{E}(v) &=& v_0\Gamma(1/k+1) \\
&=& v_0 \frac{ \sqrt {\pi}}{2} \qquad \mbox{für } k=2 \\
\mathrm{Var}(v) &=& v_0^2 \left( \Gamma(2/k+1)-\Gamma(1/k+1)^2 \right) \\
&=& v_0^2(1-\pi/4)  \qquad \mbox{für } k=2
\end{eqnarray}
$\Gamma$ ist die bekannte Gammafunktion.

\section{Turbulenz}

Neben den großskaligen Luftbewegungen, die als Wind wahrgenommen werden,
ist jede Luftströmung auch von Turbulenz bis hin zu den kleinsten
Skalen geprägt. Für die Simulation eines Ballons ist es sinnvoll,
zwei verschiedene Anteile der Turbulenz getrennt zu betrachten: Vertikale
Luftbewegungen, die das Aufsteigen des Ballons beeinflussen, sowie
Fluktuationen der horizontalen Driftbewegung.
Senkrechte Luftbewegungen beeinflussen das Aufsteigen des Ballons und
entscheiden somit, wann und ob der Ballon seine Platzgrenze erreicht.
Darüber hinaus kann eine veränderte Verweildauer in den verschiedenen
Luftschichten bei starken Windscherungen auch die Flugrichtung beeinflussen.
Das Ausmaß vertikaler Luftbewegungen hängt wesentlich von
der Stabilität der Luftschichtung ab: Nur bei labilen Wetterlagen und starker Konvektion,
in starken Windscherungen oder in der Nähe von orographischen Hindernissen
ist sie
bemerkbar\footnote{Die Stratosphäre zeigt durch die ihr eigene Temperaturinversion eine
besonders große Stabilität. Ein Kartenballon verweilt aber praktisch nur in der Troposphäre.}.
Im Mittel ist die vertikale Luftbewegung recht gering;
Messungen ergeben typische Geschwindigkeit im Bereich weniger
Meter pro Sekunde \cite{Praskovsky2002}. Verkehrsflugzeuge sind nur wenige Prozent der Flugzeit
Turbulenzen ausgesetzt\cite{ehernberger1992stratospheric}, was ebenfalls für eine ruhige
Atmosphäre spricht.
Gegenüber dieser typischen Situation, in der die vertikale Turbulenz auf
den Flug eines Ballons vernachlässigt werden kann, ist der Einfluß
bei starker Konvektion unbestreitbar. Da diese Situation oft
von Niederschlägen und Gewittern begleitet wird, ist ein
abruptes Ende des Ballonflugs aber wahrscheinlicher als eine systematische
Beeinflussung der Trajektorie. Aus diesem Grund wird
dieser Effekt vernachlässigt und später über die empirischen Daten
erschlossen.

Der horizontale Anteil der Turbulenz beschränkt nur die erreichbare Genauigkeit der Trajektorie.
Eine erste Quantifizierung der Turbulenz bietet die Korrelation der
Geschwindigkeiten $\vec v$ an zwei Orten $\vec x_1$ und $\vec x_2$ zu einer festen Zeit $t$:
\begin{eqnarray}
C_v ( \vec x_1,  \vec x_2 ) & := & \left\langle \vec v ( \vec x_1) \cdot \vec v (\vec x_2) \right\rangle \label{CvFunkt}
\end{eqnarray}
Das Mittel ist hier über verschiedene Realisierungen der Strömung zu führen.
Für die Betrachtung von Trajektorienfehlern ist es sinnvoller, die Korrelation der
Geschwindigkeitsdifferenz $\Delta \vec v$ zu betrachten (Strukturfunktion zweiter Ordnung):
\begin{eqnarray}
\Delta \vec v & := & \vec v_2 - \vec v_1 \\
\langle \Delta \vec v \cdot \Delta \vec v \rangle  &=& C_v ( \vec x_1,  \vec x_1 )  + C_v ( \vec x_2,  \vec x_2 )  - 2 C_v ( \vec x_1,  \vec x_2 )
\end{eqnarray}
Bei räumlicher Homogenität der Strömung entfällt die absolute Ortsabhängigkeit:
\begin{eqnarray}
\langle \Delta \vec v \cdot \Delta \vec v \rangle  &=& 2 \langle v^2\rangle  - 2 C_v (\vec x_2 - \vec x_1)
\end{eqnarray}
Eine alternative Charakterisierung ist die spektralen Verteilung der kinetischen Energie,
die aus dem Mittel der fouriertransformierten Geschwindigkeiten $\vec v(\vec k)$ gewonnen wird:
\begin{eqnarray}
E(\vec k) &:=& \frac{1}{2} \langle  \left| \vec v(\vec k) \right|^2 \rangle
\end{eqnarray}
Die Geschwindigkeitskorrelation steht über dem Faltungstheorem in Zusammenhang mit der
spektralen Verteilung der kinetischen Energie
\begin{eqnarray}
E(\vec k) &=&  \frac{1}{2} \tilde C_v(\vec k)
\end{eqnarray}
wobei $\tilde C_v(\vec k)$ die fouriertransformierte Korrelation aus Gl. \ref{CvFunkt} ist.
Bei räumlicher Isotropie kann das Spektrum weiter zu einer eindimensionalen Funktion reduziert werden.
Für ein zweidimensionales Problem (z.\,B. stabil geschichtete Atmosphäre) erhält man die Projektion
\begin{eqnarray}
E(k) &=&  \int E(\vec k')  \delta(k - |\vec k'|) d {k'}^2
\end{eqnarray}
wobei hier $\delta$ die Dirac-Funktion ist. Die Strukturfunktion und das Energiespektrum der Atmosphäre
sind zeit-- und ortsabhängig. Globale Mittelwerte sind aber ein robustes Maß für die Verhältnisse in
der Atmosphäre \cite{skamarock2004evaluating}. Als Basis für die weitere Betrachtung wird deshalb das
global gemittelte Ergebnis von \cite{lindborg1999can} für eine Höhe von etwa 10 km angegeben:
\begin{eqnarray}
\langle \Delta \vec v \cdot \Delta \vec v \rangle \left[ \square \metre \per \square \second \right]&=& 7,6 \times 10^{-3} r^{2/3}
+ 8,9 \times 10^{-9} r^2 - 0,59 \times 10^{-9} r^2 \ln r  \label{VCorrel} \\
E(k)\left[ \cubic \metre \per \square \second \right]  & =& 9,1 \times 10^{-4} k^{-5/3} + 3\times 10^{-10} k^{-3} \\
r & =& |\vec x_2 - \vec x_1|
\end{eqnarray}
$r$ und $k$ sind jeweils in SI-Einheiten einzusetzen.

Die in Gl. \ref{VCorrel} beschriebenen Geschwindigkeitsfluktuationen treiben zwei benachbarte Ballons
mit der Zeit auseinander. Dieser Effekt ist dabei fundamentaler Natur: Selbst wenn die Startorte der Ballons
beliebig nahe beieinander sind, entfernen sich die Ballons durch diesen diffusiven Prozess.
Dieser Aspekt deterministischen Chaos limitiert die Genauigkeit einer Trajektorienberechnung unabhängig von
der Datenqualität und dem verwendeten Algorithmus. Für die mittlere Entfernung $r$ gleichzeitig gestarteter
Ballons als Funktion der Zeit $t$ gilt das klassische Ergebnis von Richardson
\cite{richardson1926atmospheric}
\begin{eqnarray}
\langle r^2(t) \rangle &=& \frac{280}{243} (t\epsilon)^{3}  \label{MeanR} \\
&=& 0,6^2 \frac{ \mathrm{km}^2}{\mathrm{h}^3} t^{3}  \label{DiffErrorKM}
\end{eqnarray}
mit einem Wert des Parameters $\epsilon =$ 0,0187 m$^{2/3}/$s.
Die Gültigkeit dieses Gesetzes hängt von der Luftschicht
und der absoluten Größe der Ballonverteilung ab.
Auf globalen Skalen geht das Richardson-Gesetz in eine normale Diffusion über \cite{lacorata2004evidence}.

Die mittlere Entfernung in Gl. \ref{MeanR} gibt nur eine grobe Vorstellung über die Verteilung der Ballons
im Raum. Eine genauere Charakterisierung erlaubt eine Dichtefunktion $f_r(r,t)$, welche die Verteilung der
Ballons als Funktion der Zeit $t$ und dem Abstand $r$ zum Mittelpunkt der Ballonverteilung angibt.
Über die genaue Form dieser Dichtefunktion herrscht weniger Gewißheit (vergleiche
\cite{biferale2005lagrangian,boffetta2002relative,boffetta2002statistics,bourgoin2006role}).
Allgemein sind die Lösungen aber von der Form $f_r \propto \exp(-r^a)$ \cite{bouchaud1990superdiffusion},
wobei es verschiedene Ansichten über die richtige Wahl des Exponenten $a$ gibt.
Richardson gibt die eindimensionale Dichtefunktion
\begin{eqnarray}
f_r(r,t) &=& \frac{9}{4\sqrt{\pi} (t\epsilon)^{3/2}} \exp\left( -\frac{9}{4}\frac{|r|^{2/3}}{t\epsilon} \right)
\end{eqnarray}
in seiner Arbeit an. Bei einer maximalen Ballonflugzeit von 20 Stunden liefert Gl. \ref{DiffErrorKM} eine
unvermeidbare mittlere Streuung in den Trajektorien gleichzeitig gestarteter Ballons von rund 60 km.
Auch wenn dieser Wert eine kilometergenaue Berechnung von Ballontrajektorien ausschließt, ist er deutlich
kleiner als die Strecke von mehreren 100 Kilometern, die ein Kartenballon typischerweise in 20 Stunden
zurückgelegt. Eine Vertiefung dieser Fehlerbetrachtung erfolgt in Kapitel \ref{SectFehler}.

\chapter{Boden \label{ChapBoden}}

Die Fundwahrscheinlichkeit einer Ballonkarte hängt davon ab, in welcher Umgebung der
Ballon zu Boden geht. Um einen Einblick in ein typisches Landegebiet zu erhalten,
wurde mit dem Programm GoogleEarth\footnote{Siehe http://earth.google.de/index.html 20.11.2006} ein
Experiment durchgeführt. Aus einem Quadrat\footnote{Das Quadrat wird in etwa durch die Städte
Karlsruhe -- Frankfurt -- Lohr am Main -- Sulzbach an der Murr begrenzt.},
das durch den 49ten und 50ten Breitengrad sowie den 8,5$^{\circ}$(E) und 9,5$^{\circ}$(E)
Längenkreis begrenzt wird, wurden 70 Landeplätze per Zufall ausgewählt. Die Zusammensetzung
der gewählten Fläche aus Wald, Landwirtschaft und Siedlungen entspricht in etwa dem zu erwartenden Zielgebiet.

Für jeden fiktiven Fundort wurde der Abstand zum nächsten Weg (d.\,h. Straße/ Feldweg/ Waldweg/ \dots)
und zum Rand der nächsten Siedlung sowie der Typ (Stadt/Wald/Feld) der Fläche bestimmt. Die Anteile
der verschiedenen Flächentypen sind in Tabelle \ref{LandF} zusammengefaßt.
Die aus dieser kleinen Stichprobe ermittelten Daten passen gut zu den umfangreicheren Erhebungen
aus \cite{ISLModul2006}, die ebenfalls in der Tabelle enthalten sind.

\begin{figure}[t]
\begin{center}
\includegraphics[scale=0.8,angle=0]{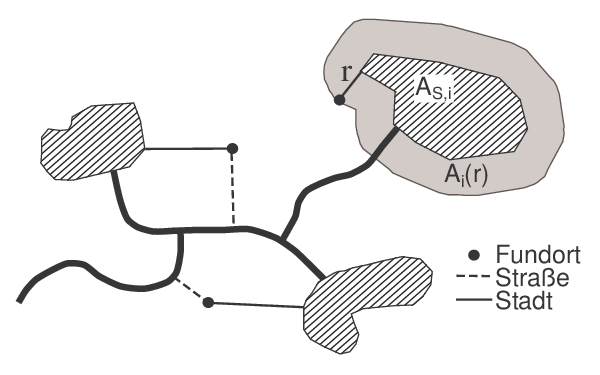} \\
\end{center}
\caption[Topologie der Fundorte]{Darstellung des kürzesten Abstandes $r$ vom Fundort
zur nächsten Straße und zum nächsten Ort. Die in Gl. \ref{pFundSingle} verwendeten Flächen sind für
einen Fundort eingezeichnet.\label{MapDist} }
\end{figure}

Für die Verteilung des Abstandes $r$ zur nächsten Straße bzw. Stadt wird eine kurze theoretische
Betrachtung durchgeführt. Abbildung \ref{MapDist} gibt eine schematische Zusammenfassung der im
Folgenden verwendeten Größen. Befindet sich eine einzelne Struktur $i$ mit der Fläche $A_{S,i}$ in einem
Gebiet mit der Fläche $A_0$, ist die Wahrscheinlichkeit, sich im Abstand $r$ oder weiter entfernt zu befinden
\begin{eqnarray}
p(>r)  &=& \left( 1 - \frac{ A_i(r) }{ A_0 - A_{S,i} } \right) \label{pFundSingle}
\end{eqnarray}
wobei die Nachbarfläche $A_i$ der Bereich aller Punkte ist, die näher als $r$ an der betrachteten Struktur sind.
Durch zufälliges Hinzufügen weiterer (nicht notwendigerweise identischer) Strukturen bei gleichzeitiger
Erweiterung der Gesamtfläche erhält man
\begin{eqnarray}
p(>r)  & = & \prod_{i=1}^n \left( 1 - \frac{A_i(r)}{n A_0 - A_{S,i}} \right)
\end{eqnarray}
Der Grenzübergang $n \rightarrow \infty$ führt schließlich auf die Poisson-Verteilung (vgl. z.\,B. \cite{casertano1985core})
\begin{eqnarray}
p(>r)     & = & \exp \left( - \Sigma \bar A(r) \right) \label{EqADistr} \\
\bar A(r) &:=& \lim_{n \rightarrow \infty} \frac{1}{n} \sum_{i=1}^n A_i(r)
\end{eqnarray}
mit der neu eingeführten Anzahldichte der Strukturen $\Sigma = 1/A_0$. Die Abstandsabhängigkeit der
Nachbarfläche steht in direktem Zusammenhang mit der Hausdorff-Dimension $D_H$ des Randes der betrachteten
Struktur:
\begin{eqnarray}
\bar A(r) & \propto & r^{2-D_H}
\end{eqnarray}
Umgekehrt kann der Längenskala $r$ eine fraktale Dimension zugeordnet werden:
\begin{eqnarray}
D_H(r) &=& 2 - \frac{d \ln(\bar A(r))}{d\ln(r)} \label{EqConDH}
\end{eqnarray}
Nach diesen Vorarbeiten werden nun die Abstandsverteilungen zur nächsten Straße und zum nächsten
Stadtrand empirisch aus dem Experiment ermittelt:
\begin{eqnarray}
p_{\mathrm{Weg}}(>r)   & \approx & \exp\left(-\frac{r}{95\mbox{ m}}\left(1+\sqrt{\frac{r}{4300\mbox{ m}}} \right)^{-1} \right) \qquad r < \mbox{600 m}  \\
p_{\mathrm{Stadt}}(>r) & \approx & \exp\left(-\left(\frac{r}{1,1\mbox{ km}} \right)^{1,5} \right)  \qquad r < \mbox{2 km}
\end{eqnarray}
Die Gültigkeit dieser Verteilungen ist durch den geringen Umfang der Stichprobe auf die
angegebenen Bereiche beschränkt. Während die funktionale Form dieser Verteilungen rein empirisch
bestimmt und daher ohne besondere Bedeutung ist, läßt sich der geometrische Gehalt gut mit Hilfe von
Gl. \ref{EqConDH} erschließen. Straßen erscheinen auf kleinen Skalen eindimensional, gehen aber auf
größeren Skalen in ein Netzwerk mit Dimension~1,5 über. Städte erscheinen als aufgelöste Punktmenge mit
Dimension~0,5, was gut den ländlichen Charakter des betrachteten Gebietes abbildet. Dieser Befund paßt
zu weitergehenden Untersuchungen zu den fraktalen Eigenschaften von
Siedlungsstrukturen \cite{dekeersmaecker2004fractal,frankhauser10comparing}, die Straßen Dimensionen
um 1,5 und ländlichen Siedlungsstrukturen Dimensionen kleiner 1,0 zuordnen.

Wenn man die regelmäßig von Menschen aufgesuchte Fläche (d.h. die Fläche, auf
der ein Kartenballon hauptsächlich gefunden wird) mit den Angaben zu Siedlung und
Verkehr in Tabelle~\ref{LandF} gleichsetzt und problematische Fundorte wie Dächer und Bäume vernachläßigt,
ergibt sich eine obere Grenze der Fundquote von etwas mehr als 10 \%. In dünn besiedelten Gebieten,
z.\,B. in Gebirgen, weiten Teilen der USA und Osteuropa oder in dichter besiedelten Regionen wie Japan ist
dieser Wert entsprechend nach oben oder unten zu korrigieren (vgl. Tab. \ref{UsedData}). Darüber hinaus
können Länder-- bzw. Sprachgrenzen den Anteil der zurückgeschickten Karten merklich
reduzieren (pers. Mitteilung M.~Kollefrath).

\begin{table}
\begin{center}
\begin{tabular}{|l|rrc|} \hline
[\%]             & B.-Würt. (2001)& D(2001) & GoogleEarth \\ \hline
Landwirtschaft   &   46,8         &  53,5 & $46 \pm 6$ \\
Wald             &   38,0         &  29,5 & $44 \pm 6$ \\
Wasser/Sonstiges &    2,0         &   4,3 &  -- \\ \hline
Siedlung u. Verkehr   & 13,2      &  12,3 & $10 \pm 4$ \\
davon: Gebäudefläche  & 53,2      &  --   &  -- \\
Erholungsfläche       &  5,0      &  --   &  -- \\
Verkehrsfläche        & 40,2      &  --   &  -- \\ \hline
\end{tabular}
\end{center}
\caption[Flächennutzung]{Flächennutzung in Baden-Württemberg und Gesamtdeutschland
nach \cite{ISLModul2006}. Die letzte Spalte enthält das GoogleEarth-Experiment. Wald bezieht
sich hier auf baumbestandene Flächen.\label{LandF}}
\end{table}

\chapter{Simulationen \label{ChapSim}}

\section{Gleichungen}

Die in den Kapiteln \ref{ChapBallonP} -- \ref{ChapWetter} vorgestellten Grundlagen  bilden einen
vollständigen Satz von Gleichungen, um die Trajektorie eines gasgefüllten Ballons zu berechnen. Die
Position des Ballons wird durch die geographische Länge $\lambda$, die Breite $\beta$ und die Höhe
über Normalnull $z$ beschrieben. Die Windgeschwindigkeit ist durch $v_u$ (ostwärts) und $v_v$ (nordwärts)
gegeben, vertikale Luftbewegungen werden vernachläßigt. Die Massenbilanz und der Auftrieb sind:
\begin{eqnarray}
M   &=& M_0 + m_G N_G + m_L N_L  \label{EqBlockA_1} \\
F_a &=& Vg\rhoAir(\lambda,\beta,z)-Mg  \label{EqBlockA_2}
\end{eqnarray}
Die Ballonmasse $M$ ergibt sich aus dem Hüllen- und Kartengewicht $M_0$, der Stoffmenge $N_G$ des
Füllgases "`G"' und der eindiffundierten Luftmenge $N_L$. $m_G$ und $m_L$ sind jeweils die Molmasse
des Füllgases und der eindiffundierten Luft.
Die Trajektoriengleichung $\dot {\vec x} = \vec v(\vec x,t)$ in den gewählten Koordinaten ist:
\begin{eqnarray}
\dot \beta   &=& \frac{v_v( \lambda,\beta,z ) }{R_{\mathrm{Erde}}+z}  \label{betadot} \\
\dot \lambda &=& \frac{v_u( \lambda,\beta,z ) }{\cos(\beta) (R_{\mathrm{Erde}}+z)} \label{lambdadot} \\
\dot z       &=& \left\{ \begin{array}{ll}
\mathrm{sign}(F_a) \sqrt{\frac{2 |F_a|}{c_w A_q\rhoAir(\lambda,\beta,z)} } & \mbox{Ballon intakt.} \\
- v_{\infty,0} \sqrt{\frac{\rhoAirRef}{ \rhoAir(\lambda,\beta,z)} } & \mbox{sonst.}
\end{array} \right.  \label{EqSteig}
\end{eqnarray}
Das Ballonvolumen $V$ und die Querschnittsfläche $A_q$ werden mit der Zustandsgleichung des Ballongases ermittelt:
\begin{eqnarray}
V &=& \frac{ (N_L+N_G)RT_L(\lambda,\beta,z)}{p_L(\lambda,\beta,z)+\Delta p}   \label{EqBlockB_1} \\
A &=& 5,0 \, V^{2/3}\\
A_q &=& \frac{A}{4,7}
\end{eqnarray}
$p_L$ und $T_L$ sind das zeit- und ortsabhängige Druck- und Temperaturprofil der Atmosphäre (siehe Kap. \ref{ChapAtm}).
Zuletzt wird die Diffusion des Füllgases bestimmt:
\begin{eqnarray}
\dot N_G &=& -\frac{A^2}{V_{\mathrm{Gummi}}} \delta_G(\lambda,\beta,z) \frac{N_G}{V}  \\
\dot N_L &=& -\frac{A^2}{V_{\mathrm{Gummi}}} \delta_L(\lambda,\beta,z) \left( \frac{N_L}{V}-n_L(\lambda,\beta,z) \right)
\end{eqnarray}
Der Ballon gilt als geplatzt, sobald das Volumen größer als das Maximalvolumen $V_{\max}$ ist oder die
Temperatur unter die Glasübergangstemperatur $T_{\mathrm{Burst}}$ des Gummis fällt.

\begin{table}
\begin{center}
\begin{tabular}{|l@{\,:\,}rl|} \hline
$\Delta p$              & 5300  & Pa \\
$V_{\max}$              &   30  & l  \\
$M_0$                   &  7,1  & g  \\
$M_{\mathrm{Gummi}}$    &  5,1  & g  \\
$c_w$                   &  0,93 &    \\
$ v_{\infty,0}$         &  1,0  & m/s\\
$T_{\mathrm{Burst}}$    & $-55$ & \celsius{} \\ \hline
\end{tabular}
\end{center}
\caption[Ballondaten]{Referenzparameter zur Beschreibung eines Kartenballons.\label{BallonP}}
\end{table}

\section{Theorie}

Zunächst wird eine analytische Betrachtung der Gleichungen durchgeführt, um die
grundlegenden Effekte zu verstehen. Zu diesem Zweck wird das rein höhenabhängige Atmosphärenmodell
in Gl.~\ref{pAtmEq}--\ref{TAtmEq} und ein kartesisches Koordinatensystem für die Ballonposition verwendet.
Die Gleichungen \ref{betadot}--\ref{lambdadot} ändern sich dadurch auf:
\begin{eqnarray}
\dot x &=& v(z) \\
\dot y &=& 0 \label{EqBlockB_2}
\end{eqnarray}
Für die folgenden Studien wird ein typischer Ballon betrachtet, dessen Parameter in Tabelle \ref{BallonP}
zusammengestellt sind. Dieser Parametersatz wird auch in Kapitel \ref{BalFlight} verwendet, wenn keine
detaillierten Informationen über die gestarteten Ballons vorlagen. Durch das mögliche Platzen des Ballons
während des Fluges können drei verschiedene Trajektorientypen unterschieden werden:
\begin{enumerate}[I]
\item  \label{Case1} Der Ballon bleibt während des gesamten Fluges intakt.
\item  \label{Case2} Der Ballon erreicht die Maximalhöhe und platzt durch den Glasübergang des Ballongummis.
\item  \label{Case3} Der Ballon bleibt unter der Maximalhöhe, platzt aber durch die Überdehnung der Hülle.
\end{enumerate}
Fall \ref{Case1} ermöglicht offensichtlich die größten Flugweiten, so daß dieses Szenario am detailliertesten
diskutiert wird. Anschließend werden die beiden letzten Szenarien angesprochen.

\subsection*{Fall \ref{Case1}}

Um ein übersichtlicheres Gleichungssystem zu erhalten, wird die Diffusion der Luft sowie der Drucksprung des
Ballons vernachläßigt ($\delta_L=0$, $\Delta p = 0$). Das Eindiffundieren von Luft vergrößert den Ballon
und erhöht dadurch indirekt den Verlust des Füllgases. Darüber hinaus verringert sich für $\Delta p >0$ der
Auftrieb, was aber nur in großen Höhen einen merklichen Beitrag liefert. Insgesamt ist der Einfluß auf die
Ballontrajektorie aber zu vernachlässigen. Der endliche Drucksprung des Ballons sorgt für einen zusätzlichen
Verlust des Auftriebs beim Aufsteigen, der aber "`gespeichert"' und beim Abstieg wieder freigesetzt wird.
Durch die stärkere Kompression in großen Höhen kann der Ballon am Boden auch viel stärker gefüllt werden, ohne
zu platzen. Dies führt zu einer deutlichen Asymmetrie der Trajektorie und ermöglicht größere Flugweiten. Trotz
des deutlichen Einflusses des Drucksprungs auf die Trajektorie, der in Kapitel \ref{SectNumInt} genauer untersucht
wird, gibt eine vereinfachte Betrachtung wertvolle Einsichten.

\clearpage

\noindent
Unter Verwendung dieser Näherungen erhält man
aus Gl.~\ref{EqBlockA_1}--\ref{EqBlockA_2} und Gl.~\ref{EqBlockB_1}--\ref{EqBlockB_2}:
\begin{eqnarray}
F_a       &=& (N_G (m_L - m_G) - M_0)g \\
\dot z    &=&  \mathrm{sign}(F_a) \sqrt{\frac{2 |F_a|}{c_w A_q\rhoAir(z)} } \\
\dot x    &=& v(z)  \\
\dot N_G &=& -\frac{A^2}{V_{\mathrm{Gummi}}} \delta_G(z) \frac{\rhoAir}{m_L}
\end{eqnarray}
Zunächst werden die Gleichungen zusammengefaßt und dimensionslos formuliert.
Aus der Leermasse $M_0$, der Luftdichte $\rhoAir$ und der Permeabilität $\delta_G$ am Boden
können alle nötigen Skalengrößen gewonnen werden:
\begin{eqnarray}
V_0   &:=& \frac{M_0}{\rhoAirZ} \frac{m_L}{m_L - m_G} \qquad  \tau_0 :=  \frac{V_{\mathrm {Gummi}}V_0}{\delta_0 A_0^2} \\
f    &:=& \frac{F_a}{M_0g} \qquad \qquad v_{z,0} :=  \sqrt{\frac{9,4 M_0 g}{c_w A_0\rhoAirZ} }
\end{eqnarray}
$V_0$ ist das Volumen eines Ballons mit $F_a=0$, $\tau_0$ die entsprechende Gasverlustzeitskala, $f$ der
Auftrieb in Einheiten des Gesamtgewichts und $v_{z,0}$ die Steiggeschwindigkeit eines Ballons mit Auftrieb $f=1$.
Die Luftdichte $\rho$ und Permeabilität $\delta$ werden in Einheiten der Werte am Boden ($z=0$) ausgedrückt:
\begin{eqnarray}
\tilde \rho   &:=& \frac{\rho(z)  }{\rho(0)} , \qquad  \tilde \delta := \frac{\delta(z)}{\delta(0)}
\end{eqnarray}
Die dimensionslosen Gleichungen sind:
\begin{eqnarray}
\dot f &=& -\frac{1}{\tau_0}(1+f)^{4/3}  \frac{\tilde \delta_G}{\sqrt[3]{\tilde \rhoAir}}\\
\dot z &=& \mathrm{sign}(f) v_{z,0} \frac{\sqrt{|f|}}{(1+f)^{1/3}{\tilde \rhoAir}^{1/6}}
\end{eqnarray}
Diese können zu
\begin{eqnarray}
\frac{\tilde \delta_G}{{\tilde \rhoAir}^{1/6}} \frac{dz}{L_0} &=& - \mathrm{sign}(f)  \frac{\sqrt{|f|}}{(1+f)^{5/3}} df
\label{GrundInt} \\
L_0 &:=& v_{z,0} \tau_0
\end{eqnarray}
zusammengefaßt werden. Die linke Seite enthält nur von $z$ abhängige Größen und kann integriert werden:
\begin{eqnarray}
\tilde z &:=& \int_0^z \frac{\tilde \delta_G}{{\tilde \rhoAir}^{1/6}}  \frac{dz'}{L_0}
\end{eqnarray}
In erster Näherung kann $\tilde z \approx z/L_0$ gesetzt werden.
Zunächst wird $f \ll 1$ untersucht.
Die Integration der Differentialgleichung ab $\tilde z=0$ liefert:
\begin{eqnarray}
\tilde z(f) &=&\frac{2}{3}\left( f_0^{3/2} - |f|^{3/2}\right) + \mathcal{O}(f^{5/3}) \label{z_fFunc}
\end{eqnarray}
Die maximale Flugweite wird erreicht, wenn der Ballon knapp unter der Maximalhöhe $h_{\max}$
bleibt und intakt zu Boden sinkt. Damit folgt für den optimalen Auftrieb:
\begin{eqnarray}
F_{a,\opt} &=& M_0 g \left( \frac{3}{2} \tilde h_{\max} \right)^{2/3}  \label{FAopt}
\end{eqnarray}

\begin{figure}[ht]
\begin{center}
\includegraphics[scale=0.4,angle=270]{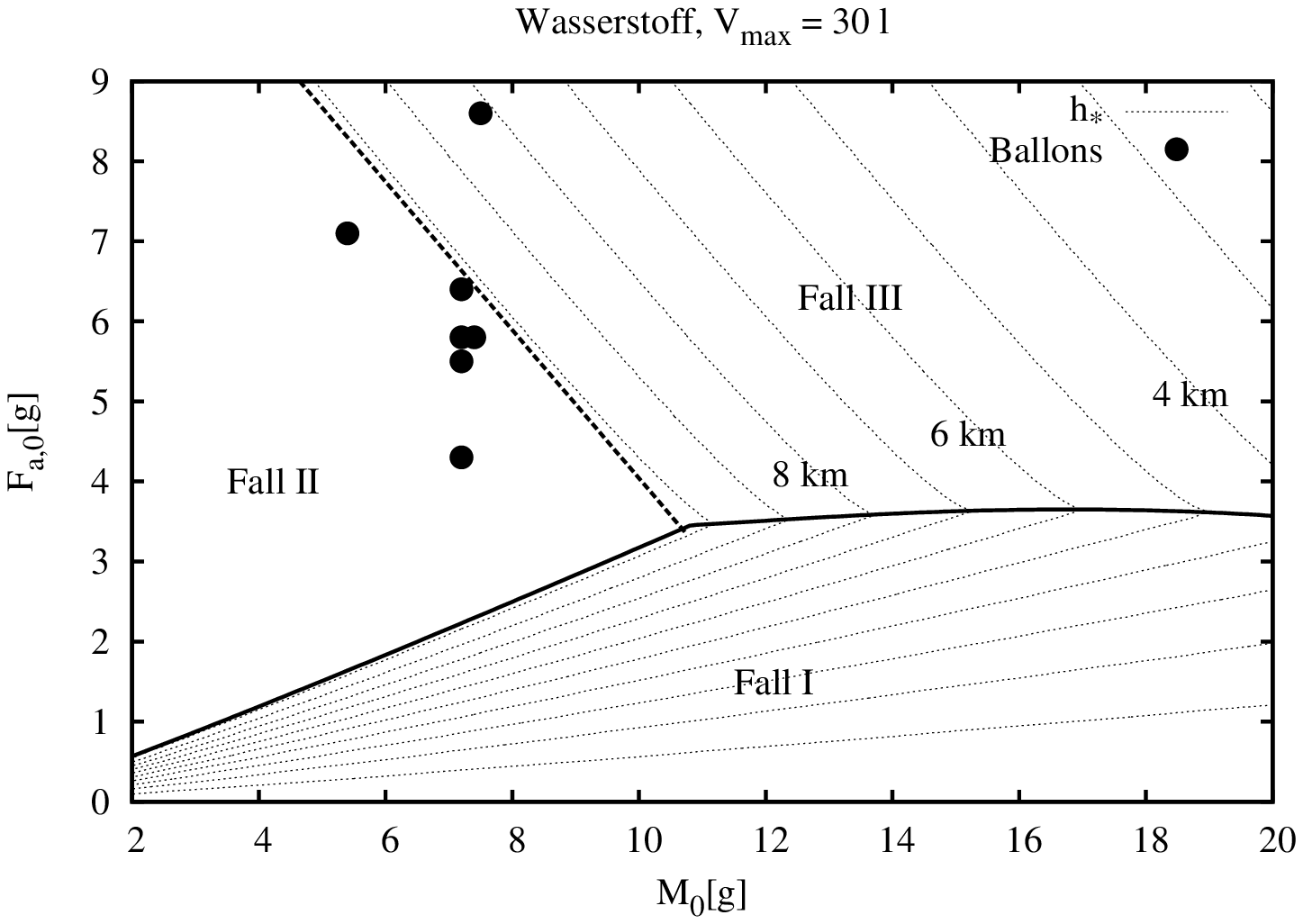} \\
\vspace*{3cm}
\includegraphics[scale=0.4,angle=270]{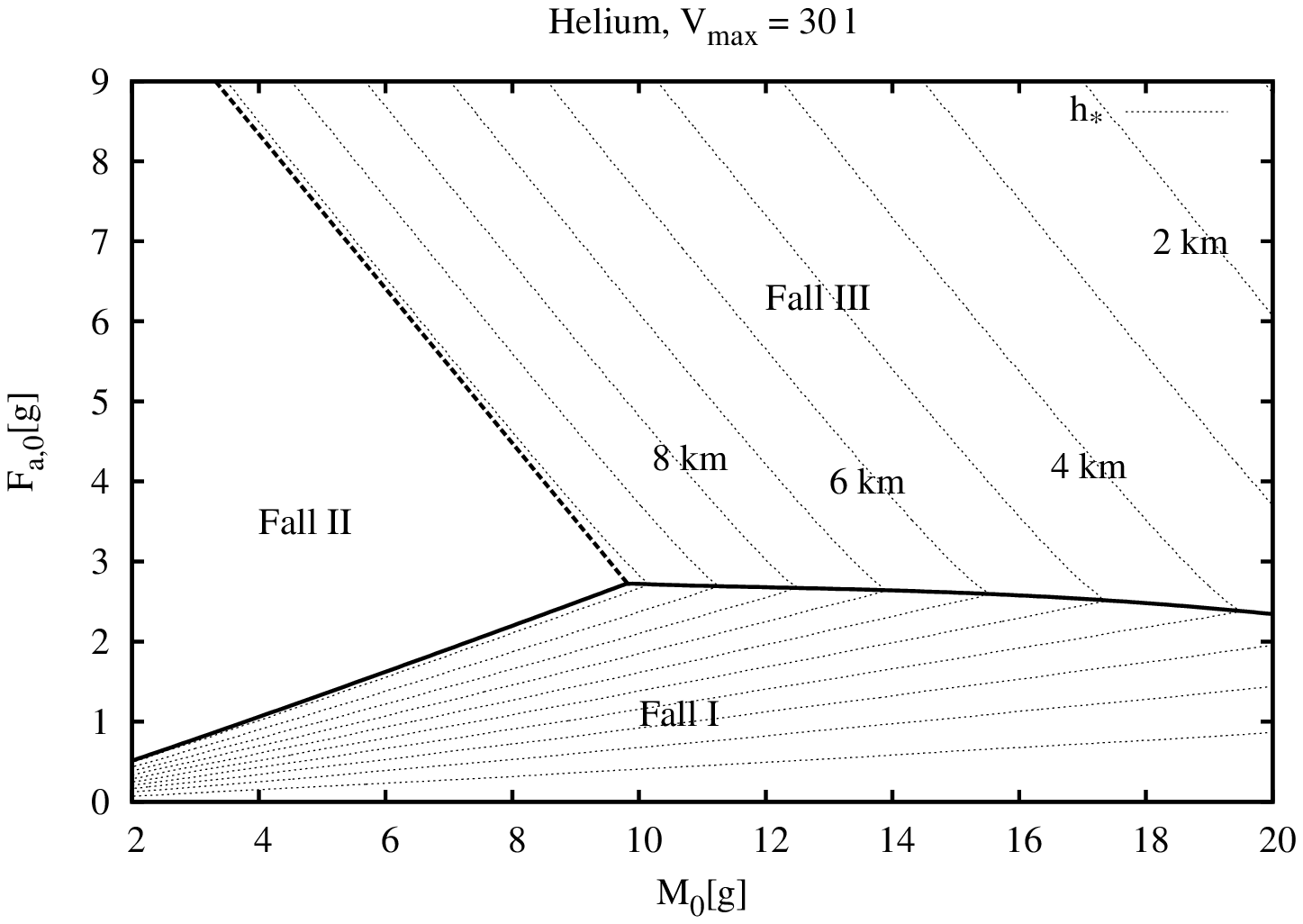}
\end{center}
\caption[Maximale Höhe für verschiedene Ballonmassen]{Ballonflüge in Abhängigkeit von der Leermasse $M_0$
und dem Auftrieb $F_a$. Die Konturlinien beschreiben die jeweils erreichte Scheitelhöhe $h^*$. Zum Vergleich
sind die Ballons aus Tabelle \ref{DataH2} mit eingetragen.\label{FlugMaps} }
\end{figure}
\clearpage

\noindent
Die Flugweite ergibt sich aus der Zeitintegration des Windprofils in Gl. \ref{VAtmEq}:
\begin{eqnarray}
d &=& \int_0^{t_{\max}} v(z(t)) dt \\
& \approx & 2 \tau_0 f_0  v(3/5 h_{\max})
\end{eqnarray}
Für den optimalen Startauftrieb nach Gl. \ref{FAopt} erhält man:
\begin{eqnarray}
d_{\max} & \approx & v(3/5 h_{\max}) \left( \frac{h_{\max}^3\rhoAirZ}{M_0}  \right)^{2/9}
\sqrt[3]{\frac{c_w V_{\mathrm{Gummi}}}{g \delta_{G,0}}} \left( \frac{m_L}{m_L-m_G}\right)^{1/9}
\end{eqnarray}
$d_{\max}$ ist die maximale Flugweite, die mit einem gegebenen Kartenballon erreicht werden kann.
Bei größeren Werten der Leermasse $M_0$ ist auch ein größerer Ballon nötig, so daß der Ballon schließlich
nicht mehr die maximale Höhe $h_{\max}$ erreichen kann und vorher platzt (Übergang zu Fall~\ref{Case3}).
In diesem Fall muß erst die Scheitelhöhe $h_*$ ermittelt werden:
\begin{eqnarray}
V_{\max}\rho_L(h_*) = M(h_*)
\end{eqnarray}
Mit dieser kann dann der optimale Auftrieb entsprechend berechnet werden:
\begin{eqnarray}
F_{a,\opt} &=& M_0 g \left( \frac{3}{2} \tilde h_* \right)^{2/3} \label{FaOptB}
\end{eqnarray}

\subsection*{Fall \ref{Case2}}

Wird der Auftrieb geringfügig gesteigert, platzt der Ballon durch die Kälte und fällt zu Boden. Die nun
erreichte Entfernung $d_{\mathrm{burst}}$ ist die maximal Flugweite für einen Ballon, der vor dem Erreichen
des Scheitelpunkts platzt und zu Boden fällt. Dieser Fall tritt ein für:
\begin{eqnarray}
0 <  F_a(h_{\max})/g & \leq & V_{\max} \rhoAir(h_{\max}) - M(h_{\max}) \label{Case2Gl} \\
h_* &=& h_{\max}
\end{eqnarray}
In diesem Regime ist die Scheitelhöhe gleich der Maximalhöhe.
Die Ungleichung \ref{Case2Gl} kann für Kartenballons mit großer Leermasse nicht erfüllt werden.
Somit ist das Erreichen der Maximalhöhe unabhängig von der Ballonfüllung
ausgeschlossen (Übergang zu Fall \ref{Case3}).

\subsection*{Fall \ref{Case3}}

Ab einem bestimmten Auftrieb platzt der Ballon bereits durch die überdehnte Hülle, bevor die Kälte
in großen Höhen den Ballon zerstört. Dies definiert eine weitere Flugweite $d_{\mathrm{burst2}}$.
Der Ballon platzt in der maximalen Höhe $h_{\max}$ (oder früher) wenn
\begin{eqnarray}
F_a(h_{\max})/g &>& V_{\max} \rhoAir(h_{\max}) - M(h_{\max})
\end{eqnarray}
gilt. Für den dimensionslosen Auftrieb in dieser Höhe bedeutet dies:
\begin{eqnarray}
f(h_{\max}) &>&  V_{\max} \rhoAir(h_{\max})/M_0 - M(h_{\max})/M_0
\end{eqnarray}
Im Bereich großer Auftriebe (Übergang zu Fall~\ref{Case2}) wird Gl. \ref{z_fFunc} ungenau und wird
durch folgende Näherung ersetzt (mit einem Fehler von 10 \% für $0,2<f<3$):
\begin{eqnarray}
\tilde z & \approx & \frac{1}{3}(f_0-f)
\end{eqnarray}
Nach dem Einsetzen dieser Näherung erhält man für den Startauftrieb am Boden
die Ungleichung:
\begin{eqnarray}
F_{a,0} & \gtrsim &  M_{\max} g - (1- 3 \tilde h_{\max}) M_0 g \label{FBurst2Burst} \\
M_{\max} & \approx &   V_{\max} \rhoAir(h_{\max})
\end{eqnarray}
Dies ist eine notwendige Bedingung für das Platzen des Ballons vor dem Erreichen der Maximalhöhe.
Nur Ballons, die leichter als $M_{\max}$ sind, können die Maximalhöhe intakt erreichen.
Der genaue Wert hängt vom Füllgas ab und liegt im Bereich von 10 -- 11 g (vgl. Abb. \ref{FlugMaps}).
Die Scheitelhöhe in diesem Regime ist die Lösung der Gleichung:
\begin{eqnarray}
F_a(h_*)/g & = & V_{\max} \rhoAir(h_*) - M(h_*)
\end{eqnarray}

\section{Numerische Integration \label{SectNumInt}}

Zur genaueren Untersuchung wurden die Modellgleichungen mit einem Runge-Kutta Verfahren zweiter
Ordnung numerisch integriert \cite{abramowitz1972handbook}. Als Zeitschritt wurden 50 Sekunden gewählt.
Durch diese Wahl können Integrationsfehler vernachläßigt werden und die erzielte Höhenauflösung
von rund 100 Metern entspricht gut der vertikalen Auflösung der interpolierten Wetterdaten.

Die in der Praxis wichtigsten (und am leichtesten beeinflußbaren) Parameter eines Kartenballons sind
die Leermasse, der Startauftrieb (bzw. die Ballonfüllung) und die Wahl des Füllgases.
Abbildung \ref{FlugMaps} zeigt die Integration des Musterballons aus Tabelle \ref{BallonP}, jeweils mit Helium und
Wasserstoff als Gasfüllung und verschiedenen Leermassen und Startauftriebskräften. Zur deutlicheren Darstellung
sind auch Leermassen kleiner 5,1 g enthalten, obwohl diese einer Nutzlast mit negativer Masse entsprechen.
Zur besseren Vergleichbarkeit wird die Auftriebskraft mit der
Erdbeschleunigung\footnote{Als Standardwert wird in dieser Arbeit $g = 9,81$ m$/$s$^2$ verwendet.}
in eine äquivalente Masse umgerechnet dargestellt.
Die bereits beschriebenen Trajektorientypen sind gut zu erkennen. Die Trennlinien zwischen
den Bereichen entsprechen den Gleichungen \ref{FAopt} (I--II), \ref{FaOptB} (I--III) und \ref{FBurst2Burst} (II--III)
und stimmen sehr gut mit dem gefundenen qualitativen Verhalten überein.
Der Übergang zwischen dem Platzen durch Kälte (II) und Überdehnung (III) wird nur
durch die Modellierung als scharfe Linie erkennbar. In der Praxis faßt man diesen Bereich besser
als einen stetigen Übergang auf. Zur Orientierung sind die Kartenballons aus der Versuchsreihe
"`Glaschke"' (siehe Tab. \ref{UsedData}) mit eingetragen. Sie zeigen deutlich, daß für die verwendeten Ballons
die maximale Flugweite von der Kälteempfindlichkeit des Ballon\-gummis abhängt, während die Platzgröße
zweitrangig ist. Dieser Befund läßt sich auch auf die bei Ballonwettbewerben verwendeten
Modelle\footnote{Ballongrößen über 25 cm $\varnothing$ sind bereits ausreichend.} übertragen.

\begin{figure}
\begin{center}
\includegraphics[scale=0.4,angle=270]{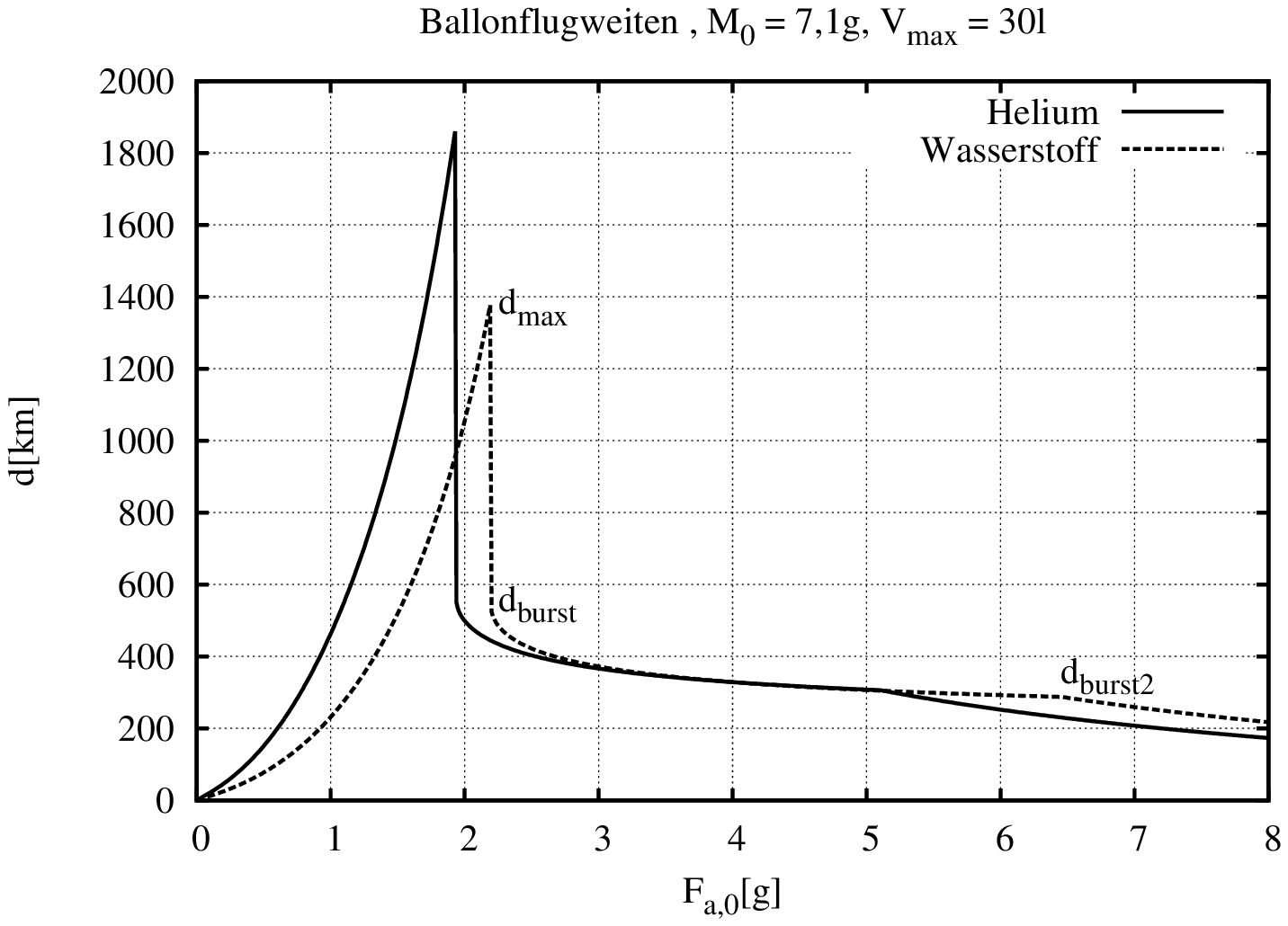} \\
\end{center}
\caption[Weite als Funktion des Auftriebs]{Flugweite als Funktion des Auftriebs für zwei Gasfüllungen.\label{DvsFa_2Gas} }
\end{figure}

Die Flugweite bei fester Leermasse und verschiedenen Startauftrieben gibt einen tieferen
Einblick in die Abhängigkeit der Flugweite von der Ballonfüllung (Abb. \ref{DvsFa_2Gas}).
Deutlich erkennbar sind die bereits angesprochenen drei ausgezeichneten Flugweiten $d_{\max}$,
$d_{\mathrm{burst}}$ und $d_{\mathrm{burst2}}$.  Die jeweils größte erreichte Flugweite wird
durch $d_{\max}$ markiert. Links davon befinden sich Trajektorien mit Ballons, die nach dem Erreichen
der Scheitelhöhe intakt absteigen. Unmittelbar nach rechts fällt die Flugweite auf $d_{\mathrm{burst}}$
ab -- hier fallen die Ballons geplatzt zu Boden. Bei noch größerem Auftrieb nimmt die Scheitelhöhe und
Flugweite weiter ab ($d_{\mathrm{burst}2}$).
Besonders deutlich wird die starke Abhängigkeit der maximalen Flugweite von der Ballonfüllung. Eine
Genauigkeit von mindestens 0,5 l beim Füllen ist notwendig, und selbst dann kann die Wetterabhängigkeit
von $F_{a,\opt}$ die genaue Lage um bis zu einem Gramm verschieben. Der Verlauf der Scheitelhöhen kann
aus einem Schnitt mit $M_0 = 7,1$ g in Abbildung \ref{FlugMaps}  entnommen werden.

\begin{figure}
\begin{center}
\includegraphics[scale=0.5,angle=270]{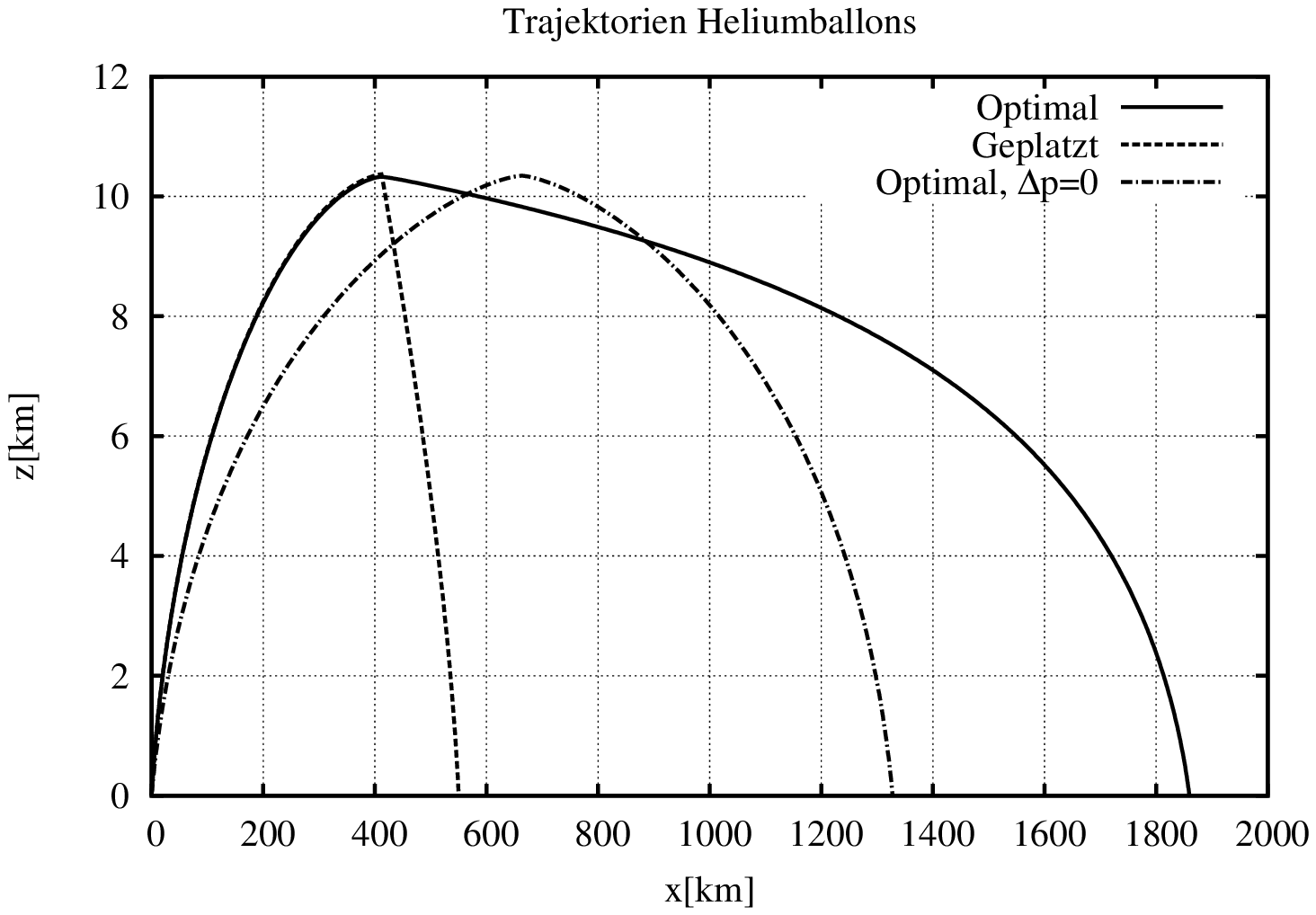}
\end{center}
\caption[Ballontrajektorien]{Trajektorien für einen optimal gefüllten Ballon, einen leicht übervollen
Ballon und einen optimal gefüllten Ballon ohne Drucksprung. \label{BTrajekt} }
\end{figure}

\begin{figure}
\begin{center}
\includegraphics[scale=0.5,angle=270]{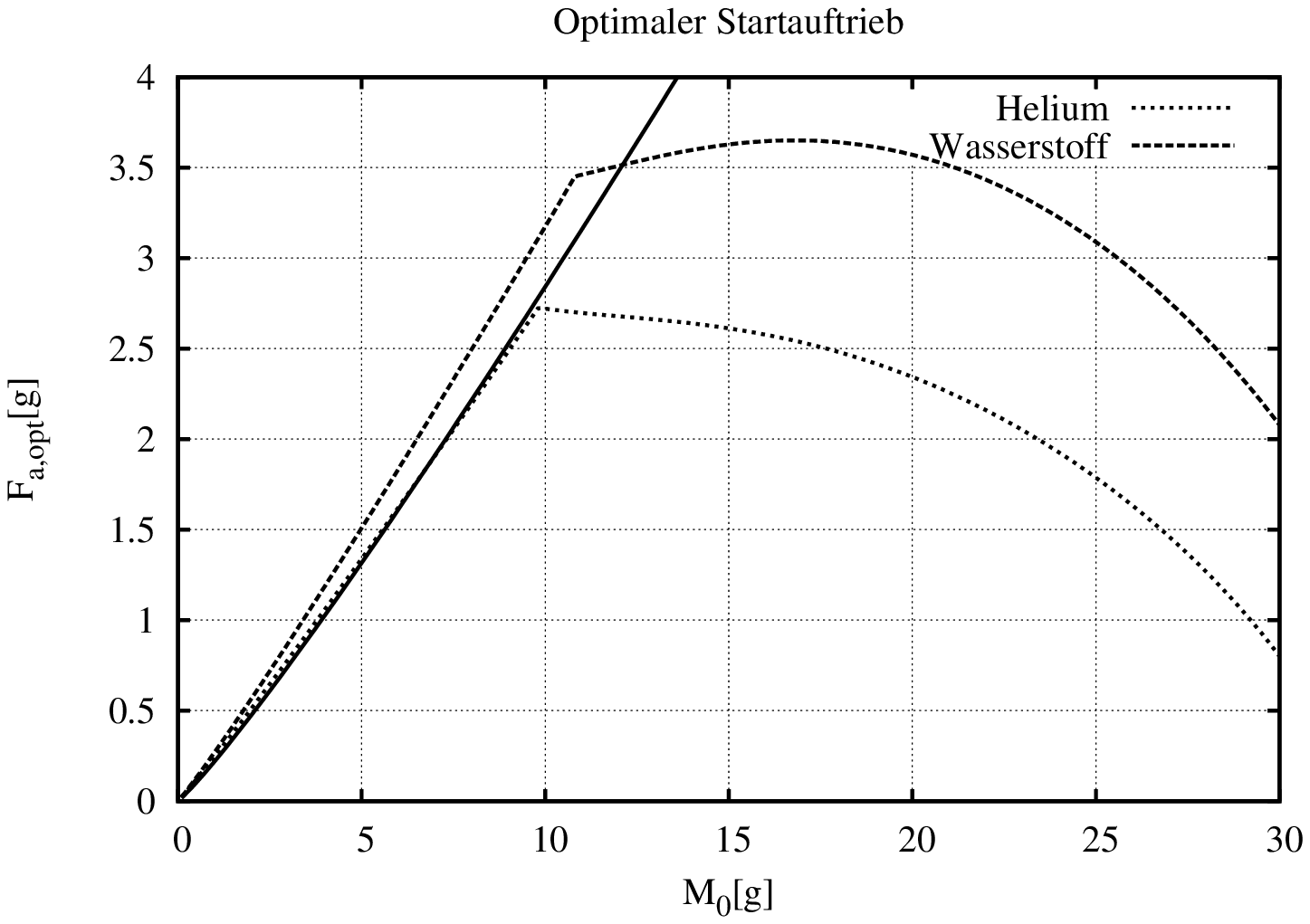}
\end{center}
\caption[$F_{a,\opt}$ als Funktion der Leermasse]{$F_{a,\opt}$ als Funktion der Leermasse.
Die durchgezogene Linie ist Gl. \ref{FaOptTheo}. \label{FaOptM0} }
\end{figure}

Die numerische Lösung der Differentialgleichungen bietet auch die Möglichkeit, den Einfluß des vernachlässigten
Drucksprungs auf die Ballontrajektorie zu diskutieren. Zum Vergleich wurde ein optimal gefüllter Heliumballon mit
und ohne Drucksprung gerechnet sowie ein leicht überfüllter Ballon (Abb. \ref{BTrajekt}). Der endliche Drucksprung
$\Delta p$ macht sich in einer ausgeprägten Asymmetrie des Ballonaufstiegs bemerkbar: Beim Aufsteigen wird
der Ballon stärker komprimiert (relativ zu einem Ballon mit $\Delta p=0$), so daß der Auftrieb schneller abnimmt
und der Ballon regelrecht gebremst wird. Beim anschließenden Abstieg wird dieser "`gespeicherte"' Auftrieb wieder
freigesetzt und verlangsamt den Sinkflug. Dieser Effekt wird zusätzlich durch die starke Temperaturabhängigkeit
der Permeabilität verstärkt\footnote{Roberts \cite{roberts1995dynamics} geht von einer anderen
Temperaturabhängigkeit der Permeabilität aus und findet nur eine schwach ausgeprägte Asymmetrie.}. Dieser
Mechanismus erklärt auch, warum die Flugweite $d_{\mathrm{burst}}$ weit geringer ist als die Hälfte von
$d_{\max}$. Trotz dieser starken Vereinfachung der analytischen Rechnung behalten die abgeleiteten
Skalierungsgesetze ihre Gültigkeit (vgl. Abb. \ref{FaOptM0}).

Abschließend wird noch der optimale Auftrieb für verschiedene Leermassen in Abbildung \ref{FaOptM0}
dargestellt. Leermassen kleiner 10 g entsprechen Fall~I und sind für das Erreichen großer Flugweiten
besonders zu empfehlen. Anhand der numerischen Kurven kann auch die Proportionalitätskonstante in
Gl. \ref{FAopt} bestimmt werden:
\begin{eqnarray}
F_{a,\opt}[\mbox{g}] & \approx & 0,22 \left( \frac{M_0}{1 \mbox{ g} }\right)^{10/9} \label{FaOptTheo}
\end{eqnarray}
Als Faustformel gilt, daß der Auftrieb rund 20 \% der Leermasse betragen sollte.

\chapter{Ballonflüge \label{BalFlight}}

\section{Einleitung}

Der Entwurf eines Trajektorienmodells muß mit Experimenten verglichen werden,
um die verwendeten Annahmen beurteilen zu können und Hinweise auf unbekannte
Phänomene zu erhalten.
Die praktische Durchführung von Experimenten wird durch mehrere Faktoren
erschwert: Kartenballons haben nur
einen geringen Auftrieb, der es unmöglich macht Messtechnik anzubringen,
ohne die Ballonparameter zu beeinflussen. Dies schließt detaillierte Informationen
über die Flugbahn aus. Werden vor dem Start eines Ballons alle Parameter
erfaßt, sind die einzigen zugänglichen Informationen die Balloneigenschaften
beim Start und der Landeplatz. Dieser Ansatz wird jedoch durch die geringe
Fundquote von im besten Fall 10 \% (vgl. Tab. \ref{UsedData}) erschwert,
die mehrere tausend Ballons erfordert, um statistisch signifikante Aussagen
treffen zu können.

Aus diesem Grund wurde zunächst eine Vorstudie unter kontrollierten Bedingungen
durchgeführt (siehe Kapitel \ref{H2Ballons}) und durch Daten von Ballonwettbewerben ergänzt.
Der Vorteil von Ballonwettbewerben besteht in den umfangreichen Daten, die oft über
mehrere Jahre mit mehreren Dutzenden gefundenen Karten pro Wettbewerb anzugeben sind.
Der Nachteil besteht in den unbekannten Ballonparametern, zumal die wissenschaftliche
Auswertung erst im Nachhinein erfolgt und typischerweise nicht Teil der Wettbewerbsplanung ist. Eine
besondere Datenquelle besteht in der Zusammenarbeit mit Herrn Haberlandt \cite{Haberlandt},
der Ballonstarts über Jahre privat durchgeführt hat und genaue Angaben zu den Startbedingungen
machen kann.

Die Eckdaten jedes gefundenen Kartenballons bestehen somit aus Startort und Startzeit sowie dem Fundort.
Einblicke in den Ballonflug und die beteiligte Physik können nur indirekt gewonnen werden,
indem jeder Fundort mit einer individuell berechneten Trajektorie verglichen wird. Unstimmigkeiten
geben dann einen Hinweis auf vernachlässigte Effekte oder Fehler des Trajektorienmodells.

Die Anzahl von mehreren tausend Funden macht eine automatische Auswertung zwingend erforderlich.
Zunächst wurden alle verfügbaren Daten einheitlich formatiert und die
entsprechenden Trajektorien automatisch berechnet. Die Berechnung und Auswertung
der Trajektorien erfolgte mit einem selbstentwickelten
Computerprogramm\footnote{Anhang \ref{UsedProg} stellt das entwickelte Programm kurz vor.}.

\section{Funddaten}

Alle verwendeten Daten sind in Tabelle \ref{UsedData} zusammengefaßt.
Angegeben ist jeweils der Zeitraum, in dem die Ballonstarts erfolgt sind,
die Anzahl der Starttage in dieser Zeit sowie die
Gesamtzahl der gefundenen Ballons. In den letzten Spalten ist das verwendete Füllgas und
die Fundquote angegeben.

"`Campingkirche"' und "`Trostberg"' sind regelmäßig stattfindende
Ballonflugwettbewerbe, die auch aktuell noch durchgeführt
werden\footnote{Daten nach 2008 waren teilweise erst nach dem Beginn der Auswertung verfügbar und
sind in dieser Arbeit noch nicht berücksichtigt.}.
Sie bilden über Jahre einen homogenen Datensatz. Die Daten "`Trostberg"'
enthalten auch Informationen zu der Zeit, die zwischen dem Start eines Ballons und
dem Fund der Karte vergangen ist.

Die umfangreichste Datenquelle ist "`Haberlandt"'. Die Vielfalt der durchgeführten
Starts machte eine Aufteilung nach den Angaben des Experimentators in drei Blöcke
nötig: Block A enthält explorative Versuche mit
Ballonvolumina\footnote{Es sind jeweils die geometrischen Volumina angegeben.}
bis zu 17 l, die schrittweise auf etwa 8,2 l reduziert wurden. Als Mittelwert werden 13 l
angenommen. Die Ballons wurden über den Tag verteilt bis in den frühen Abend gestartet.
Block B sind Starts mit der als optimal erkannten Füllung von 8,2~l und in Block C
wurden Wettervorhersagen des Windes und des Niederschlags mit einbezogen, um die
Flugweite weiter zu erhöhen und die Verlustrate durch ungünstiges Wetter zu
minimieren. Starts erfolgten hier nur noch am frühen Abend.

"`Glaschke"' repräsentiert eigene Ballonexperimente, in denen neben der Uhrzeit,
dem Gewicht und Auftrieb auch die Ballons selbst photographisch erfaßt wurden.

Die letzten beiden Zeilen sind eigenständige Forschungsarbeiten, die Kartenballons
zur Untersuchung von Luftströmungen eingesetzt haben. Eine genaue Untersuchung der
Ballontrajektorien selbst erfolgte im Rahmen dieser Arbeiten jedoch nicht.
Bereits der Umfang der Ballondaten macht eine Neuanalyse sehr attraktiv, bedingt durch
den zeitlichen Abstand und der teilweise geringen Verfügbarkeit von meteorologischen
Daten wurde diese Perspektive aber vorerst nicht weiterverfolgt.
Die Daten von Sakagami zeigen beeindruckend, wie stark der Anteil der zurückgeschickten
Ballons von der Bevölkerungsdichte (hier Japan) abhängt.

\begin{table}[t]
\begin{center}
\begin{tabular}{|l|r@{ -- }rrr|r|r@{\,$\pm$}r|c|} \hline
Referenz      &  \multicolumn{2}{c}{Zeitraum}    &  Tage    & Ballons & Gas & \multicolumn{2}{c}{Fundquote}  & Quelle \\ \hline
Campingkirche & 2000    &    2008    &  46       & 881 & He    & \multicolumn{2}{r|}{ 10\,--\,15 \%}  & \cite{Campingkirche2008}  \\  \hline
Glaschke      & 1998    &    2001    &  7        &   7 & H$_2$ & 11  & 4   \%    &   \\  \hline
Haberlandt A  & 2003    & 10.2004    &  60       & 162 & He    & 4,2 & 0,3 \%    & \cite{Haberlandt}  \\
Haberlandt B  & 10.2004 & 08.2007    &  215      & 640 & He    & 3,6 & 0,1 \%    & \cite{Haberlandt}  \\
Haberlandt C  & 09.2007 &    2008    &  72       & 281 & He    & 4,3 & 0,3 \%    & \cite{Haberlandt}  \\ \hline
Trostberg     & 1998    &    2010    &  13       & 240 & He    & 9,2 & 0,6 \%    & \cite{Schwarzau}   \\ \hline \hline
Sakagami (1961)& \multicolumn{2}{c}{1960} & 16   & 11\,152        & H$_2$& 22,8 & 0,2  \% & \cite{sakagami12diffusion} \\
Stocker (1990) & \multicolumn{2}{c}{1986} &  1   & $\approx$ 8600 & He   & 4,5  & 0,05 \% & \cite{stocker1990characteristics} \\  \hline
\end{tabular}
\end{center}
\caption[Verwendete Daten]{Auflistung aller verwendeten Ballonwettbewerbe mit Angaben zu den gefundenen Ballonkarten.
Geschätzte Fundquoten sind ohne statistischen Fehler angegeben. \label{UsedData}}
\end{table}

\section{Eigene Experimente \label{H2Ballons}}

\begin{table}[hb]
\begin{center}
\begin{tabular}{|ccccrr||rr|} \hline
Nr        &  Datum     &  Startzeit  &  $F_a$[g] & $M_0$[g] & $v_{z,0}$[km/h] & $d$[km] & $t_\mathrm{Flug}$ \\ \hline
1 & 19.09.1998 & 18:25  & 7,1 & 5,4 & 5,0 & 181 & ?    \\
2 & 03.08.1999 & 14:23  & 8,6 & 7,5 & 5,4 &  30 & $<2$ h \\
3 & 16.08.1999 & 14:28  & 5,8 & 7,2 & 4,6 & 105 & ?    \\
4 & 18.09.1999 & 19:45  & 5,8 & 7,4 & 4,8 & 143 & ?    \\
5 & 21.09.1999 & 16:13  & 5,5 & 7,2 & 4,6 & 136 & ?    \\
6 & 01.01.2000 & 14:46  & 6,4 & 7,2 & 4,8 &  21 & 34 min \\
7 & 28.01.2001 & 12:35  & 4,3 & 7,2 & 4,2 & 267 & ?    \\ \hline
\end{tabular}
\end{center}
\caption[]{Startdaten der zurückgeschickten Ballonkarten. \label{DataH2}}
\end{table}
Tabelle \ref{DataH2} enthält die Startdaten aller gefundenen Kartenballons.
Die letzten beiden Spalten enthalten die Flugweite und
Informationen über die Flugzeit, soweit sie aus den Angaben des Finders ableitbar waren. Die Ballons
erreichten teilweise eine Flugweite von mehr als 250 km, wobei die Weite mit
abnehmendem Auftrieb des Ballons zunimmt.

Die bekannten Parameter der Ballons erlauben es, eine vollständige Simulation
jeder Ballontrajektorie durchzuführen. Als einziger freier Parameter wurde
die Höhe, in der der Ballon durch unbekannte Einflüsse platzt, verwendet.
Mit Flugweiten um 100 km bewegen sich die Trajektorien noch unter der
Auflösungsgrenze des Windmodells (siehe Kap. \ref{ChapWetter}), so daß die freie Platzhöhe
solange verändert wurde, bis die Modellflugweite dem Experiment
entspricht\footnote{Den Fundort exakt anzupassen ist im Rahmen der Modellgenauigkeit nicht möglich.}.
Die abgeleiteten Platzhöhen sind zusammen mit der Temperatur in dieser
Höhe und der Gesamtflugzeit in Tabelle \ref{DataSimH2} enthalten.
Die Platzhöhen fallen grob in zwei Gruppen: Ballons, die in rund 10 km
Höhe bei Temperaturen um $-50$\celsius{} mit einem Volumen von etwa 30 l platzen
und Ballons, die deutlich früher platzen. Die erste Gruppe entspricht den
Erwartungen an die Haltbarkeit des Ballongummis (Kap. \ref{ChapBallonP}),
so daß keine weiteren äußeren Einflüsse angenommen werden müssen.
Ballons der zweiten Gruppe platzen bereits nach wenigen Stunden, insbesondere bei
Starts am frühen Vormittag, was die Zerstörung der Ballonhülle durch die
Sonneneinstrahlung nahelegt.

\begin{figure}[t]
\begin{center}
\includegraphics[scale=0.4,angle=270]{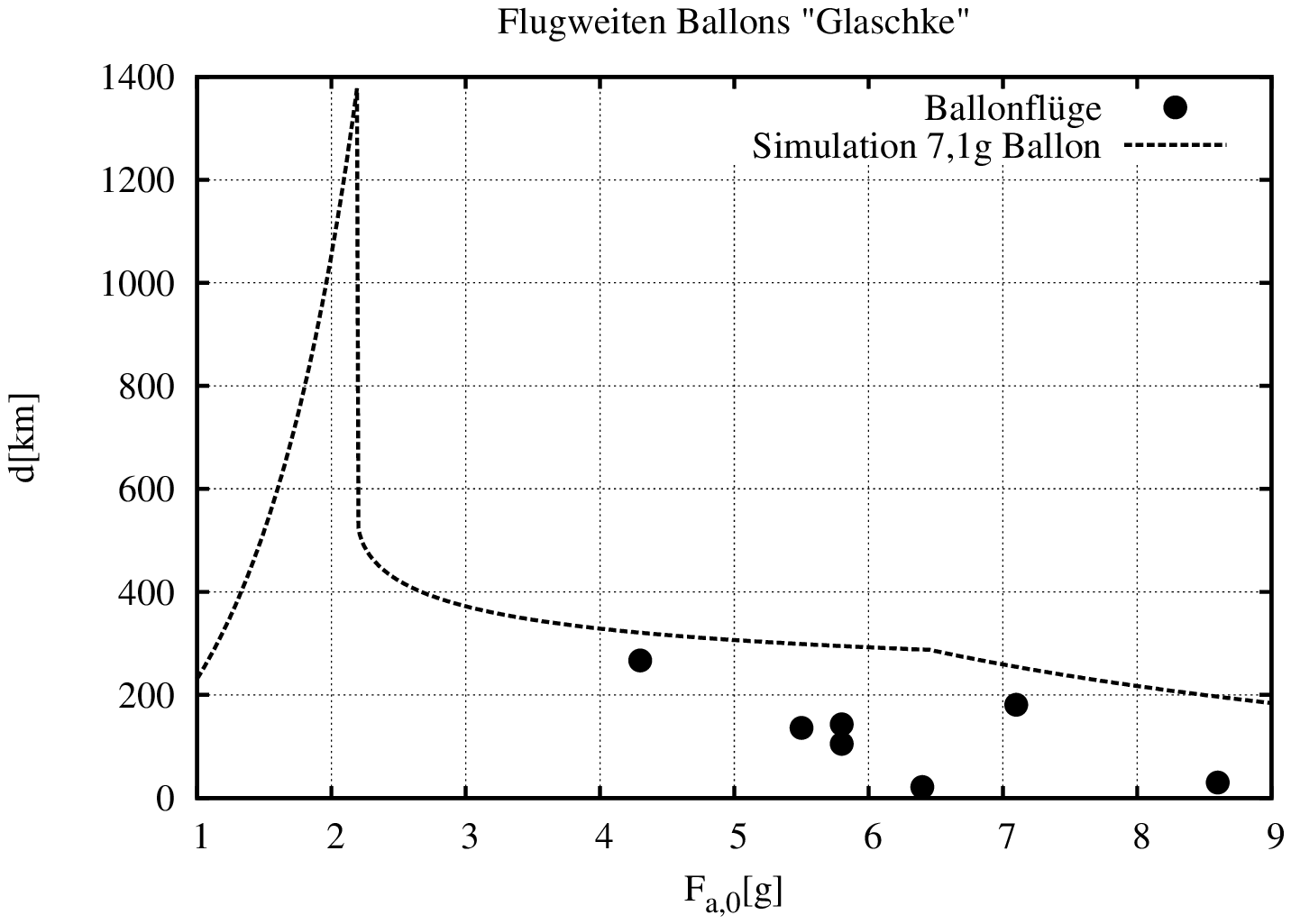}
\end{center}
\caption[Vergleich eigener Experimente mit Simulation]{Vergleich der Flugweiten der
Ballons aus Tabelle \ref{DataH2} mit Simulationen des Referenzballons.\label{VglSimDat} }
\end{figure}

Die Simulationen zeigen aber auch weitergehende Unterschiede in den beiden
Fällen mit bekannter Flugzeit:
\begin{itemize}
\item Bei Ballon 6 konnte der Finder die Landung der Karte beobachten, so daß die Flugzeit genau bekannt ist.
Aus der Flugzeit von 34 Minuten folgt eine maximale Höhe von etwa 1,2 Kilometern, was deutlich unter der berechneten Höhe ist.
Die längere Flugzeit des berechneten Ballons deutet auf eine Unterschätzung der Winde durch die interpolierten
Radiosondendaten hin.
\item Die Flugzeit von Ballon 2 weicht leicht von der gemessenen Flugzeit ab, die in diesem
Fall aber nur eine Obergrenze darstellt (wann der Ballon genau vor dem Fund gelandet ist, ist nicht
bekannt). Im Rahmen der erwarteten Genauigkeit ist der Fehler akzeptabel.
\end{itemize}
Abbildung \ref{VglSimDat} gibt eine abschließende Übersicht aller Flugweiten zusammen mit einer Simulationsreihe
des Standardballons aus Tabelle \ref{BallonP}. Hier wird besonders deutlich, daß
alle gestarteten Ballons weit vom berechneten Optimum entfernt sind. Der Wert dieser Versuchsreihe
besteht vor allem in der Absicherung der Bedingungen, unter denen ein Ballon platzt.

\clearpage

\begin{table}[t]
\begin{center}
\begin{tabular}{|crrrr|} \hline
Nr        &  $h_{\max}$[km] & $T_{\min}$[$^{\circ}$C] & $t_\mathrm{Flug}$[h] & $V_{\max}$[l]   \\ \hline
1 & 10,0 &  $-47$  & 4,0 & 27,7 \\
2 &  5,8 &  $-14$  & 2,3 & 24,0 \\
3 &  5,3 &  $-14$  & 2,3 & 19,0 \\
4 & 10,8 &  $-54$  & 4,4 & 31,0 \\
5 &  7,6 &  $-32$  & 3,2 & 22,1 \\
6 &  4,0 &  $-12$  & 1,7 & 17,1 \\
7 &  9,8 &  $-53$  & 4,4 & 27,5 \\ \hline
\end{tabular}
\end{center}
\caption[]{Auswertung der zurückgeschickten Ballonkarten. \label{DataSimH2}}
\end{table}

\section{Startdynamik\label{SectStartDyn}}

Für den erfolgreichen Start eines Ballons ist das Windprofil in Bodennähe
von großem Einfluß. Während der Wind Ballons um geschlossene Hindernisse fast
störungsfrei mitnimmt, stellen durchlässige Objekte wie Bäume und Antennen
mögliche Hindernisse nach dem Start dar.

\begin{figure}[ht]
\begin{center}
\includegraphics[scale=0.4,angle=270]{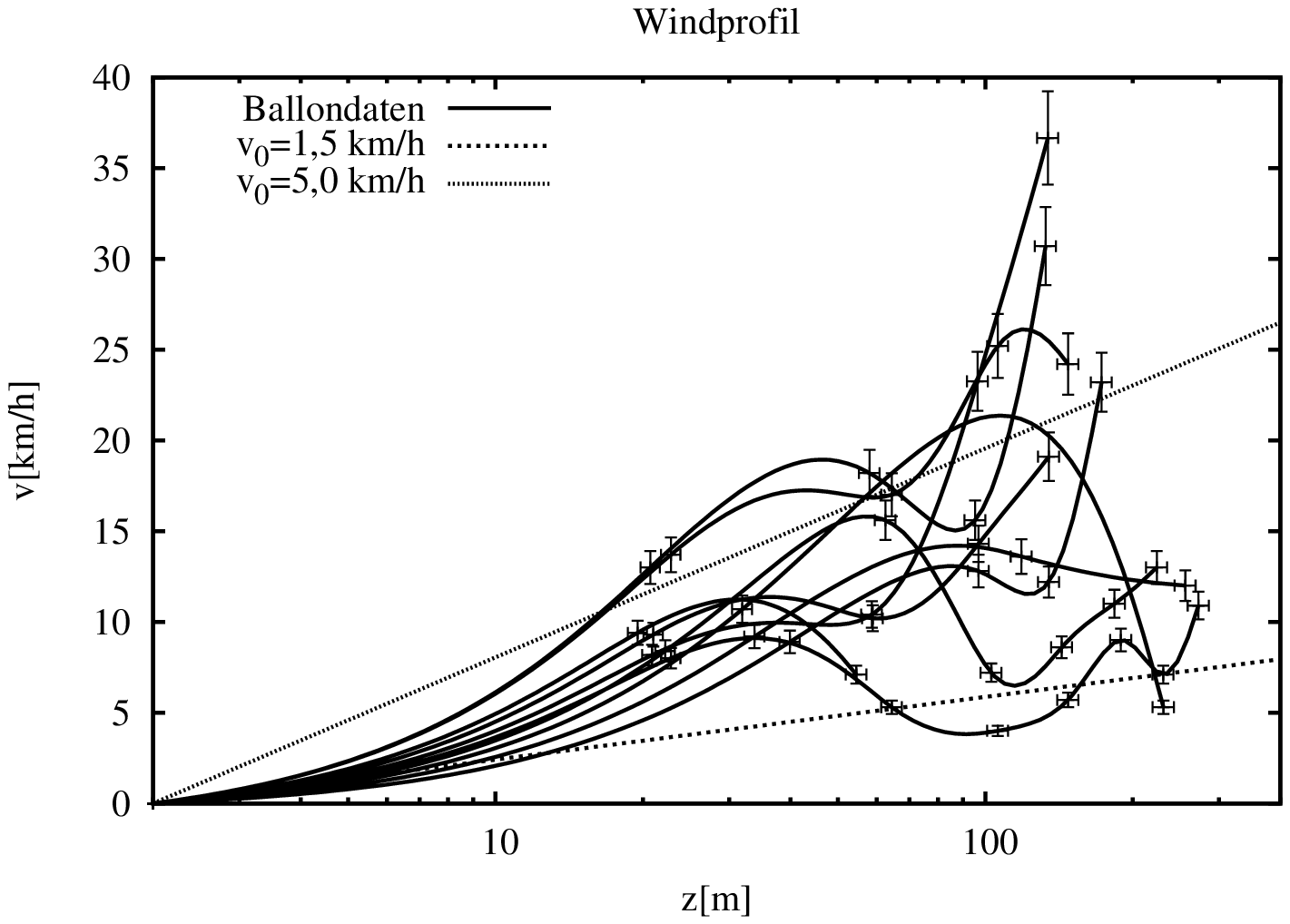}
\end{center}
\caption[Windprofil in Bodennähe]{Windprofil in Bodennähe. Jede Kurve entspricht einem Ballonstart.
Die durch Gl.~\ref{vLogRange} beschriebenen Profile sind durch zwei Geraden angedeutet.
\label{WindProfil} }
\end{figure}

Zur genaueren Beurteilung wurde ein Teil der in dem vorhergehenden Kapitel geschilderten
Ballonstarts verwendet, um das bodennahe Windprofil zu bestimmen. Die Vorgehensweise
entspricht der Windmessung durch Pilotballone: Nach dem Start wird der Ballon in regelmäßigen
Abständen angepeilt und der Höhen-- und Azimutwinkel notiert. Aus der bekannten
Steigrate $v_z$ des Ballons kann dann die dreidimensionale Flugbahn und somit die
Windgeschwindigkeit bestimmt werden.
Als Modellfunktion wird das bereits vorgestellte logarithmische Windprofil (Gl. \ref{vLogProfile})
verwendet, da keine Quantifizierung der atmosphärischen Stabilität durchgeführt werden konnte:
\begin{eqnarray}
v(z)  &=&  v_{0} \ln(z/z_0)  \label{vLogSimple}
\end{eqnarray}
Zur kompakteren Darstellung wurden alle Vorfaktoren zu einer Skalengeschwindigkeit $v_0$
zusammengezogen. Mit Hilfe dieses Windprofils kann die Flugbahn des Ballons im Raum
als Funktion der Höhe $z$ explizit bestimmt werden:
\begin{eqnarray}
r(z)  &=&  \frac{v_{0}}{v_z} \left( z \ln(z/z_0)-(z-z_0) \right)
\end{eqnarray}
$r$ ist der horizontal gemessene Abstand zum Startort. Für die Abschätzung der Kollisionsgefahr
mit umstehenden Objekten ist der Abflugwinkel $\phi$ besser geeignet:
\begin{eqnarray}
\tan(\phi ) &=& \frac{z}{r(z)} \\
& \approx & \frac{1}{ \ln(z/z_0)-1 } \frac{v_z}{v_{0}} \\
\phi &\approx & 20^{\circ}\frac{v_z}{v_{0}}
\end{eqnarray}
Zur Bestimmung der Skalengeschwindigkeit $v_0$ wurden für mehrere Ballonstarts
Geschwindigkeitsprofile erstellt (Abb. \ref{WindProfil}) und mit Gl. \ref{vLogSimple}
verglichen. Da eine einzelne Trajektorie keine gemittelte, repräsentative Messung darstellt,
ist ein logarithmisches Windprofil nicht für alle Messungen klar erkennbar. Dies trifft
besonders auf Starts mit böigem Wind zu. Aus diesem
Grund konnte nur ein Parameterbereich ermittelt werden:
\begin{eqnarray}
v(z) &=& (1,5\dots 5,0) \times \ln \left(\frac{z}{2 \mbox{ m}}\right) \quad  \frac{\mbox{km}}{\mbox{h}} \label{vLogRange}
\end{eqnarray}
Der Wertebereich deckt sich mit unabhängigen Bestimmungen des Windprofils
durch Radiosonden \cite{parlange1990radiosonde}.
Der zu erwartende kleine Abflugwinkel von optimal gefüllten Kartenballons macht eine
besondere Sorgfalt beim Start erforderlich.

\section{Ballonwettbewerbe}

Die Auswertung der weiteren Daten aus Tabelle \ref{UsedData} erfolgt auf eine andere
Weise. Da die genauen Parameter der gestarteten Ballons (insbesondere das Ballongewicht,
der Startauftrieb und in vielen Fällen die Uhrzeit) nicht bekannt sind,
ist eine detaillierte Auswertung von jedem Start nicht möglich. Statt dessen
wird der folgende Ansatz gewählt: Für alle Starts wird pauschal ein Kartenballon
mit der Masse 5,1 g und einer Karte von 2 g angenommen. Unbekannte Startzeiten
werden durch den Schätzwert 15:00 Uhr ersetzt. Der unbekannte Startauftrieb wird durch
einen Ensembleansatz umgangen: Für jeden Ballonstart wird eine ganze Reihe
virtueller Flüge mit $F_{a,0}=1 \dots 4$~g durchgeführt und die jeweils am besten passende
Trajektorie zur Fehlerberechnung verwendet. Dieses Vorgehen führt im Prinzip zu einer
Überschätzung der Trajektoriengenauigkeit, aber in der Mehrzahl der Fälle ist der Einfluß
des Auftriebs auf die Flugrichtung so gering, daß der wahre Fehler kaum beeinflußt wird.
Zusätzlich wird auch für jeden Start der optimale Auftrieb berechnet, der die Flugweite maximiert.
Die folgenden zwei Abschnitte gehen näher auf die Flugweiten und Modellfehler
ein. Abschließend werden starke Abweichungen der Trajektorien diskutiert.

\section{Ballonflugweite \label{BallonDistance}}

\begin{figure}[ht]
\begin{center}
\includegraphics[scale=0.4,angle=270]{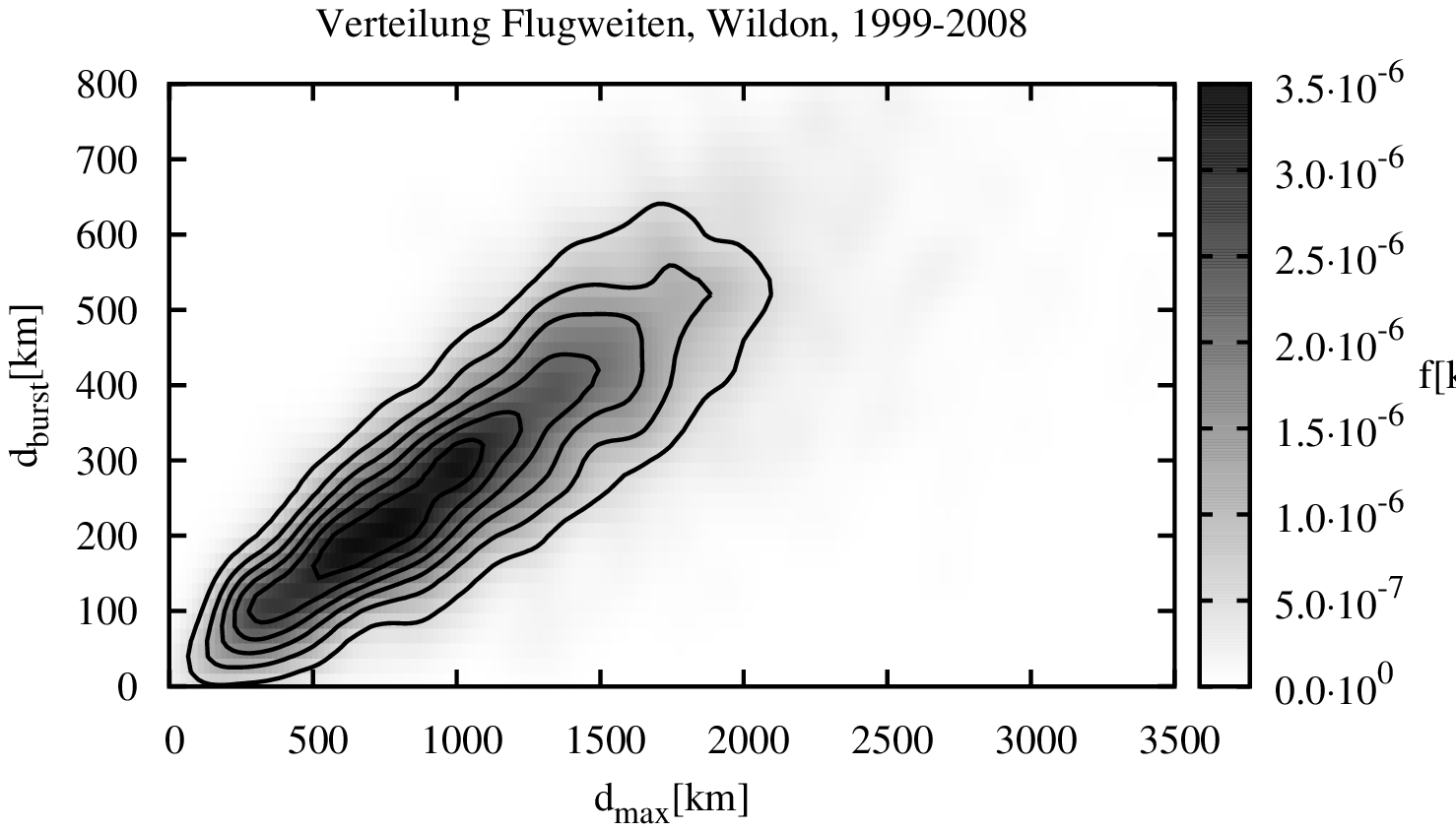}
\end{center}
\caption[Verteilung der charakteristischen Flugweiten]{Verteilung der charakteristischen
Flugweiten $d_{\max}$ und $d_{\mathrm{burst}}$.\label{DMaxDBurstDist} }
\end{figure}

Die individuellen Flugweiten $d$ der zur Verfügung stehenden Kartenballons bilden
die Grundlage, um die in den vorangegangenen Kapiteln beschriebenen Effekte zu
untersuchen und quantitativ zu erfassen. Hierbei muß das Augenmerk auf einer
sorgfältigen Zusammenführung der verschiedenen Flugtage und Wettbewerbe liegen,
da die pro Flugtag verfügbaren Daten nur in den seltensten Fällen ausreichen,
um eine aussagekräftige Statistik zu erstellen.

Die Schwierigkeit besteht in den individuell verschiedenen Wetterbedingungen
jeden Flugtags, die einen einfachen Vergleich der Daten ausschliessen. Am prominentesten
ist die zeitliche Veränderung des Windfelds, welche direkt die Flugweite eines
Kartenballons beeinflußt. Die Eigenschaften des Windfelds erlauben aber eine
elegante Lösung dieses Problems. Zum einen ist die Mehrzahl der Trajektorien
nur schwach gekrümmt -- dies zeigen berechnete Trajektorien und Kartenballonfundorte --
so daß die Betrachtung auf das vertikale Windprofil $v(z)$ beschränkt werden
kann. Zum anderen wird eben dieses Windprofil im Mittel gut durch eine lineare
Funktion beschrieben (vgl.~Abb.~\ref{MeiningenD}). Dies läßt erwarten, daß sich
der Einfluß des Windes durch eine einfache Skalierung der Flugweite korrigieren
läßt. Um diesen Ansatz weiter zu stützen, wurden die Trajektorienrechnungen
aus Kapitel \ref{Jahresstat} verwendet, um die Verteilung der charakteristischen
Flugweiten $d_{\max}$ und $d_{\mathrm{burst}}$ zweidimensional darzustellen.
Trotz der unterschiedlichen Trajektorien, die mit den jeweiligen charakteristischen
Flugweiten verknüpft sind (vgl. Abb. \ref{BTrajekt}), zeigt die Verteilung in
Abbildung~\ref{DMaxDBurstDist} eine sehr starke Korrelation.

Diese Argumente erlauben die Einführung einer skalierten Flugweite $\tilde d$,
die aus den individuell berechneten Flugweiten $d_{\max}$ und $d_{\mathrm{burst}}$
für jeden Kartenballon berechnet wird. Die stückweise linear skalierte Entfernung
$\tilde d$ ist definiert durch:
\begin{eqnarray}
\tilde d &:=& \left\{
\begin{array}{ll}
0,3  \frac{d}{ d_{\mathrm{burst}} } & \mbox{falls } d < d_{\mathrm{burst}}  \\
\frac{d}{ d_{\max} } & \mbox{falls } d > d_{\max}  \\
0,3 + 0,7\frac{d-d_{\mathrm{burst}}}{d_{\max}-d_{\mathrm{burst}}} & \mbox{sonst.}
\end{array}
\right.   \label{Weite_Scal}
\end{eqnarray}
Die Wahl des kritischen Verhältnisses $d_{\mathrm{burst}}/d_{\max} = 0,3$ wurde über ein Mittelwert
aus vielen Simulationen (vgl. Abb. \ref{DMaxDBurstDist}) bestimmt. Die Skalierung der
Flugweite $d_{\mathrm{burst}}$ auf einen festen Wert wird durch die Aussagekraft
dieses Wertes motiviert: Flugweiten größer als $d_{\mathrm{burst}}$ fordern eindeutig, daß der
Ballon den Scheitelpunkt der Trajektorie intakt passiert hat, während kleinere Werte
wahrscheinlich (aber nicht notwendigerweise) auf einen geplatzten Ballon hindeuten. Normierte
Flugweiten größer Eins sollten im Sinne dieser Definition nicht vorkommen. Eine genaue Analyse
dieses Falls erfolgt mit einer ausführlichen Fehlerbetrachtung in den Kapiteln \ref{SectFehler}
und \ref{SectAusr}.

Die Verwendung einer skalierten Flugweite erlaubt, windbedingte Unterschiede in den Flugweiten
zu korrigieren, was die Kombination von Kartenballondaten verschiedener Flugtage ermöglicht.
Darüber hinaus ist die Mittelung verschiedener Flugtage eine Möglichkeit, Aussagen unabhängig
und ohne genaue Kenntnis des Wettergeschehens der einzelnen Starttage treffen zu können.

Die im Folgenden vorgestellten Verteilungen der skalierten Flugweite erlauben auch die Umgehung eines
weiteren Problems: Wenn der Startauftrieb eines Ballons nicht bekannt ist (z.\,B. im Falle der
Ballonwettbewerbe), kann aus der Flugweite eines einzelnen Ballons nicht eindeutig auf die Flugzeit
geschlossen werden (vgl. Abb. \ref{DvsFa_2Gas}). Bei der Betrachtung einer Verteilung kann aber
jeder skalierten Flugweite $\tilde d$ im Mittel eine zuverlässigere Flugzeit $t_{\mathrm{Flug}}$ zugeordnet werden:
\begin{eqnarray}
t_{\mathrm{Flug}} & \approx & 20 \mbox{ h} \times \left\{
\begin{array}{ll}
0,45 \times (\tilde d/0,3)^{2/3} & \mbox{falls } \tilde d < 0,3  \\
0,45 + 0,55\frac{\tilde d-0,3}{0,7} & \mbox{sonst.}
\end{array}
\right.   \label{t_dScal}
\end{eqnarray}
Diese Gleichung wurde ebenfalls aus einem geeigneten Mittel über viele Simulationen gewonnen.

\begin{figure}
\begin{center}
\includegraphics[scale=0.5,angle=270]{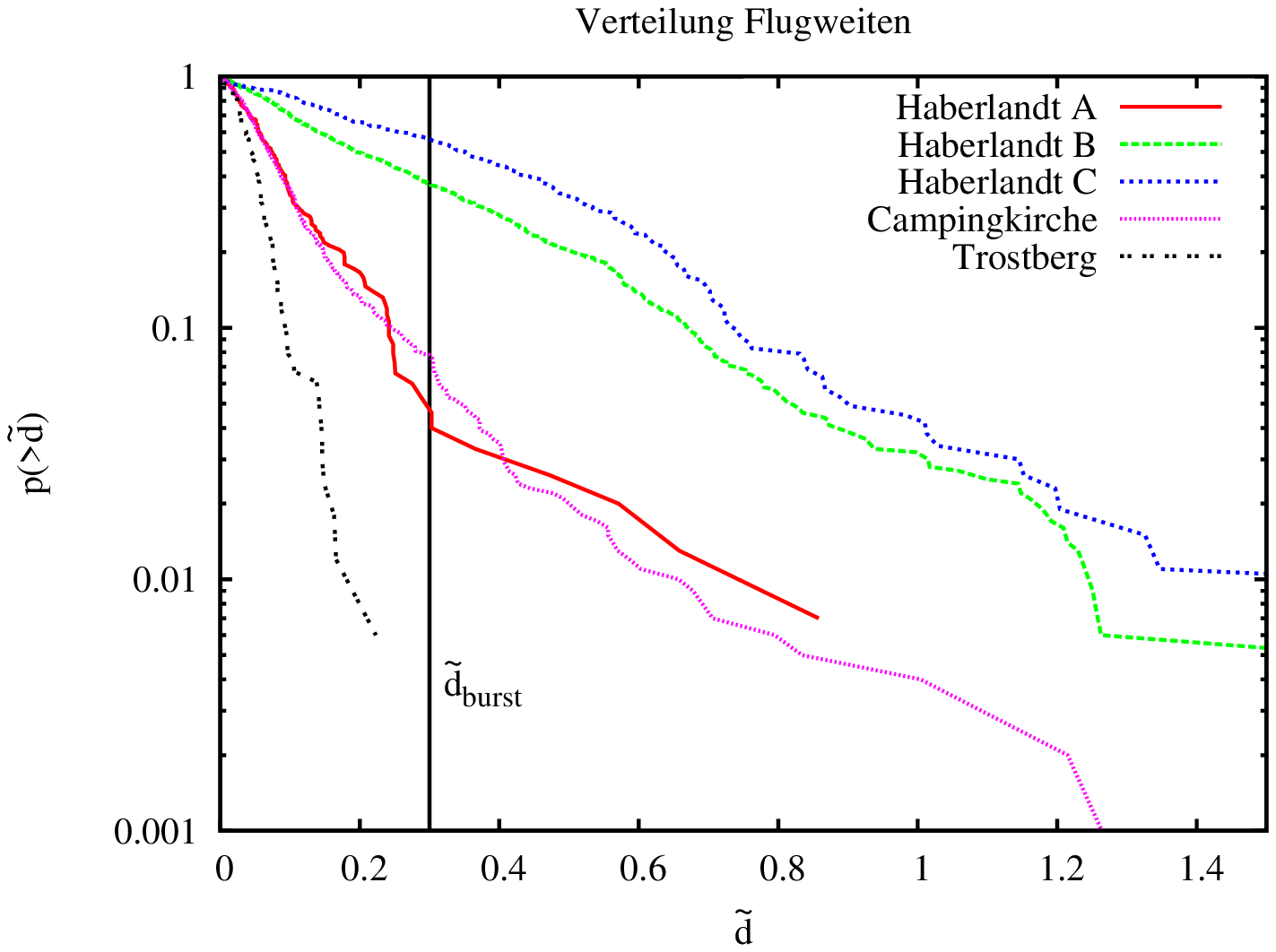}
\end{center}
\caption[Verteilung skalierte Flugweiten]{Kumulative Verteilungen der normierten Flugweite.\label{Cum_Weite_Scal} }
\end{figure}

\begin{figure}
\begin{center}
\includegraphics[scale=0.5,angle=270]{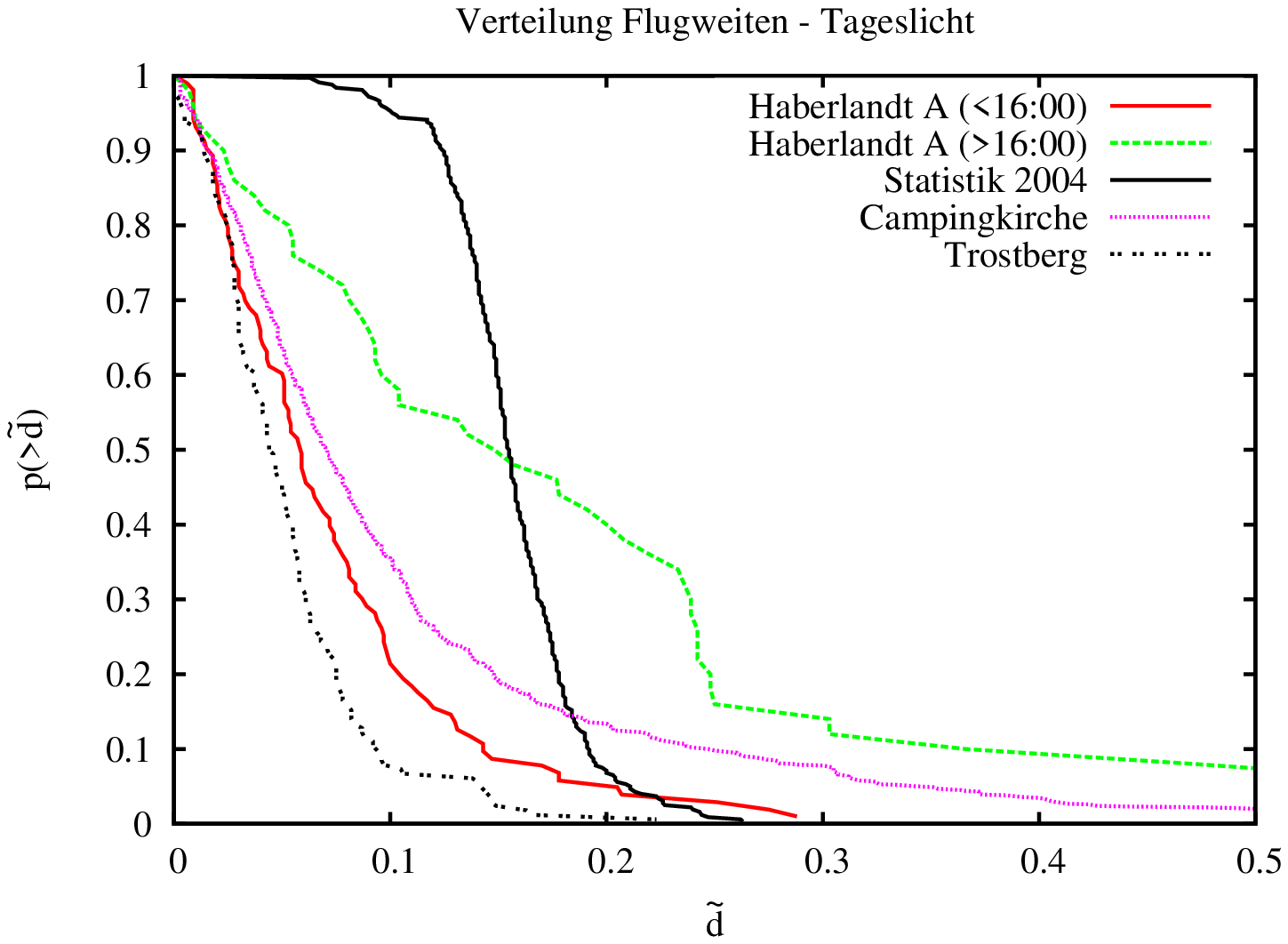}
\end{center}
\caption[Verteilung Flugweiten am Tag]{Kumulative Verteilungen der normierten Flugweite am Tag. \label{Weite_A_Day} }
\end{figure}

Zunächst wird die Verteilung der normierten Flugweiten in Abbildung \ref{Cum_Weite_Scal} untersucht.
Der Idealfall, d.\,h. alle Ballons erreichen die maximal mögliche Flugweite, ist gegeben durch
\begin{eqnarray}
p \left(> \tilde d \right) &=& \Theta \left( 1-\tilde d \right) \label{IdealpD}
\end{eqnarray}
mit der Heaviside-Sprungfunktion $\Theta$. Eine erste Inspektion der Verteilungen zeigt bereits,
daß dieses idealisierte Szenario nicht einmal annähernd eine treffende Beschreibung gibt. Zur weiteren
Untersuchung wurden die Verteilungen anhand ihrer Ähnlichkeit in zwei Gruppen aufgeteilt:
"`Haberlandt B"' und "`C"' (Gruppe 2) sowie die verbleibenden Datensätze (Gruppe 1). Die meisten
Flugweiten in Gruppe~1 liegen deutlich unter $d_{\mathrm{burst}}$, so daß hier in Übereinstimmung mit
den Angaben zu "`Haberlandt A"' von sehr stark gefüllten Ballons auszugehen ist. Überraschend
ist die starke Streuung der Flugweiten, die sich nicht alleine mit unterschiedlich gefüllten
Ballons erklären läßt. Gerade im Bereich stark gefüllter Ballons variiert die Flugweite
nur schwach mit dem Auftrieb.

Gruppe 2 zeigt deutlich größere normierte Flugweiten, was durch die gezielte
Optimierung des Startauftriebs zu erwarten war. Aber welche Verteilung ist in diesem Fall
überhaupt zu erwarten? Das Volumen der Ballons in Gruppe 2 bewegt sich im Rahmen des
technisch möglichen dicht am optimalen Wert 8,2~l. Dennoch ist es unwahrscheinlich, daß
alle Ballons die maximale Flugweite erreichen, da der optimale Bereich sehr schmal
ist und die genaue Lage vom Wettergeschehen abhängig ist.
In diesem Licht ist es sinnvoll, eine annähernde Gleichverteilung der Ballonflugweiten
zu postulieren:
\begin{eqnarray}
p \left(> \tilde d \right) &=& 1 - \tilde d \label{EqGroup2}
\end{eqnarray}
Aber selbst diese vielleicht etwas zu pessimistische Verteilung überschätzt den Anteil großer
skalierter Flugweiten.

Es müssen also weitere Einflußfaktoren vorliegen, die bisher vernachläßigt wurden
und weitgehend unabhängig von den Balloneigenschaften operieren. Der universelle, annähernd
exponentielle Verlauf aller Verteilungen deutet in die gleiche Richtung. Bereits genannte
Kandidaten sind UV/Ozon (Kap. \ref{SectUV}) und Regen (Kap. \ref{SectRegen}).
Zur genaueren Analyse dieser Einflußfaktoren wurde für die Daten "`Haberlandt A"' (hier sind die
Startzeiten bekannt) jeweils eine eigene Statistik für die Starts vor 16:00~Uhr und danach erstellt.
Um einen quantitativen Vergleich der Verteilungen in Gruppe 1 mit einem idealisierten Szenario zu
ermöglichen, wurde zusätzlich eine Jahresstatistik für den Startort
Wildon\footnote{Startort der Daten "`Haberlandt"'} berechnet. An jedem Tag des Jahres 2004 wurde
ein 13~l Heliumballon (vgl. Tab. \ref{BallonP}) virtuell um 14:00 Uhr(UT) gestartet.
Abbildung \ref{Weite_A_Day} stellt noch einmal alle Verteilungen aus Gruppe~1 und die berechnete
Jahresstatistik dar. Obwohl ein konstantes Ballonvolumen die in der Praxis auftretenden Variationen
nicht berücksichtigt, ist die Diskrepanz zu den empirisch bestimmten Verteilungen signifikant.
Alleine "`Haberlandt A$>$16:00"' zeigt einen mit der Jahresstatistik vergleichbaren Einbruch
in der Verteilung bei $\tilde d=0,25$ durch zu stark gefüllte Ballons.

Es ist hilfreich, den bereits geschilderten Befund für eine genauere Diskussion zu quantifizieren.
Dazu wird der Ballonflug in zwei Prozesse zerlegt: Den ungestörten (d.\,h. idealisierten) Ballonflug,
der alleine durch die Balloneigenschaften und die Windströmung bestimmt wird, und einen von außen
eingreifenden Mechanismus, der den Ballon zerstört bzw. zu Boden bringt.
Unter der Annahme, daß beide Prozesse unabhängig voneinander sind, können die jeweiligen
Überlebensfunktionen einfach miteinander multipliziert werden:
\begin{eqnarray}
F(\tilde d) &=& F_{\mathrm{Traject}}(\tilde d) \times  F_{\mathrm{Decay}}(\tilde d)  \label{FMultEq}
\end{eqnarray}
Im hier betrachteten Fall entspricht die Gesamtüberlebensfunktion $F$ der Verteilung der skalierten
Entfernungen $p(>\tilde d)$, und die Teilfunktion $F_{\mathrm{Traject}}$ bezieht sich auf den ungestörten
Ballonflug. $F_{\mathrm{Decay}}$ beschreibt die Zerstörung des Ballons durch die bereits angesprochenen
Umwelteinflüsse. Die treffende Modellierung dieser Effekte erscheint auf den ersten Blick unmöglich,
da keine detaillierten Informationen über die zugrundeliegenden Mechanismen vorliegen. Durch die
Mittelung verschiedener Flugtage und die chaotische Natur des Wettergeschehens ist es aber sinnvoll
anzunehmen, daß diese Prozesse unabhängig und homogen in Zeit und Ort operieren. In diesem Fall erhält
man zwangsläufig einen exponentiellen Zerfall
\begin{eqnarray}
F_{\mathrm{Decay}}(\tilde d) &=& \exp(- t_{\mathrm{Flug}}(\tilde d)/\tau) \label{BalDecay}
\end{eqnarray}
wobei die Flugzeit $t_{\mathrm{Flug}}$ gemäß Gl. \ref{t_dScal} eingesetzt wird. Alles relevante Wissen
über die externen Effekte ist in der Lebenszeit $\tau$ enthalten. Es ist klar, daß Gl. \ref{BalDecay}
nur eine erste Näherung ist, die durchaus vorhandene Korrelationen wie z.\,B. den Tagesrhythmus
außer Acht läßt. Durch die bereits vorhandenen Unsicherheiten der Daten ist aber eine weitergehende
Interpretation nicht sinnvoll.

\begin{table}
\begin{center}
\begin{tabular}{|l|r@{\,--\,}r|l|} \hline
Referenz               & \multicolumn{2}{c|}{$\tau$[h]} & Kommentar     \\ \hline
Trostberg              & 1,5 & 2,5  & Starts Nachmittags, Juni, Juli \\ \hline
Haberlandt A $<$ 16:00 &   3 & 4    & Starts ab Morgens  \\ \hline
Campingkirche          &   4 & 5    & Starts Nachmittags \\ \hline
Glaschke               &   4 & 6    & Starts Nachmittags \\ \hline
Haberlandt A $>$ 16:00 &  10 & 12   & Starts bis in den Abend      \\ \hline
Haberlandt B           &  12 & 16   & Starts Abends                \\ \hline
Haberlandt C           &  20 & 26   & Starts Abends, Wetterbericht \\ \hline
\end{tabular}
\end{center}
\caption[Ballonlebensdauern]{Ermittelte Lebensdauern für verschiedene Datensätze.\label{TauBallons}}
\end{table}

Das Einsetzen von Gl. \ref{BalDecay} in Gl. \ref{FMultEq} zur Bestimmung einzelner Lebenszeiten
$\tau$ erfordert die angemessene Wahl einer ungestörten Überlebensfunktion $F_{\mathrm{Traject}}$.
Hierfür muß notwendigerweise auf Simulationen zurückgegriffen werden, die durch ihre Konstruktion
keine externen Einflüsse beinhalten. Für Gruppe~1 wird die bereits genannte Jahresstatistik 2004
verwendet, während für die deutlich verschiedene Gruppe~2 Gl. \ref{EqGroup2} eingesetzt wird.
Die Durchführung einer regelrechten Ausgleichsrechnung zur Bestimmung der Lebenszeiten würde den
bereits genannten Limitierung nicht gerecht werden, so daß statt dessen jeweils von Hand mögliche
Wertebereiche ermittelt wurden.

In Tabelle \ref{TauBallons} sind alle ermittelten Lebenszeiten zusammengestellt. Im Zweifel wurden
großzügigere Wertebereiche angegeben, um vor allem die Unsicherheit von $F_{\mathrm{Traject}}$
besser abzubilden. Durch die genau bekannten Ballonparameter des Datensatzes "`Glaschke"' konnte
die Lebenszeit direkt aus Tabelle \ref{DataSimH2} ermittelt werden. Trotz der unterschiedlichen
Datenquellen und Methoden zeigen "`Haberlandt A$<$16:00"', "`Campingkirche"' und "`Glaschke"' mit
vergleichbaren Startbedingungen eine bemerkenswerte Übereinstimmung der Lebenszeit von 4--5 Stunden.
"`Haberlandt A$>$16:00"' sowie "`B"' mit Starts am späten Nachmittag bzw. frühen Abend sind
ebenfalls in zufriedenstellender Übereinstimmung, vor allem wenn man den deutlichen Unterschied
in den Ballonfüllungen und Flugzeiten bedenkt. Mit dem Datensatz "`Haberlandt C"', der zusätzlich
nur Starts bei trockenem Wetter enthält, wird die größte Lebenszeit erreicht. Mit "`Trostberg"' ist
die kürzeste Lebenszeit im Stundenbereich gegeben.

Abschließend werden die Angaben in Tabelle \ref{TauBallons} verwendet, um jeweils den einzelnen
Effekten Lebensdauern zuzuordnen. In erster Näherung ergeben sich die aufgeführten Lebensdauern
aus einer Summe der verschiedenen Beiträge:
\begin{eqnarray}
\frac{1}{\tau} & \approx & \frac{1}{\tau_{\mathrm{Licht}}} + \frac{1}{\tau_{\mathrm{Regen}}} + \frac{1}{\tau_{\mathrm{Rest}}}
\end{eqnarray}
Aus dem Vergleich der beiden "`Haberlandt A"'-Blöcke ergibt sich der durch Tageslicht bestimmte
Anteil zu $\tau_{\mathrm{Licht}} \approx 10$ h. Vergleiche mit "`Haberlandt B"' und
"`C"' erlauben die Zuordnung für Regen $\tau_{\mathrm{Regen}} \approx 20$ h und die verbleibenden
(d.\,h. unbekannten) Effekte $\tau_{\mathrm{Rest}} \approx 20$ h. Solange diese Aufteilung nicht
durch eine direkte Verknüpfung der Ballonflugweiten mit Wetterdaten bestätigt wird
(vgl. Kap. \ref{SectRegen}), sind diese Angaben vorläufige Schätzwerte.

In Anbetracht dieser Aufteilung bleibt alleine der Datensatz "`Trostberg"' rätselhaft, der eine
deutlich kleinere Lebensdauer aufzeigt. Nach Rücksprache mit dem Veranstalter wurde die Auswertung
noch einmal mit einem veränderten Ballonmodell wiederholt. Das Ergebnis änderte sich aber nur
geringfügig, ohne diesen Befund erklären zu können. Möglicherweise führten die Startzeiten in den
Monaten Juni und Juli zu einem besonders starken Einfluß der Sonneneinstrahlung, aber auch die Nähe
des Startorts zu den Alpen ist ein Umstand, der von dem Trajektorienmodell nicht detailliert
berücksichtigt wird.

\begin{table}
\begin{center}
\begin{tabular}{|l|r|} \hline
Referenz      &  \% Intakt \\ \hline
Campingkirche & 7,8    \\ \hline
Glaschke      & 0,0    \\ \hline
Haberlandt A  & 4,8    \\ \hline
Haberlandt B  &  37,4  \\ \hline
Haberlandt C  &  56,3   \\ \hline
Trostberg     & 0,0   \\ \hline
\end{tabular}
\end{center}
\caption[Intakt gelandete Ballons]{Anteil intakt gelandeter Ballons.\label{IntaktBallons}}
\end{table}

\section{Intakte Ballons}

Die kumulativen Verteilungen in Abbildung \ref{Cum_Weite_Scal} erlauben auch eine Abschätzung
des Anteils der Ballons, die intakt gelandet sind. Ballons, die weiter als $d_{\mathrm{burst}}$
geflogen sind, müssen den Scheitelpunkt intakt passiert haben. Da die Hülle beim Abstieg wieder
schrumpft und dadurch zunehmend entlastet wird, ist es sehr wahrscheinlich, daß ein Großteil
dieser Ballons ebenfalls intakt landet. Der aus den kumulativen Verteilungen entnommene Anteil
dieser Ballons ist in Tabelle \ref{IntaktBallons} als Schätzwert zusammengestellt.

\section{Fehler\label{SectFehler}}

Das vorgestellte Modell muß durch eine Fehlerbetrachtung geprüft
und beurteilt werden. Unstimmigkeiten zwischen dem Fundort eines
Kartenballons und der korrespondierenden Trajektorie können zwei prinzipiell
verschiedene Ursachen haben: Zum einen kann das verwendete Trajektorienmodell
selbst fehlerbehaftet sein, und zum anderen können die mit dem gefundenen Kartenballon
verknüpften Daten ungenau sein.

Der Fundort eines Kartenballons wird selten präzise angegeben. In der Regel wird die
nächste Stadt genannt, was im Mittel eine Genauigkeit von wenigen Kilometern
erlaubt (vgl. Kap. \ref{ChapBoden}). Wenn ein Kartenballon jedoch ins Meer oder in
ein fließendes Gewässer fällt, hat auch ein genauer Fundort nur noch eine begrenzte
Aussagekraft: Durch den zusätzliche Transport geht jede Information über den wahren
Landeplatz verloren. In diesem Fall sind fast beliebig große Abweichungen
möglich sind. Oft ist auch die genaue Startuhrzeit eines Kartenballons nicht bekannt.
Hier muß mit Schätzwerten gearbeitet werden, die aber um bis zu mehrere Stunden falsch
sein können. Besonders bei dynamischen Wettersituationen ist dann ein deutlicher Einfluß
auf die berechnete Trajektorie zu erwarten. Zuletzt besteht die Möglichkeit,
daß beim Aufzeichnen der Daten Fehler aller Art auftreten. Das können Verwechslungen und
Zahlendreher bei der Archivierung sein, Probleme bei der Identifizierung des Fundorts
oder gezielte Fälschungen durch den Finder oder den Teilnehmer eines Ballonwettbewerbs.

Das Trajektorienmodell besteht aus mehreren Komponenten, die unterschiedlich stark
die Genauigkeit einer Trajektorie beeinflussen. Die wichtigsten Teile sind die zugrunde
liegenden Gleichungen (Kap. \ref{ChapSim}), die verwendeten Wetterdaten (Kap. \ref{ChapWetter})
und die numerische Integration des Gesamtsystems. Von zentraler Bedeutung sind
Fehler in den Grundgleichungen. Diese können jedoch nur indirekt erschlossen werden,
indem geprüft wird, ob der Gesamtfehler alleine durch die begrenzte Genauigkeit der
Wetterdaten erklärt werden kann. Das numerische Lösungsverfahren und der Zeitschritt
sind bereits so gewählt, daß der Integrationsfehler vernachlässigbar ist.

Die begrenzte Genauigkeit des Wettermodells ist nicht alleine auf fehlerhafte Einzelmessungen
zurückzuführen, sondern vor allem auf die geringe Abtastung der Wetterwerte in
Zeit und Ort \cite{stohl1998computation}. Windbewegungen variieren bereits im Minuten--
und Stundenbereich \cite{van1957power}, was deutlich kürzer als das übliche Startintervall
der Radiosonden von 12 Stunden (bzw. in seltenen Fällen 6 Stunden) ist. Diese Unterabtastung
verursacht zwei Fehlerbeiträge: Zum einen können schnell variierende Windbeiträge
gar nicht durch das interpolierte Wettermodell wiedergegeben werden (Auflösungs-Effekt),
zum anderen werden schnelle Fluktuationen durch die Unterabtastung in niedrigere Frequenzbereiche
verschoben (Alias-Effekt) und ändern so ebenfalls das Modell. Während die hochfrequenten
Beiträge im Laufe einer Trajektorienberechnung gemittelt und damit gedämpft werden,
summieren sich langperiodischen Beiträge systematisch auf.

\begin{figure}
\begin{center}
\includegraphics[scale=0.8]{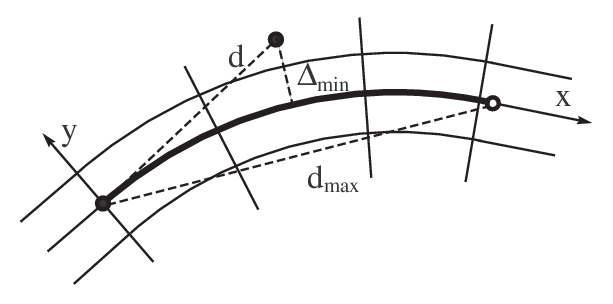}
\end{center}
\caption[Trajektorienkoordinatensystem]{Das durch die Trajektorie mit minimalem $\Delta_{\min}$ aufgespannte
Koordinatensystem. Der offene Kreis ist der Endpunkt der Trajektorie. Der korrespondierende Fundort
befindet sich im Abstand $d$ zum Startort.\label{TrajectCoord} }
\end{figure}

Die gleiche Argumentation kann auf die Abtastung im Ort angewendet werden. Bei einem
mittleren Abstand der verwendeten Radiosondenstationen von bis zu 200 km und
typischen Korrelationslängen des Windfelds von mehreren 100 bis 1000 km \cite{kahl1986uncertainty}
ist der Einfluß aber etwas geringer ausgeprägt. Zu beachten ist, daß sich bei kurzen
Trajektorien die kleinskaligen Windstrukturen nicht herausmitteln können und damit den
Fehler dominieren.

Darüber hinaus wird die erreichbare Genauigkeit durch fehlende Wetterdaten fundamental limitiert:
Interpolationsverfahren können die Auswirkungen fehlender Daten mildern, aber keinesfalls aufheben.
In diesem Licht sind eventuell durch das Interpolationsverfahren selbst eingeführte Fehler zu vernachlässigen.

Der Gesamtfehler $\Delta$ einer Trajektorie als Funktion der Flugzeit $t$ ergibt sich aus der
Summe des Auflösungsfehlers $\Delta_{\mathrm{Res}}$ und des Alias-Fehlers $\Delta_{\mathrm{Alias}}$:
\begin{eqnarray}
\Delta(t) &=& \Delta_{\mathrm{Res}}(t) + \Delta_{\mathrm{Alias}}(t)
\end{eqnarray}
Bedingt durch den systematischen Charakter wächst $\Delta_{\mathrm{Alias}}$ ebenso
wie die Flugweite $d$ näherungsweise linear mit der Zeit:
\begin{eqnarray}
\Delta_{\mathrm{Alias}} & \approx & t \langle \Delta v \rangle    \label{ErrModel}  \\
d(t) & \approx & t \langle v \rangle
\end{eqnarray}
$\langle \Delta v \rangle$ ist der mittlere Geschwindigkeitsfehler und $\langle v \rangle$
die mittlere Windgeschwindigkeit. Da die Ursache des Windfehlers $\langle \Delta v \rangle$ eine
Verschiebung im Zeit- bzw. Ortsspektrum ist, ist $\langle \Delta v \rangle \propto \langle v \rangle $ zu erwarten.
Für die meisten Trajektorien liefert $\Delta_{\mathrm{Res}}$ nur einen kleinen Beitrag, weshalb
auf die genaue Analyse des zeitliche Verlaufs verzichtet wird.

Zur Fehlerauswertung wird der Fundort mit der Vorhersage der Simulation über ein spezielles Koordinatensystem in
Verbindung gebracht (siehe Abb. \ref{TrajectCoord}). Die $y$-Koordinate wird aus
der Länge des Lots auf die Trajektorie bestimmt (positiv für eine Abweichung nach links vom
Startort aus gesehen), wobei die Trajektorie an den Enden über ihre Tangente verlängert wird.
Die Länge der Trajektorie bis zum Fußpunkt ist die in Flugrichtung positiv gezählte
$x$-Koordinate. Ein einfaches und robustes Fehlermaß ist die kürzeste Entfernung $\Delta_{\min}$ des
Kartenfundorts $\vec x_f$ zur berechneten Trajektorie, die in der Abbildung durch eine
gestrichelte Linien dargestellt ist. Aus den vorangegangenen Ausführungen erscheint der relative
Fehler des Fundorts besonders geeignet, der ebenfalls berechnet wird:
\begin{eqnarray}
\Delta_{\min}  &:=& \min \{ |\vec x_f - \vec x'| \quad | \vec x' \in \mathrm{Trajektorie}  \} \label{DeltaMin} \\
\Delta_{\mathrm{Rel}} &:=& \frac{\Delta_{\min}}{d} \label{RelError}
\end{eqnarray}
Durch die Definition von $\Delta_{\min}$ ist der relative Fehler beschränkt: $\Delta_{\mathrm{Rel}} \leq 1$.

\begin{figure}
\begin{center}
\includegraphics[scale=0.5,angle=270]{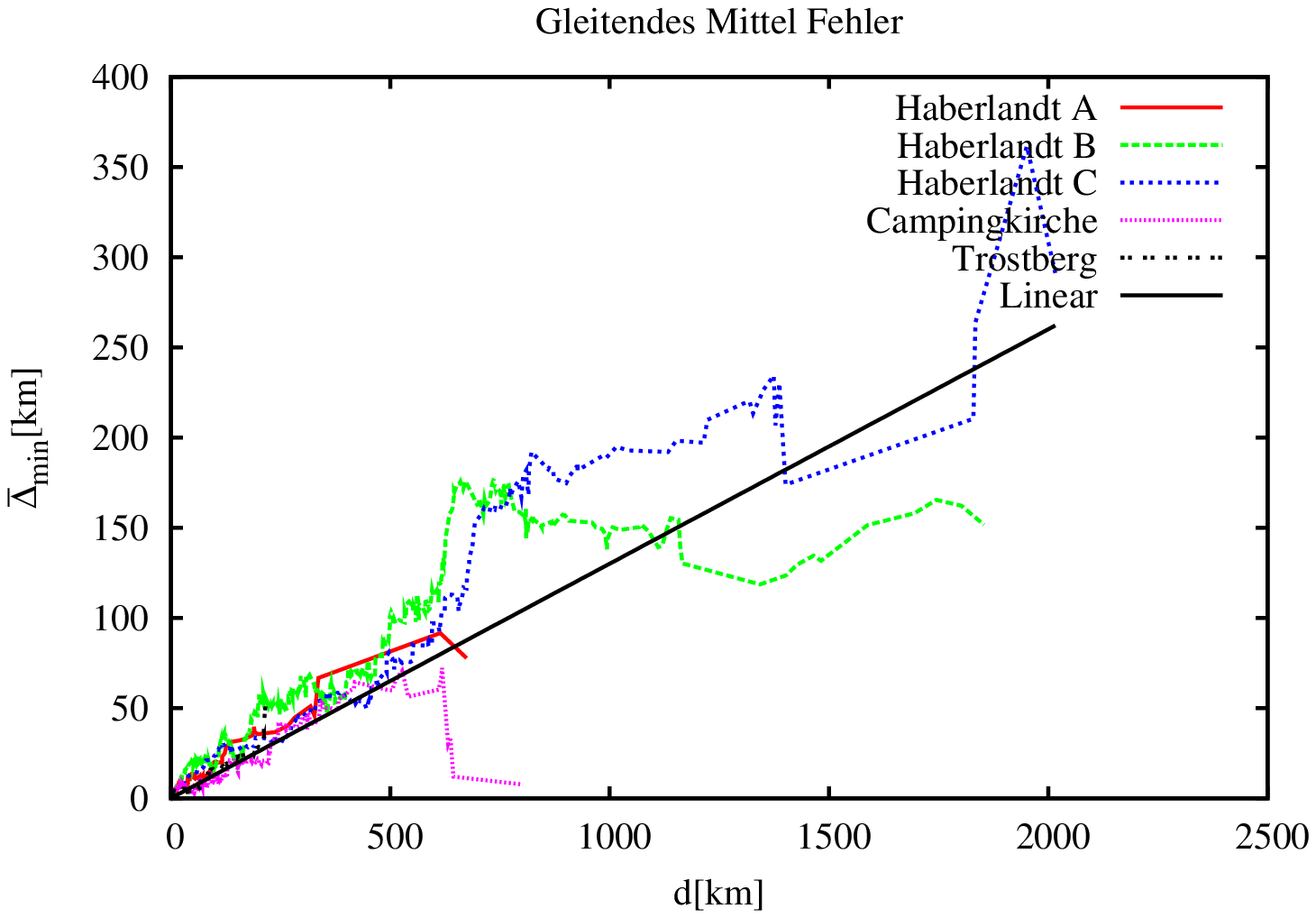}
\end{center}
\caption[Mittlerer Fehler]{Gleitendes Mittel des absoluten Trajektorienfehlers $\Delta_{\min}$. Gl.~\ref{AvgErrorD}
ist als Gerade mit eingetragen. \label{Trans_Fehler} }
\end{figure}

Nun werden die in Gl. \ref{DeltaMin} und Gl. \ref{RelError} angegebenen Fehlermaße berechnet
und statistisch ausgewertet. Zuerst wird die Abhängigkeit von $\Delta_{\min}$ von
der Flugweite $d$ untersucht. Um statistische Schwankungen zu reduzieren, werden die einzelnen
Fehlerwerte $\Delta_{\min,k}$ aufsteigend sortiert und ein gleitendes Mittel angewendet:
\begin{eqnarray}
\bar \Delta_{\min,i} = \frac{1}{2l+1} \sum_{k=-l}^l  \bar \Delta_{\min,i+k}
\end{eqnarray}
Der Mittelungsbereich umfaßt 41 Werte ($l=20$) und wird an den Rändern symmetrisch verkürzt.
Abbildung \ref{Trans_Fehler} faßt alle Fehler zusammen. Trotz der unterschiedlichen Datenquellen zeigt
sich ein universelles lineares Ansteigen des absoluten Fehlers mit der Flugweite. Dies
bestätigt die bereits ausgeführte Überlegung, daß der Alias-Effekt in der Windfeldabtastung
einen systematischen Fehler bedingt. Das Fehlerverhalten läßt sich gut durch
\begin{eqnarray}
\bar \Delta_{\min} (d) &=& 0,13\,d  \label{AvgErrorD}
\end{eqnarray}
beschreiben. Der mittlere Fehler von 13 \% paßt gut zu den Ergebnissen von
\cite{rolph1990sensitivity,stohl1995interpolation}, wobei zu berücksichtigen ist, daß sich
Gl.~\ref{AvgErrorD} nur auf einen eindimensionalen Fehler senkrecht zur Trajektorie bezieht.

\begin{figure}
\begin{center}
\includegraphics[scale=0.5,angle=270]{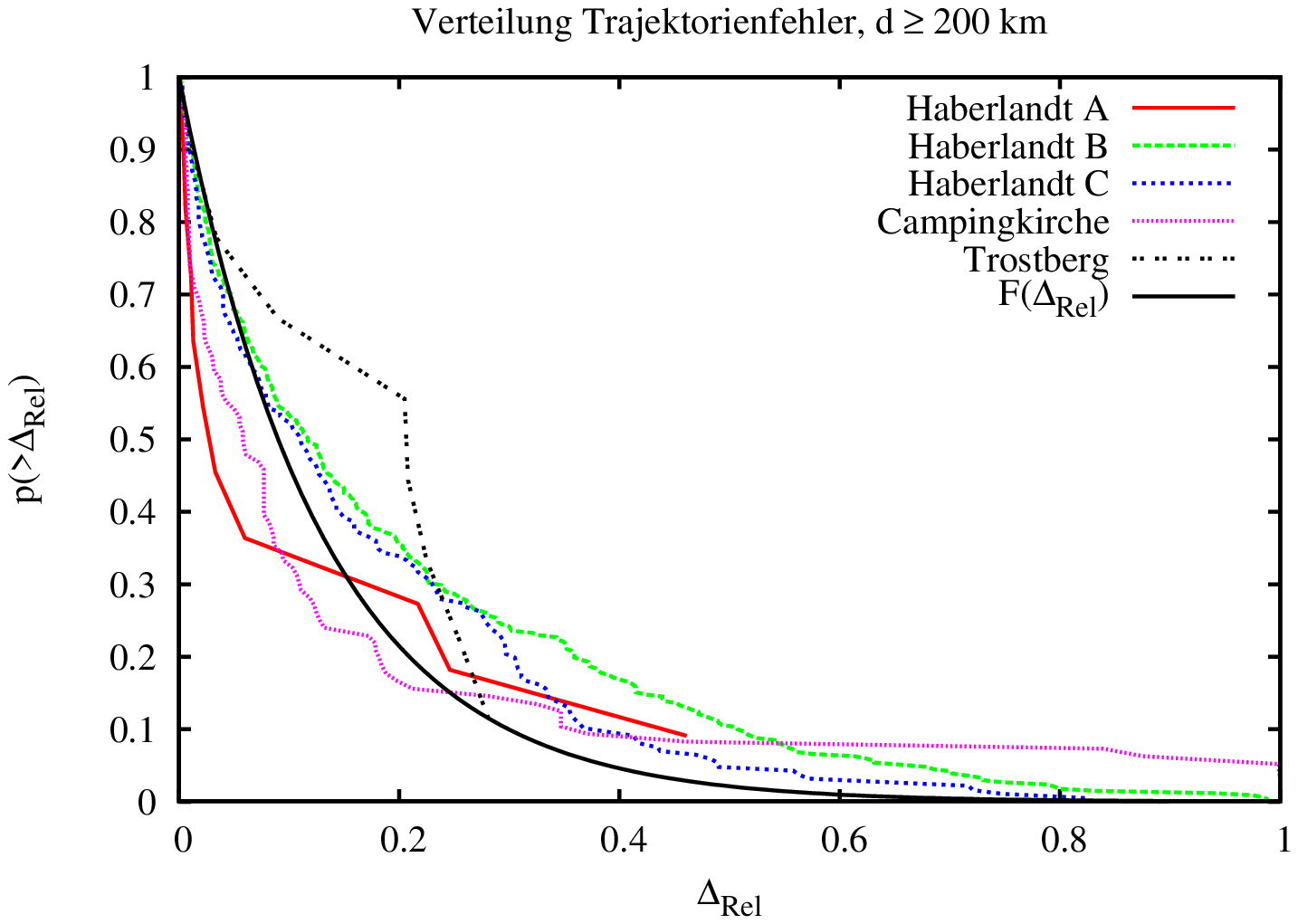}
\end{center}
\caption[Relativer Fehler]{Relativer Fehler der einzelnen Trajektorienrechnungen. Gl. \ref{FRelError}
ist mit eingetragen.\label{Rel_Fehler}}
\end{figure}

Da der Alias-Effekt von dem Anteil hoher Frequenzen in den Windfluktuationen abhängt,
ist es durchaus möglich, daß Wettersituationen vorliegen, die auch mit einer Abtastung
von 12~h relativ gut zur rekonstruieren sind. Um einen Einblick in die Fehlerverteilung zu erhalten,
wird der relative Fehler nach Gl. \ref{RelError} für jeden Kartenballon ausgewertet und
in einer kumulierten Verteilung aufgetragen. Trajektorien kürzer als 200~km wurden ausgeschlossen,
um einer Verzerrung der Verteilungen durch den vernachlässigten Auflösungsfehler vorzubeugen.

Wie die Schwankungen im mittleren Fehler erwarten lassen,
sind hier die Ergebnisse weit uneinheitlicher (Abb. \ref{Rel_Fehler}). Qualitativ folgen aber alle Verteilungen
annähernd der exponentiellen Verteilung
\begin{eqnarray}
F\left( \Delta_{\mathrm{Rel}} \right) &=& \exp\left(- \frac{   \Delta_{\mathrm{Rel}} }{ 0,13 } \right) \label{FRelError}
\end{eqnarray}
die mit dem Mittelwert in Gl. \ref{AvgErrorD} konsistent ist. Die Abnahme der verfügbaren
Daten mit zunehmender Flugweite erschwert die Beurteilung, ob diese Verteilung wirklich
unabhängig von der Flugweite $d$ ist. Zumindest im Bereich $d < 1000$~km unterstützen
die Daten diese Annahme.

Da keine zeitlich höher aufgelösten Winddaten zur Verfügung stehen, läßt sich dieses Fehlerverhalten
nicht eindeutig mit dem Windspektrum in Verbindung bringen. Einzelne Messungen und theoretische Arbeiten
deuten aber in diese Richtung \cite{schlax2001sampling,daoud2003synoptic}.

\section{Ausreißer\label{SectAusr}}

Ballons, die deutlich weiter fliegen als berechnet oder deren Trajektorie
einen großen Fehler aufzeigt, sind mögliche Hinweise auf Schwächen oder
Fehler in dem vorgestellten Trajektorienmodell. Um zu entscheiden, ob es sich
in diesen Fällen nur um Extremwerte der beschriebenen Fehlerverteilung (Abb. \ref{Rel_Fehler})
oder wirkliche Probleme handelt, wurden alle auffälligen Ballons noch einmal genauer untersucht.
Ballons mit einer normierten Flugweite $\tilde d$ über 1,1 wurden in einer Gruppe zusammengefaßt.
Aus den verbleibenden Ballons wurden alle Ballons mit einem Fehler $\Delta_{\min}$ über 200 km in einer
weiteren Gruppe zusammengefaßt. Um der Ursache der Abweichungen auf die Spur zu kommen,
wurden für jeden Ballon noch einmal Trajektorien "`von Hand"' berechnet und die
Struktur des interpolierten Windfelds untersucht. Anhand dieser Ergebnisse wurden
vier Fehlerkategorien erstellt:
\begin{enumerate} \label{ErrorKatTab}
\item
Vertretbare Abweichung. Bei Gruppen sind die Ergebnisse für die
restlichen Ballons gut oder die Flugweite war nahe 200 km.
\item \label{Error2}
Schwierige Windfeldrekonstruktion (Divergenzen, Scherströmungen, keine
dominante Hauptströmung). Dies umfaßt sowohl Orts-- als auch Zeitinterpolationsprobleme.
\item \label{Error3}
Starke Abweichung von der berechneten Trajektorie und anderen Ballons (falls verfügbar).
Flug gegen die vorherrschende Windrichtung ohne erkennbaren Grund.
\item \label{Error4}
Verbleibende starke Abweichungen.
\end{enumerate}
Tabelle \ref{UnusualData} enthält die Klassifizierung aller auffälligen Ballons. Der Großteil
der Abweichungen ist dabei der ungenauen Rekonstruktion des
Windfelds anzulasten (Fall \ref{Error2}). Für die Gruppe der Ballons mit normierten Flugweiten
über 1,1 deckt diese Erklärung alle Flüge ab, so daß aus den Daten kein Widerspruch
zur berechneten maximalen Flugweite abgeleitet werden kann.

\begin{table}[tb]
\begin{center}
\begin{tabular}{|l|r|r|rrrr||r|rrrr|} \hline
Referenz      & Ballons & $\tilde d >1,1$ & 1) & 2) & 3) & 4) &  $\Delta_{\min}>200$ km & 1) & 2) & 3) & 4) \\ \hline
Campingkirche &     881 &  2  &  1 & 1  &  0 &  0 &  6 &  0 &  1 &   4 &   1 \\ \hline
Glaschke      &       7 & --  & -- & -- & -- & -- & -- & -- & -- &  -- &  -- \\ \hline
Haberlandt A  &     156 & --  & -- & -- & -- & -- & -- & -- & -- &  -- &  -- \\
Haberlandt B  &     636 & 13  &  4 &  9 &  0 &  0 & 31 &  7 & 17 &   3 &   4 \\
Haberlandt C  &     266 &  7  &  7 &  0 &  0 &  0 & 19 &  0 & 15 &   0 &   4 \\ \hline
Trostberg     & 211 & --  & -- & -- & -- & -- & -- & -- & -- &  -- &  -- \\ \hline
\end{tabular}
\end{center}
\caption[Auffällige Daten]{Auffällige Daten. Die detaillierte Beschreibung der Kategorien 1--4 befindet
sich auf Seite \pageref{ErrorKatTab}. \label{UnusualData}}
\end{table}

Es verbleibt aber ein Teil Ballonflüge mit unklaren Ursachen (Fall \ref{Error4}). In einigen Fällen
könnten die Ballons an Küsten angespült worden sein, aber auch durch starke Regenfälle o.\,ä.
veränderte Trajektorien könnten in diesen Fällen eine Rolle spielen.

Besonders rätselhaft sind die Fälle, in denen Ballons gegen die vorherrschende
Windrichtung gedriftet sind, ohne daß die Windfeldrekonstruktion problematisch erscheint  (Fall \ref{Error3}).
Hier könnten auch einfache Datenfehler\footnote{Bei der Formatierung der Ballonflugdaten wurden je nach Datenquelle
in 0,1 \% bis 1 \% der Fälle Fehler erkannt (Zahlendreher, namensgleiche Fundorte u.\,ä.).},
fehlerhaft ausgefüllte Ballonkarten oder Verwechslungen bei der Bestimmung des
Fundorts vorliegen. Bei Ballonwettbewerben bleibt schließlich auch die Möglichkeit der gezielten Fälschung.
Herr Haberlandt nennt eine Fälschung (oder einen Scherz)
durch den Finder [sic] der Karte. Aber selbst wenn man Fall~\ref{Error3} durchgängig als gezielte Fälschungen betrachtet,
bleibt der auf die Teilnehmer der Ballonwettbewerbe hochgerechnete Anteil sehr gering.
Auch wenn eine abschließende Beurteilung zu diesem Zeitpunkt nicht möglich ist, stellen
diese Fälle mit einem Anteil um 0,5 \% nur einen Bruchteil aller Kartenballons dar.

\section{Jahresstatistiken \label{Jahresstat}}

Die absolute Flugweite eines Kartenballons wurde bis jetzt kaum diskutiert, da
sie in hohem Maße von den Wind-- und Wetterbedingungen abhängt und nur
wenig über die Gültigkeit eines Trajektorienmodells aussagt. Diese
für die Ausrichter von Ballonwettbewerben und natürlich deren Teilnehmer
wichtige Größe soll nun näher beleuchtet werden.

Mit dem erfolgreich validierten Modell wurden "`virtuelle"' Ballontrajektorien und
Landepunkte berechnet, die eine genaue Vorstellung von den erreichbaren Flugweiten
ermöglichen. Dieser Ansatz erlaubt eine weit bessere Statistik, die mit echten Ballons
kaum mit vertretbarem Aufwand möglich wäre.

Als Startpunkt dieser virtuellen Flüge dient Wildon\footnote{Wildon/Steiermark,
46\degree 54\arcminute N 15\degree 33\arcminute E}, das auch der
Startort des größten Datensatzes "`Haberlandt A--C"' ist. Starts wurden für jeden Tag der
Jahre 1999~--~2008 mit der Startzeit 14:00 Uhr(UT) durchgeführt. Bei einer
Flugzeit von rund 20 Stunden liegt der Zeit\-ab\-lauf der Trajektorie symmetrisch zu den Wetterdaten
um 00:00 Uhr(UT), wodurch systematische Einflüsse der Zeitinterpolation reduziert werden. Da das
Atmosphärenmodell deterministisch ist, ergibt ein Ballon pro Startzeitpunkt jeweils eine eindeutige
Trajektorie. Für jeden Startzeitpunkt wurde der Ballonauftrieb iterativ verändert, um die
charakteristischen Flugweiten $d_{\max}$ und $d_{\mathrm{burst}}$ automatisch zu bestimmen.

\begin{figure}
\begin{center}
\includegraphics[scale=0.5,angle=270]{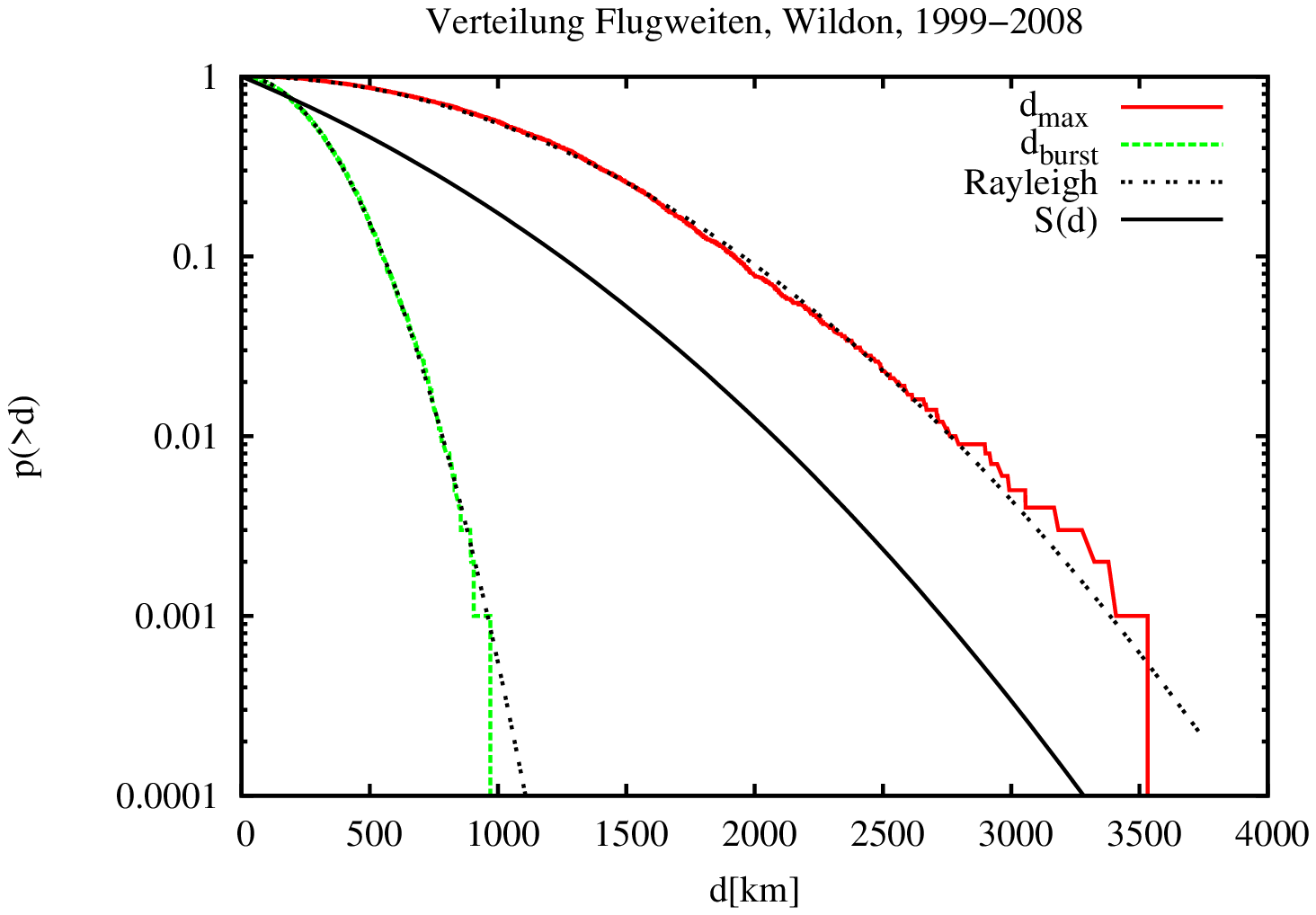}
\end{center}
\caption[Jahresstatistik]{Berechnete Jahresstatistik. Zum Vergleich zeigt $S(d)$ die
Verteilung in Gl. \ref{SdGleich} \label{YearStatistics} }
\end{figure}

Die zweidimensionale Verteilung der Fundorte mit maximaler Flugweite $d_{\max}$ auf der Erdoberfläche
wird gut durch eine bivariate Gauß-Verteilung beschrieben. Bedingt durch die Westwinddrift
liegt der Modalwert 660 km weiter östlich bei
(46\degree 15\arcminute  N 24\degree 4\arcminute E).
Trotz dieser Asymmetrie lassen sich die in Abbildung \ref{YearStatistics} zusammengefassten
kumulierten Abstandsverteilungen gut durch einfache Rayleigh-Verteilungen annähern:
\begin{eqnarray}
S_{\max}(>d) &=& \exp\left(-\frac{d^2}{2 \sigma_{\max}^2} \right)
\qquad   \sigma_{\max} = 911 \mbox{ km} \\
S_{\mathrm{burst}}(>d) &=& \exp\left(-\frac{d^2}{2 \sigma_{\mathrm{burst}}^2} \right)
\qquad   \sigma_{\mathrm{burst}} = 258 \mbox{ km}
\end{eqnarray}
Die Form der Abstandsverteilung geht im wesentlichen auf die zugrundeliegenden rayleighverteilten
Windgeschwindigkeiten zurück. Bei den größten Abständen weichen die Daten etwas von dieser einfachen
Verteilung ab, aber dieser Bereich wird bereits durch starke statistische Schwankungen dominiert.
Die gewonnenen Verteilungen spiegeln bis jetzt nur den Idealfall wieder -- jeder Ballon erreicht nach
dem Start die maximal mögliche Flugweite. Mit einer bekannten Überlebensfunktion
$F$(siehe Kap.~\ref{BallonDistance}), bezogen auf die normierte Flugweite, kann diese Verteilung aber
näherungsweise korrigiert werden, um ein realistischeres Bild zu erhalten:
\begin{eqnarray}
S(>d)  &=& - \int_0^1 S(d / x ) \frac{dF(x)}{dx} d x
\end{eqnarray}
Exemplarisch sei hier das Ergebnis für den Fall gleichverteilter normierter Flugweiten gegeben:
\begin{eqnarray}
S_{\max}(>d) &=&  \exp\left(-\frac{d^2}{2 \sigma_{\max}^2} \right)
- \sqrt{\frac{\pi}{2}} \frac{d}{\sigma_{\max}} \mathrm{erfc} \left(\frac{d}{\sqrt{2} \sigma_{\max}} \right)  \label{SdGleich} \\
&=& \frac{\sigma_{\max}^2}{d^2} \exp\left(-\frac{d^2}{2 \sigma_{\max}^2} \right)
+ \mathcal{O}\left( \frac{\sigma_{\max}^4}{d^4} \right)
\end{eqnarray}
Bereits dieses ebenso idealisierte Szenario zeigt eine deutlich geringere Wahrscheinlichkeit
großer Flugweiten.

Der Vergleich der Kartenballondaten mit dieser Betrachtung zeigt, daß alle Ballonwettbewerbe
weit hinter der Modellfunktion Gl. \ref{SdGleich} zurückbleiben. Nur Haberlandt B und besonders
Haberlandt C erreichen beachtliche Flugweiten. Hier ist aber zu beachten, daß gezielt Tage
mit hohen Windgeschwindigkeiten zum Start ausgewählt wurden, während in der berechneten Jahresstatistik
über alle Tage des Jahres gleich gemittelt wurde.

Diese Statistik schließt immer wieder gemeldete extreme Flugweiten wie Flüge über den Atlantik
mit Sicherheit aus. Alte Ausgaben des Guinness-Buch der Rekorde verzeichnen
noch "`Rekordflugweiten"' deutlich über 10\,000 km \cite{roberts1995dynamics}, aber in neueren Auflagen
ist diese Kategorie folgerichtig nicht mehr
enthalten\footnote{Diese Kategorie wird mindestens seit 1995 nicht mehr geführt, Mitteilung
der Redaktion des Guinness World Records Buches vom 23.2.2010.}.
Auf der Basis der hier durchgeführten Rechnungen liegt die maximal erreichbare Distanz
(zumindest in Europa) wenig über 3000 km.

\begin{figure}
\begin{center}
\includegraphics[scale=0.5,angle=270]{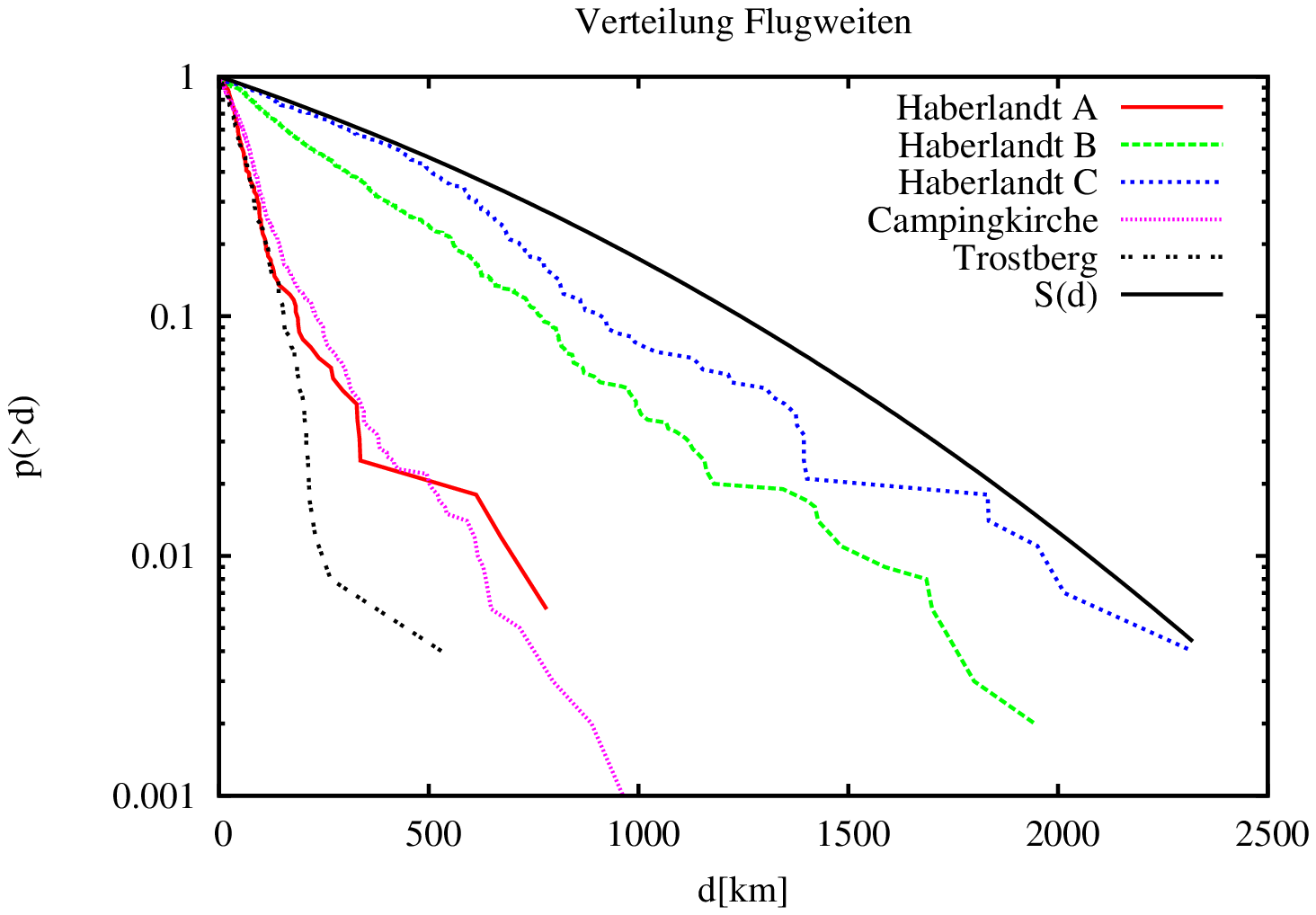}
\end{center}
\caption[Erreichte Flugweiten]{Erreichte Flugweiten.
Gl. \ref{SdGleich} dient hier ebenfalls als Vergleichspunkt.\label{Weite_km} }
\end{figure}

\section{Zeit bis zum Fund}

Abschließend wird noch die Zeit betrachtet die vergeht, bevor eine Karte gefunden
und zurückgeschickt wird. Unter der Annahme, daß die Fundwahrscheinlichkeit an einem
Landeplatz zeit\-un\-abhängig ist, erhält man eine exponentielle Verteilung des Anteils
der noch nicht gefundenen Karten $  F_{\mathrm{fund}} $:
\begin{eqnarray}
F_{\mathrm{fund}}(t) &=&\exp(-t/\tau) \label{SimplepFund}
\end{eqnarray}
Die mittlere Liegezeit $\tau$ bis zum Fund der Karte beträgt für die Daten von Glaschke
6,7 Tage, nach den Daten Trostberg ergibt sich eine Lebensdauer von 6~--~18 Tagen. Werden
nur kurze Zeiten im Wochenbereich betrachtet, liefert Gl. \ref{SimplepFund} eine sehr
gute Beschreibung, auch wenn signifikante Abweichungen erkennbar sind. Besonders
deutlich wird das bei den Daten von Haberlandt, die Karten beinhalten, die nach mehr als
einem Jahr zurückgeschickt wurden. Hier erfordert die Heterogenität der Fundorte, von
einer ganzen Verteilung von mittleren Liegezeiten auszugehen:
\begin{eqnarray}
F_{\mathrm{fund}}(t) &=& \int_0^{\infty} \exp(-t/\tau) f(\tau) d\tau \label{ComplexpFund}
\end{eqnarray}
Im konkreten Fall der Daten von Haberlandt sind bereits
zwei Liegezeiten ausreichend, um eine gute Näherung zu erhalten:
\begin{eqnarray}
F_{\mathrm{fund}}(t) & \approx & 0,85 \exp\left(-\frac{t}{40 \mbox{ d}} \right)   +0,15 \exp\left(-\frac{t}{220 \mbox{ d}} \right)
\end{eqnarray}
Es ist zu erwarten, daß die Fundwahrscheinlichkeit mit der Bevölkerungsdichte und der
Siedlungsstruktur in Zusammenhang gebracht werden kann. Diese Idee wurde aber aufgrund
der unzureichenden Datenlage nicht weiter verfolgt.

\chapter{Diskussion}

Das in dieser Arbeit entwickelte Modell ermöglicht vielfältige Einblicke in die Dynamik
eines Kartenballons. Der Vergleich der durchgeführten Simulationen mit Ergebnissen von
Ballonwettbewerben erlaubt fundierte Aussagen über die typische Ballontrajektorie:
Ballons erreichen regelmäßig Höhen über 12 km, widerstehen dabei Temperaturen bis
zu $-60$ \celsius{} und können bis zu 24 Stunden und länger in der Luft bleiben.
Als wichtigster Parameter bei der Flugweitenmaximierung hat sich die Ballonfüllung
herausgestellt. Der berechnete optimale Startauftrieb von rund 20 \% der Leermasse
ist dabei in guter Übereinstimmung mit dem unabhängig experimentell ermittelten
optimalen Ballonvolumen von 8,2~l \cite{Haberlandt}.
Ist die Ballonfüllung optimiert, haben alle anderen Balloneigenschaften
bemerkenswerterweise nur noch einen geringen Einfluß auf die Flugweite. Alleine
die maximal erreichbare Flughöhe zeichnet sich durch einen fast quadratischen
Einfluß aus. Regen und UV-Strahlung (besonders im Zeitraum Juni -- Juli) erscheinen als die
stärksten externen Einflüsse, die die Lebenszeit eines Ballons auf wenige Stunden
beschränken können.

In der Praxis sind unter diesen Umständen Flugweiten bis zu 3000 km möglich, was die Angaben
in \cite{roberts1995dynamics} deutlich nach oben korrigiert. Nach der Kenntnis des Autors liegt
der in dieser Arbeit durch Trajektorienrechnungen bestätigte Rekord bei 2321~km \cite{Haberlandt}.
Der auf den ersten Blick gering
erscheinende Drucksprung eines Ballon trägt besonders zu einer langen Verweilzeit in großen
Höhen und damit zu großen Flugweiten bei. Leider bleiben Ballonwettbewerbe durch zu
stark gefüllte Ballons und Starts tagsüber im Sommer weit unter diesen Werten.

Erfreulich sind die im Vergleich zu den Betrachtungen in Kapitel \ref{ChapBoden}
sehr hohen Fundquoten (siehe Tab. \ref{UsedData}), die vermuten lassen, daß die
Mehrheit der gefundenen Karten auch anschließend zurückgeschickt wird.

Trotz der erfolgreichen Validierung des Kartenballonmodells besteht noch Raum für
Verbesserungen. Die Qualitätskontrolle und Integration der primären Radiosondendaten
könnte stärker systematisiert werden \cite{collins2001operational, collins2001operational2}.
Prinzipiell ist auch die Einbindung eines digitalen Geländemodells in das Atmosphärenmodell
denkbar (z.\,B. \cite{GTOPO30}). Ein solcher Schritt erfordert aber zwangsläufig
eine konsistente Integration des Windfelds, was weit über die in dieser Arbeit verwendete
einfache Interpolation hinausgehen würde.

Alle Aussagen über das Ballonverhalten im Flug sind nur indirekt erschlossen. Direkte
Einblicke würden nur Flugsimulationen in einer Klimakammer bringen (in eingeschränkter
Form von \cite{Kofoed1992} verwendet), die auch definitive Aussagen über das Materialverhalten
von Gummi bei tiefen Temperaturen und starker Dehnung liefern würden. Ebenso wären
Laborexperimente nötig, um das Zusammenspiel von UV-Strahlung und Oxidantien wie
Ozon bei der Wirkung auf den Ballongummi zu untersuchen.

Zur Steigerung der Zuverlässigkeit ist auch die Verwendung von Ballongespannen möglich,
die weitaus vielfältigere Möglichkeiten zur Optimierung bieten. Aber auch bei einer weitgehenden
Optimierung eines Kartenballons bleibt ein Problem bestehen: Ein normaler Kartenballon verfügt
über keine Regelmechanismen, die das Aufsteigen über die maximal mögliche Höhe und damit
das Platzen aktiv verhindern.

\thispagestyle{plain}
\section*{Danksagung}

\addcontentsline{toc}{chapter}{Danksagung}

Jede Theorie steht und fällt mit dem Experiment. Ohne die Zusammenarbeit mit
den Veranstaltern verschiedener Ballonwettbewerbe wäre diese Arbeit in der
vorliegenden Form nicht möglich gewesen.

Ich bedanke mich herzlich bei Herrn Dr. Michael Kollefrath (Campingkirche FCO~Stollhofen)
und Herrn Hans Aitl (Kolpingfamilie Trostberg/Schwarzau) für die Unterstützung dieses Projekts
mit ihren Ballonwettbewerben.
Ganz besonders bedanke ich mich bei Herrn Oskar Haberlandt, der durch seine
reichhaltigen Ballonexperimente und die geduldige Diskussion aller damit in
Zusammenhang stehenden Fragen diese Arbeit ungemein bereichert hat.

Mein Dank gilt auch Frau Dr.~Michaela~Kleinert und Herrn Dr.~Andreas~Ernst,
die tatkräftig alle Entwürfe durchgesehen und mit Ihren Anmerkungen
verbessert haben. Herr Dr. Mathias Schott hat besonders den experimentellen Teil
dieser Arbeit unterstützt.

\begin{appendix}

\chapter{Nützliche Konstanten\label{UsefulConst}}

\begin{tabular}{l@{\qquad:\qquad}ll}
$g$       &  9,81 m/s$^2$                               & Erdbeschleunigung \\
$R_{\mathrm{Erde}}$ &  6371 km                          & Mittlerer Erdradius \\
$e$       & $ 1,602\, 176 \, 487(40) \times 10^{-19}$ C & Elementarladung \\
$k_B$     & $ 1,380\,6504(24)\times 10^{-23}$ J/K       &  Boltzmann Konstante \\
$R$       & $ 8,314\, 472(15)  $ J mol$^{-1}$ K$^{-1}$  & Gaskonstante  \\
$\sigma $ & $ 5,670 \, 400 (40)\times 10^{-8} \frac{\mathrm{W}}{\mathrm{m}^2\mathrm{K}^4}$ & Stefan-Boltzmann-Konstante  \\
$c$       & $299\, 792\, 458$ m/s                       & Lichtgeschwindigkeit \\
$h$       & $6,626\,068\,96(33) \times 10^{-34}$ Js     & Plancksches Wirkungsquantum \\
$N_A$     & $ 6,022 \, 141 \,79(30) \times 10^{23} $ mol$^{-1}$  & Avogadro Konstante \\
$V_m$     & $24,464 \, 0424 \,$ l mol$^{-1}$            & Molares Volumen bei Standardbedingungen \\
$T_0$     & $- 273,15^{\circ}$C                         & Absoluter Nullpunkt   \\
NTP       & 25$^{\circ}$C \quad  101\,325 Pa            & Standardbedingungen (nicht genormt) \\
STP       & \,\, 0$^{\circ}$C \quad  101\,325 Pa        & Normbedingungen  \\ \hline
\end{tabular}

\section*{Materialeigenschaften}

\begin{tabular}{lrrrrr}
& $\rho$[kg/m$^3$] & \%[Luft] &  $m_{\mathrm{Mol}}$[g] &  $\lambda[ \frac{\mathrm{W}}{\mathrm{m\,K}}]$ & $c_p[\frac{\mathrm{J}}{\mathrm{g\,K}}]$ \\ \hline
Luft   &  1,204   & 100,000 & 28,964  & 0,0240 & 1,005 \\
He     &  0,1785  & --      &  4,0026 & 0,15   & 5,19412 \\
H$_2$  &  0,0837  & --      &  2,016  & 0,1861 & 14,3 \\
N$_2$  &  1,166   & 78,084  & 28,013  &        & 1,040 \\
O$_2$  &  1,332   & 20,942  & 31,999  &        & 0,9196 \\
CO$_2$ &  1,842   & 0,038   & 44,010  &        & 0,8504 \\
Gummi  & 1000?    &  --     &  --     & 0,16?  & 1,4? \\
\end{tabular}

\vspace{5mm} \noindent
Dichten jeweils bei 20\celsius{} NP.
\begin{tabular}{l@{\qquad:\qquad}ll}
Kinematische Viskosität von Luft   & $ 1,5 \times 10^{-5}$ & m$^2$/s \\
\end{tabular}

\vspace*{5mm}
\noindent
Spezifische Dichte verschiedener Materialien:
\begin{center}
\begin{tabular}{|l|cl|} \hline
PE-Folie                          & 0,12 & g/dm$^2$ \\
Holzstab $\varnothing$ 2 mm       & 0,44 & g/dm \\
Al-Folie (Schokoladenverpackung)  & 0,30 & g/dm$^2$ \\
Al-Folie                          & 0,37 & g/dm$^2$ \\
Strohhalm                         & 0,11 & g/dm \\
Papier                            & 0,83 & g/dm$^2$ \\
Butterbrotpapier                  & 0,43 & g/dm$^2$ \\ \hline
\end{tabular}
\end{center}

\chapter{Datenverarbeitung\label{UsedProg}}

\hspace*{1.2cm} \includegraphics[scale=0.3]{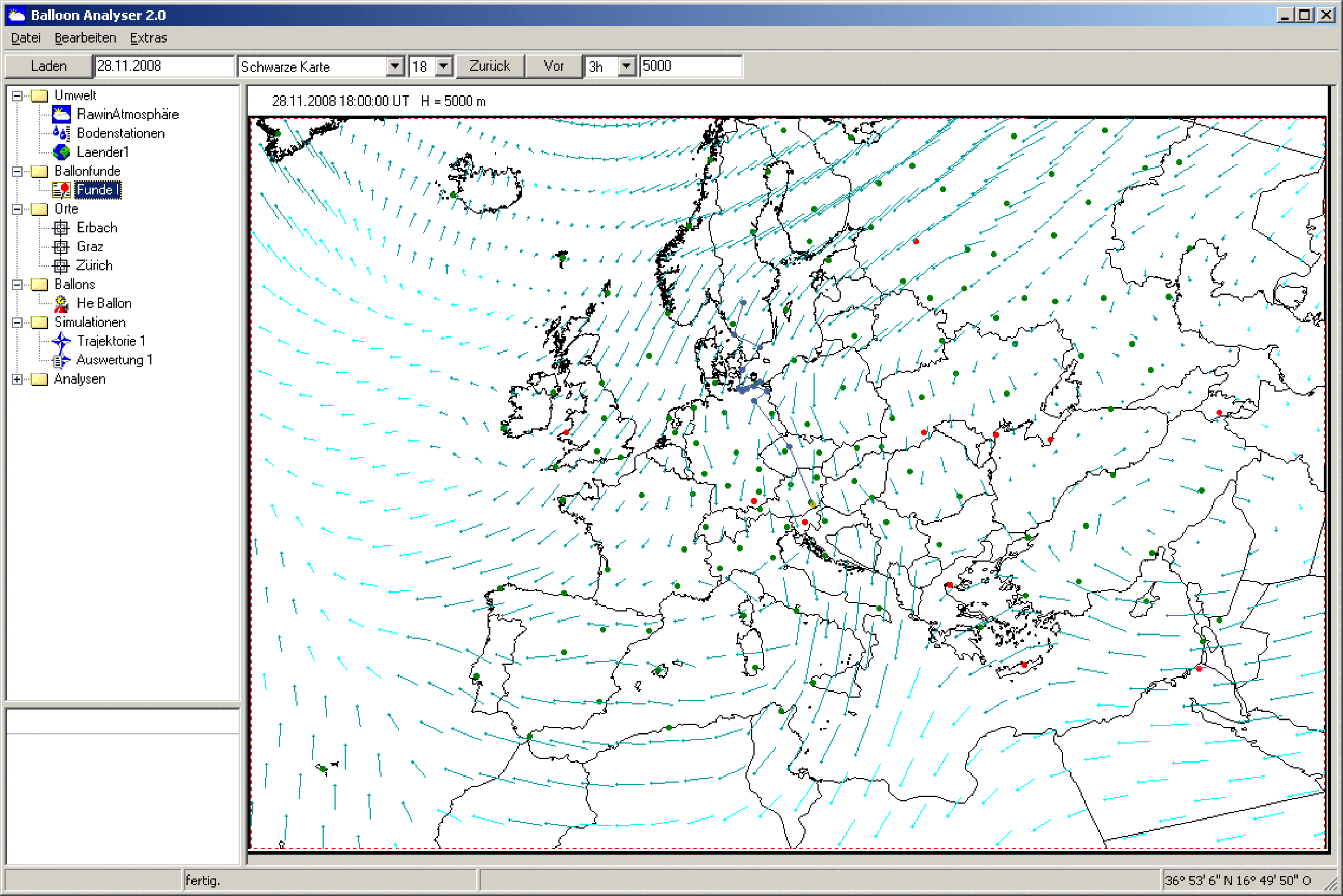}

\vspace*{5mm}

\noindent
Benutzeroberfläche des selbst entwickelten Auswertungsprogramms {\em Balloon Analyser}. Alle innerhalb
des Programms referenzierten Objekte sind links in einem Baum übersichtlich zusammengestellt. Aufgeführt
werden externe Wetterdaten und Kartenmaterial (Ordner  {\em Umwelt}), importierte Dateien mit Ballonfunden
(Ordner {\em Ballonfunde}), selbstdefinierte Startorte für unabhängige Simulationen (Ordner  {\em Orte}),
benutzerdefinierte Ballons zur Durchführung virtueller Flüge (Ordner  {\em Ballons}), Simulationen zum
Erstellen von Statistiken und Fundauswertungen (Ordner  {\em Simulation}) und schließlich Analysen zum
automatischen Erstellen verschiedener Diagramme (Ordner  {\em Analysen}).

Die Objekte im Baum können zu einem Ablaufschema verknüpft werden, das automatisch alle Simulationen berechnet
und auswertet. Neuberechnungen mit veränderten Parametern oder aktualisierten Daten sind dadurch bequem
möglich.

Das Abspeichern aller Einstellungen in einer Projektdatei dokumentiert die aufgesetzten Simulationen
und ermöglicht das spätere Nachvollziehen der durchgeführten Berechnungen.

\chapter{Formelzeichen}

\vspace*{5mm}\noindent

\begin{tabular}{l@{:\qquad}ll}
$V$          & & Volumen \\
$A$          & & Oberfläche \\
$A_q$        & & Querschnittsfläche \\
$R$          & & Radius / Gaskonstante  \\
$R_0$        & & Radius druckfreier Ballon \\
$D$          & $2R$ & Durchmesser / Diffusivität \\
$\varepsilon$ & & Volumenkorrekturfaktor  \\
$V_{\max}$   & & Maximales Ballonvolumen \\
$V_{\mathrm{Gummi}}$   & & Gummivolumen der Ballonhülle \\
$M_{\mathrm{Gummi}}$   & & Gewicht der Ballonhülle \\
$\rho$       & & Dichte  \\
$N$          & & Stoffmenge [mol]  \\
$T$          & & Temperatur  \\
$t$          & & Zeit  \\
$\mathcal{F}$& & Strahlungsfluß\\
$\lambda$    & & Wärmeleitfähigkeit / Geographische Länge\\
$\beta$      & & Geographische Breite \\
Nu           & & Nusselt-Zahl \\
Re           & $v L/\nu$          & Reynolds-Zahl \\
Pr           & $\eta c_p/\lambda$ & Prandtl-Zahl\\
Gr           & & Grashof-Zahl\\
$\nu$        & & Kinematische Viskosität \\
$\eta$       & & Dynamische Viskosität \\
$c_w$        & & Widerstandsbeiwert\\
$F_a$        & & Nettoauftrieb ($=$ Auftrieb $-$ Gewichtskraft) \\
$Q$          & & Permeabilität (bezogen auf Partialdruck) \\
$\delta$     & & Permeabilität (bezogen auf Stoffmenge) \\
$k_H$        & & Henry-Konstante\\
$n$          & & Teilchendichte \\
$p(>x)$      & & Komplementäre Verteilungsfunktion \\
$m$          & & Molekularmasse \\
$\Delta$     & & Fehler \\
$v$          & & Geschwindigkeit \\
$r$          & & Abstand \\
\end{tabular}

\chapter{Reibungskoeffizient \label{FrictApp}}

\begin{center}
\includegraphics[scale=0.5,angle=270]{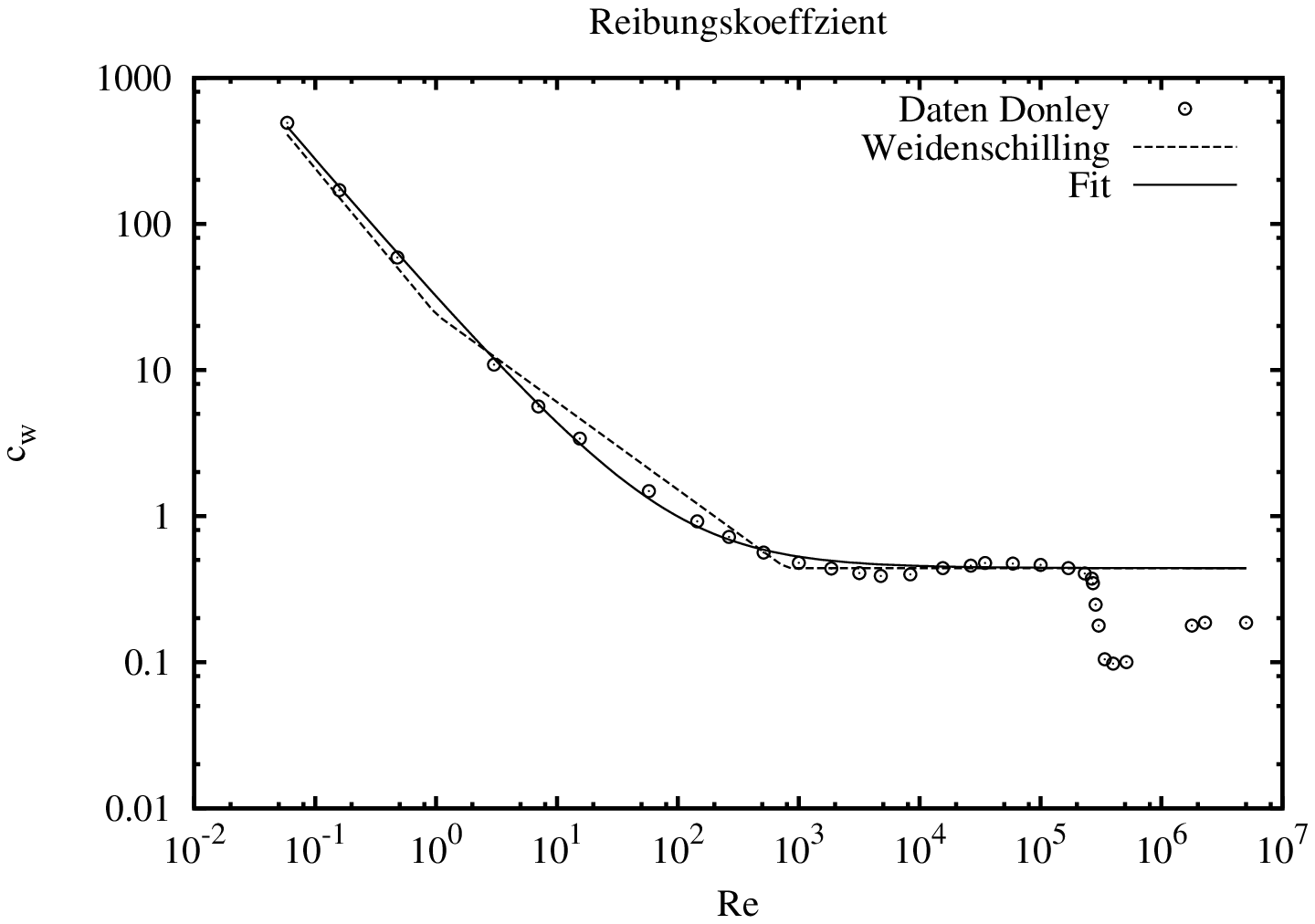}
\end{center}

\noindent
Formel von Weidenschilling \cite{weidenschilling1977aerodynamics}:
\begin{eqnarray}
c_w &=& \left\{ \begin{array}{lcl}
24/\mathrm{Re}       & \mbox{für} &  \mathrm{Re}<1 \\
24\,\mathrm{Re}^{-0,6} & \mbox{für} &   1<\mathrm{Re}<800 \\
0,44                 & \mbox{für} &   800<\mathrm{Re}
\end{array} \right. \qquad \mathrm{Re} = \frac{2R v}{\nu}
\end{eqnarray}
Fit der Daten von \cite{Donley2008}:
\begin{eqnarray}
c_w &=& \frac{24}{\mathrm{Re}} + 7,6 \, \mathrm{Re}^{-0,693} +0.44
\end{eqnarray}

\chapter{Mittleres Atmosphärenprofil\label{AtmApp}}

\begin{table}[h]
\begin{center}
\begin{tabular}{|rrrrr|} \hline
Höhe[km] & T[$^{\circ}$C] &  $v$[m/s] &  Druck[hPa] & Dichte [$\rho/\rho_0$] \\ \hline
0,02 &   $8,6$   &  5 & 1013,8 &  1,00 \\
0,50 &   $6,8$   &  8 &  956,5 &  0,95 \\
1,00 &   $4,7$   &  9 &  899,1 &  0,90 \\
2,00 &   $0,0$   & 10 &  795,1 &  0,81 \\
3,00 &   $-5,1$  & 12 &  700,4 &  0,73 \\
4,00 &   $-10,7$ & 14 &  616,3 &  0,65 \\
5,00 &   $-16,8$ & 17 &  540,3 &  0,59 \\
6,00 &   $-23,2$ & 20 &  473,6 &  0,53 \\
7,00 &   $-30,0$ & 23 &  410,9 &  0,47 \\
8,00 &   $-37,0$ & 26 &  354,8 &  0,42 \\
9,00 &   $-44,2$ & 29 &  306,8 &  0,37 \\
10,00 &   $-51,6$ & 32 &  264,1 &  0,33 \\ \hline
\end{tabular}
\end{center}
\caption[Atmosphärendaten]{ Mittlere Daten der Atmosphäre als Funktion der Höhe nach \cite{Broeckelmann1909}.\label{AtmData}}
\end{table}

\end{appendix}

\newpage
\nocite{*}
\addcontentsline{toc}{chapter}{Literaturverzeichnis}

\bibliography{WorkB}
\bibliographystyle{plain}

\end{document}